\documentclass[a4paper,11pt]{article}
\usepackage{heppub}
\pdfoutput=1
\usepackage{tikz}

\tikzset{
    position/.style args={#1:#2 from #3}{
        at=(#3.#1), anchor=#1+180, shift=(#1:#2)
    }
}
\usepackage{bm}
\usepackage{graphicx}
\usepackage{subcaption}
\usepackage{hyperref}
\hypersetup{
    linktocpage,
     colorlinks,
     citecolor=darkgreen,
     linkcolor= darkgreen,
     urlcolor=darkgreen
}
\definecolor{col1}{rgb}{0.81, 0.85, 0.91}
\definecolor{col2}{rgb}{0.96, 0.88, 0.74}
\definecolor{col3}{rgb}{0.87, 0.91, 0.76}
\definecolor{col4}{rgb}{0.98, 0.82, 0.76}
\definecolor{col5}{rgb}{0.86, 0.84, 0.91}
\definecolor{col6}{rgb}{0.93, 0.83, 0.73}

\newcommand\evenoverset[2]{\overset{\scriptstyle #1\mathstrut}{\scriptstyle #2}}

\usepackage{xspace}
\usepackage{placeins}
\usepackage{comment}
\usepackage{wasysym}
\usepackage{slashed}
\usepackage{empheq}
\usepackage{marginnote}
\usepackage{xcolor}
\usepackage[normalem]{ulem}
\usepackage{amsmath}
\usepackage{mathrsfs}
\nopagebreak

\newcommand{\abs}[1]{\lvert#1\rvert}

\newcommand{\ord}[1]{\mathcal{O}(#1)}

\newcommand{\ahat} {\hat{\alpha}}
\newcommand{\nn}{\nonumber}

\newcommand{\bn}{{\bar n}}

\newcommand{\sjt}{ \widetilde{\phantom{\,}sj} }

\newcommand{\etal}{{{\eta_{\ell}}}}
\newcommand{\etah}{{{\eta_{h}}}}

\newcommand{\as}{\alpha_s}

\newcommand{\GeV}{\,\mathrm{GeV}}

\newcommand{\can}{\text{can}}
\newcommand{\pro}{\text{pro}}

\newcommand{\cusp}{\mathrm{cusp}}

\newcommand{\re}{\text{Re}}

\allowdisplaybreaks[3]

\definecolor{darkyellow}{rgb}{0.5, 0.5, 0.0}
\definecolor{darkpurple}{rgb}{0.5, 0.2, 0.8}
\definecolor{darkblue}{rgb}{0.0, 0.0, 0.8}
\definecolor{darkgreen}{rgb}{0.0, 0.4, 0.0}
\definecolor{darkred}{rgb}{0.5, 0.0, 0.0}

\title{\boldmath NNLL Resummation of Sudakov Shoulder Logarithms in the Heavy Jet Mass Distribution}

\author[a]{Arindam Bhattacharya,}
\emailAdd{arindamb@g.harvard.edu}

\author[b]{Johannes K.~L.~Michel,}
\emailAdd{jklmich@mit.edu}

\author[a]{Matthew D.~Schwartz,}
\emailAdd{schwartz@g.harvard.edu}

\author[b]{\\Iain W.~Stewart,}
\emailAdd{iains@mit.edu}

\author[a]{and Xiaoyuan Zhang}
\emailAdd{xiaoyuanzhang@g.harvard.edu}

\affiliation[a]{Department of Physics, Harvard University, Cambridge, MA 02138, USA}
\affiliation[b]{Center for Theoretical Physics,\,Massachusetts Institute of Technology,\,Cambridge,\,MA\,02139,\,USA}

\abstract{%
The heavy jet mass event shape has large perturbative logarithms near the leading order kinematic threshold at $\rho = \frac{1}{3}$. Catani and Webber named these logarithms Sudakov shoulders and resummed them at double-logarithmic level. A resummation to next-to-leading logarithmic level was achieved recently. Here, we extend the resummation using an effective field theory framework to next-to-next-to-leading logarithmic order and show how to combine it with the resummation of dijet logarithms. We also solve the open problem of an unphysical singularity in the resummed momentum space distribution, in a way similar to how it is resolved in the Drell-Yan $q_T$ spectrum:  through a careful analysis of the kinematics and scale-setting in position space.
The heavy jet mass Sudakov shoulder is the first observable that does not involve transverse momentum for which position space resummation is critical.
These advances may lead to a more precise extraction of the strong coupling constant from $e^+ e^-$ data. 
}

\date{June 13, 2023}

\preprint{\vbox{%
\hbox{MIT-CTP 5570}
}
}

\begin{document}

\maketitle
\newpage

\section{Introduction}
\label{sec:intro}

Event shapes in electron-positron collisions have long been valuable
testing grounds for quantum field theory and quantum
chromodynamics. High quality event shape data predominantly from the Large
Electron-Positron Collider (LEP) have allowed for a detailed comparison
between theory and experiment and engendered improvements on both fronts. On
the theory side, inspiration from data has lead to higher-precision fixed-order
computations, up to the current state of the art of NNLO ($\alpha_s^3$)~\cite{Gehrmann-DeRidder:2007nzq, Gehrmann-DeRidder:2007vsv, Weinzierl:2008iv,  Weinzierl:2009ms, DelDuca:2016csb, DelDuca:2016ily}. Data has also inspired exponential progress in resummation,
up to N$^3$LL precision using Soft-Collinear Effective
Theory (SCET)~\cite{Bauer:2003pi,Bauer:2000yr,Bauer:2001ct,Bauer:2001yt,Bauer:2002nz} in the dijet region for thrust, heavy jet mass, and the $C$
parameter~\cite{Schwartz:2007ib,Becher:2008cf,Chien:2010kc,Hoang:2014wka}, and NNLL$^\prime$ precision on other event shapes~\cite{Bell:2018gce}. 
The comparison between theory and data has also motivated improvements in the treatment of non-perturbative corrections from hadronization, including
using matrix elements defined in quantum field theory~\cite{Korchemsky:1999kt,Hoang:2007vb,Abbate:2010xh}, an improved understanding of universality relations~\cite{Dokshitzer:1995zt,Lee:2006nr}, and the inclusion of hadron mass corrections~\cite{Salam:2001bd,Mateu:2012nk}. 

Until recently, however, there has been very little understanding of the trijet region, with 3 hard jets, in $e^+ e^-$ events. This region has traditionally been
assumed to be well-modeled in fixed-order perturbation theory, since it is well-separated from the dijet region where large threshold logarithms spoil the convergence of the perturbation expansion in $\alpha_s$. However, fixed-order perturbation theory  fails
spectacularly right at the 3-jet threshold where it predicts kinks and
discontinuities. 
The failure of perturbation theory in the trijet region was
understood by Catani and Webber~\cite{Catani:1997xc}, and also Seymour~\cite{Seymour:1997kj}, as due to the appearance of
Sudakov shoulder logarithms. Although the resummation of the Sudakov shoulder logarithms at the double-logarithmic level was achieved analytically for the $C$ parameter, little additional progress was made in the last 25 years. Recently, Sudakov shoulder resummation was revisited using SCET where a factorization theorem was derived and next-to-leading
logarithmic (NLL) resummation was achieved for thrust and heavy jet mass~\cite{Bhattacharya:2022dtm}. In addition, there has
been recent progress in understanding power corrections near the 3-jet region using 
renormalon methods~\cite{Caola:2021kzt,Caola:2022vea} and the dispersive approach~\cite{Nason:2023asn}.
In this paper, we focus on understanding and
improving perturbative Sudakov shoulder resummation for heavy jet mass. Heavy jet mass is special in that it has a Sudakov shoulder on the left of the trijet threshold, so the shoulder logarithms propagate into the region usually used for fits to $\alpha_s$. In contrast, thrust only has a right shoulder, so that large shoulder logarithms do not impact the region used for such fits.

Thrust is defined in the center-of-mass frame of an $e^+ e^-$ collision as
\begin{equation}
  T \equiv \max_{\vec{n}} \frac{\sum_j | \vec{p}_j \cdot \vec{n} |}{\sum_j |
  \vec{p}_j |}
  \label{tdef}
  \,,
\end{equation}
where the sum is over all particles in the event and the maximum is over
3-vectors $\vec{n}$ of unit norm. The vector $\vec{n}$ that maximizes thrust is known as the {\it thrust axis}. The thrust axis splits the event into two hemispheres. One can then add all the momenta in each hemisphere and compute the hemisphere invariant masses $M_L$ and $M_R$. Heavy jet mass is defined as 
\begin{equation}
    \rho = \frac{1}{Q^2} \max(M_L^2, M_R^2)
   \,.
\end{equation}
Conventionally one also defines $\tau = 1-T$. So at zeroth order with two massless partons $\tau=\rho=0$. We call the region near $\tau=\rho=0$ the dijet region. At leading order in $\alpha_s$ the thrust and heavy jet mass distributions are the same (for massless particles) with a kinematic maximum at $\tau=\rho=\frac{1}{3}$. 

Both thrust and heavy jet mass have non-analytic behavior near $\tau=\rho=\frac{1}{3}$. To be concrete, the leading
non-analytic behavior of thrust near $\tau = \frac{1}{3}$ at NLO was computed
in~\cite{Bhattacharya:2022dtm} and found to be in
\begin{equation}
  \frac{1}{\sigma_{0}} \frac{d \sigma}{d \tau} \approx
  \frac{\alpha_s}{4 \pi} 48 C_F \,  
  \left[ 3 - \left(\frac{\alpha_s}{4 \pi}\right)
  \frac{3}{2} (2 C_F + C_A) \Gamma_0 \ln^2 t +
  \cdots \right] t\theta(t) 
  \label{tnear13}
 \,,
\end{equation}
where $t \equiv \tau - \frac{1}{3}$, with $\sigma_0$ denoting the Born cross section. Thrust only has a right shoulder: its behavior for $\tau \lesssim
\frac{1}{3}$ is smooth. To get the full leading order thrust distribution in this region one must add to Eq.~\eqref{tnear13} the analytic linear function $-\frac{\alpha_s}{4 \pi} 144 C_F t$ which effectively turns  $t \theta(t)$ into $(-t)\theta(-t)$ with support only for $\tau < \frac{1}{3}$. 
For heavy jet mass
near $\rho = \frac{1}{3}$ we define $r = \frac{1}{3} - \rho$. The leading non-analytic behavior of its distribution near $r = 0$ is
\begin{align}
  \frac{1}{\sigma_{0}} \frac{d \sigma}{d\rho}  \approx \frac{\alpha_s}{4 \pi} \,48C_F\, 
  \Bigl[ 3 &-  \Bigl(\frac{\alpha_s}{4 \pi}\Bigr)\frac{1}{2} (2 C_F + C_A) \Gamma_0  \ln^2 r + \cdots \Bigr]\,r\, \theta(r)
  \nn \\
&+ \Bigl(\frac{\alpha_s}{4 \pi}\Bigr)^2 \, 48 C_F \Bigl[  -(2 C_F + C_A) \Gamma_0 \ln^2 (- r) + \cdots \Bigr](-r)\,\theta(-r)
  \,.
\end{align}
Heavy jet mass has large logarithms on both sides of $\rho=\frac{1}{3}$, giving it a left $(r > 0)$ and right ($r < 0$) Sudakov shoulder.

Physically, the origin of the large logarithms can be understood as follows.
At leading order (three partons) there is only a single phase space point with
$\rho = \frac{1}{3}$, namely the ``trijet configuration'' where the partons
have energy $\frac{Q}{3}$ and are spaced $\frac{2 \pi}{3}$ apart; two partons are clustered into
one hemisphere and one parton remains in the other. The thrust axis can be aligned with
any of the partons. As these three partons emit soft and collinear gluons, the
cross section becomes logarithmically enhanced. These soft and collinear
logarithms in the cross section in the trijet region become Sudakov shoulder logarithms after the thrust or heavy jet mass measurement function is applied.

A careful analysis of soft and collinear kinematics near the trijet region
leads to a factorization formula for Sudakov shoulders. We can treat collinear
emissions as transforming the three massless partons into jets with masses $M_{1,2,3}$. The thrust axis will
align with one of the jets, but which one it aligns with depends on the
exact kinematics. If $M_1$ is the mass of the jet in
the light hemisphere, and $M_2$ and $M_3$ are the masses of the jets in the heavy hemisphere, then, defining $m_i = \frac{M_i}{Q}$, the bound is
\begin{equation}
  m_1^2 < r + m_2^2 + m_3^2
  \,.
\end{equation}
As $m_1$ gets larger or as $m_2$ or $m_3$ gets smaller, the thrust axis flips and a different
parton is in the light hemisphere. Soft contributions can be analyzed
similarly. The soft and collinear emissions into the light and heavy
hemispheres can be grouped into effective light-hemisphere and heavy-hemisphere mass-shift variables $m_{\ell}^2 \approx m^2_1$
and $m_h^2 \approx m_2^2 + m_3^2$ and the factorization formula for heavy jet
mass written as
\begin{equation} \label{massshift}
  \frac{d \sigma}{d \rho} = \int_0^{\infty} d m_h^2  \int_0^{\infty} d m_{\ell}^2
  \frac{d^2 \sigma}{d m_{\ell}^2 d m_h^2} (r + m_h^2 - m_{\ell}^2) \theta (r +  m_h^2 - m_{\ell}^2) \,.
\end{equation}
The double differential distribution $\frac{d^2 \sigma}{d m_{\ell}^2 d m_h^2}$ can then be resummed using standard
techniques, see~\cite{Fleming:2007xt}.

A key feature of the heavy jet mass Sudakov shoulder factorization formula in Eq.~\eqref{massshift} that
distinguishes it from typical event shape factorization formulas is the relative
minus sign between $m_h^2$ and $m_{\ell}^2$. In particular, for small $r$, the
region where $m_{\ell}^2 \gg 1$ and $m_h^2 \gg 1$ contributes, even though
this region is not described by soft and collinear physics. Such contributions
could generically lead to non-global logarithms, or worse, a failure of the
factorization theorem to describe the threshold behavior. However, in this
case, as explained in~\cite{Bhattacharya:2022dtm}, we still expect all the shoulder logs to be
described correctly. The reason is that when $m_h$ and $m_{\ell}$ are large,
the point $r = 0$ is not special: contributions to this region should go
smoothly from $r < 0$ to $r > 0$. The kinks and large logarithms in the
shoulder region emerge only when $m_h$ and $m_{\ell}$ are small so that the
lower bounds $m_h > 0$ and $m_{\ell} > 0$ become relevant. Despite this, when one uses standard resummation methods
with canonical scale choices, 
spurious ``Sudakov Landau pole'' singularities result. Similar poles have been seen also in the small $q_T$
Drell-Yan spectrum \cite{Frixione:1998dw,Chiu:2012ir,Becher:2010tm,Becher:2011xn,Monni:2016ktx,Ebert:2016gcn}.
A main result of this paper is a demonstration that these Sudakov
Landau poles are avoided by carrying out the resummation with canonical scales in position space. Additional results are the extension of the Sudakov shoulder resummation to NNLL, and smoothly combining the dijet resummation, shoulder resummation, and fixed-order corrections for the heavy jet mass. 

An outline of the paper is as follows. In Section~\ref{sec:fact}, we review the factorization theorem for the Sudakov shoulder region, and describe the necessary ingredients for NNLL resummation. Following that, in Section~\ref{sec:scale_setting}, we review the appearance of spurious Sudakov Landau poles when setting scales in momentum ($r$) space, and address how these can be carefully removed by setting scales in position (Fourier) space. In Section~\ref{sec:integration_and_matching}, we describe our resummation procedure for Sudakov shoulder logs, and set up a matching procedure to account for dijet logs as well, thereby forming a consistent picture of resummation across both the dijet ($\rho \gtrsim 0$) and trijet region ($\rho\sim 1/3$) in heavy jet mass. In Section~\ref{sec:numerics}, we provide our numerical results which combine NNLL shoulder resummation and NNLL$^{\prime}$ dijet resummation matched to NLO.
Finally, we conclude in Section~\ref{sec:conclusions}.

\section{Factorization formula \label{sec:fact}}

In this section we review the factorization formula from~\cite{Bhattacharya:2022dtm}.
In the trijet limit, there are three independent channels associated with a gluon jet in the light hemisphere, a quark jet in the light hemisphere or an anti-quark jet in the light hemisphere. The contribution of these channels to the total cross section add incoherently, leading to a form
\begin{equation}
    \frac{d\sigma_\text{sh}}{d\rho} =\sum_{i}\frac{d\sigma_i}{d\rho} = \frac{d\sigma_g}{d\rho} + 2 
    \frac{d\sigma_q}{d\rho} 
  \,,
\end{equation}
where we have used symmetry under charge-conjugation to identify the quark and antiquark channels. In our notation, when $\sigma$ has a $g$ or $q$ subscript it automatically identifies it as a leading power factorized contribution near the shoulder.

\subsection{Factorization formula}

We first focus on the gluon channel for concreteness. In~\cite{Bhattacharya:2022dtm}, it was shown that the factorization formula for heavy jet mass in the shoulder region has the form 
\begin{align}
    \frac{1}{\sigma_{\text{LO}}}\frac{d \sigma_g}{d \rho}
  &=  \Pi_g\,
    \int_0^\infty d m_\ell^2 \int_0^\infty  d m_h^2\,
    \left.
    \sjt_g \cdot
    \left( \frac{Q m_l^2}{\mu_{s\ell}} \right)^{a}
    \left( \frac{Q m_h^2}{\mu_{sh}} \right)^{b}\frac{1}{m_l^2 m_h^2}  
    \frac{e^{-\gamma_E (a+b)}}{\Gamma(a) \Gamma(b)}  \right|_{\evenoverset{a=\eta_\ell}{b=\eta_h}}
    \label{dsdrform}
    \nn \\
&\qquad \times (r + m_h^2 - m_{\ell}^2) \theta (r +
  m_h^2 - m_{\ell}^2)  
 \,,
\end{align}
with $\sigma_{\text{LO}}=\frac{\alpha_s}{4\pi}48 C_F \sigma_0$, where $\sigma_0$ is the Born cross-section for $e^+e^-\to q\bar q$. 
Here $m_h^2$ and $m_\ell^2$ are the (dimensionless) mass-shift variables, representing the change in the heavy or light jet masses-squared due to soft and collinear emissions relative to $Q^2$;
the evolution kernel is,
\begin{align}
   \Pi_g
    &=  \exp \Big[ 
    4 C_F \left(S (\mu_h, \mu_{jh} )
    + S (\mu_{sh}, \mu_{jh})\right)
    + 2 C_A \left(S (\mu_h, \mu_{j\ell}) 
    +  S (\mu_{sl}, \mu_{j\ell})\right)\Big]
    \nn\\
   &\quad 
   \times
    \exp \Big[ 
    2 A_{\gamma sg} (\mu_{s\ell}, \mu_h) 
    + 2 A_{\gamma sqq} (\mu_{sh}, \mu_h) 
    + 2 A_{\gamma j g} (\mu_{j\ell}, \mu_h) 
    + 4 A_{\gamma j q} (\mu_{jh}, \mu_h) \Big]
   \nn\\
   &\quad \times H (Q, \mu_h)   \left(\frac{Q^2}{\mu_h^2}\right)^{-2C_F A_\Gamma (\mu_h,\mu_{jh})-C_A A_\Gamma (\mu_h,\mu_{j\ell})} \label{Pig}
  \,,
\end{align}
the fixed-order matching operator is
\begin{equation}
\sjt_g =     \widetilde{j}_q\left( \partial_b + \ln \frac{Q \mu_{sh}}{\mu_{jh}^2} \right) 
     \widetilde{j}_{\bar{q}} \left( \partial_b+ \ln \frac{Q \mu_{sh}}{\mu_{jh}^2} \right) \widetilde{j}_g \left( \partial_a + \ln \frac{Q \mu_{s\ell}}{\mu_{j\ell}^2} \right) \widetilde{s}_{q \bar{q};g} (\partial_a,\partial_b) 
  \,,
\end{equation}
and after taking derivatives, $a$ and $b$ are set to
\begin{equation}
a = \eta_\ell = 2 C_A A_{\Gamma} (\mu_{j\ell}, \mu_{s\ell}),\qquad
b = \eta_h= 4 C_F A_{\Gamma} (\mu_{jh}, \mu_{sh})
  \,.
\end{equation}
Here we have allowed for different jet and soft factorization scales in the 
heavy hemisphere ($\mu_{jh}$, $\mu_{sh}$) and light hemisphere ($\mu_{j\ell}$, $\mu_{s\ell}$),
slightly generalizing the expression in~\cite{Bhattacharya:2022dtm}.
As we discuss in Section~\ref{sec:complexscales} below, this separation is not sensible beyond NLL, and ultimately we will set $\mu_{s\ell}$ and  $\mu_{s h}$ to a common scale  $\mu_s$.
For the quark channel the evolution kernel $\Pi_g$ is replaced by
\begin{align}
&\Pi_q=\exp \Big[ 
    (2C_F+2C_A) \left(S (\mu_h, \mu_{jh} )
    + S (\mu_{sh}, \mu_{jh})\right)
    + 2 C_F \left(S (\mu_h, \mu_{j\ell}) 
    +  S (\mu_{sl}, \mu_{j\ell})\right)\Big] 
  \nn \\
 & \times
    \exp \Big[ 
    2 A_{\gamma sq} (\mu_{s\ell}, \mu_h) 
    + 2 A_{\gamma sqg} (\mu_{sh}, \mu_h) 
    + 2 A_{\gamma j q} (\mu_{j\ell}, \mu_h) 
    + 2 A_{\gamma j q} (\mu_{jh}, \mu_h)+ 2 A_{\gamma j g} (\mu_{jh}, \mu_h) \Big]
  \nn \\
  &   \times H (Q, \mu_h)   \left(\frac{Q^2}{\mu_h^2}\right)^{-(C_F+C_A) A_\Gamma (\mu_h,\mu_{jh})-C_F A_\Gamma (\mu_h,\mu_{j\ell})} 
  \,,
\end{align}
and $\sjt_g$ is replaced by
\begin{equation}
 \sjt_q= \widetilde{j}_q \left( \partial_b + \ln \frac{Q \mu_{sh}}{\mu_{jh}^2} \right) 
    \widetilde{j}_{g} \left( \partial_b+ \ln \frac{Q \mu_{sh}}{\mu_{jh}^2} \right)
    \widetilde{j}_q \left( \partial_a + \ln \frac{Q \mu_{s\ell}}{\mu_{j\ell}^2} \right)  
    \widetilde{s}_{q g; q} (\partial_a, \partial_b)
   \,,
\end{equation}
with
\begin{equation}
    a =\eta_\ell = 2 C_F A_{\Gamma} (\mu_{j\ell}, \mu_{s\ell}),\qquad
b =\eta_h = 2(C_F+C_A) A_{\Gamma} (\mu_{jh}, \mu_{sh})
 \,.
\end{equation}
For both channels, the resummation is expressed in terms 
of~\cite{Neubert:2004dd} 
\begin{equation}
	S_\Gamma(\nu,\mu)=-\int_{\alpha_s(\nu)}^{\alpha_s(\mu)} d\alpha\ \frac{\Gamma_{\text{cusp}}(\alpha)}{\beta(\alpha)}\int_{\alpha_s(\nu)}^{\alpha} \frac{d\alpha^\prime}{\beta(\alpha^\prime)}
 \,,
 \end{equation}
 and
 \begin{align}
	A_\Gamma(\nu,\mu)&=-\int_{\alpha_s(\nu)}^{\alpha_s(\mu)} d\alpha\ \frac{\Gamma_{\text{cusp}}(\alpha)}{\beta(\alpha)},\quad & 
	A_{\gamma}(\nu,\mu)&=-\int_{\alpha_s(\nu)}^{\alpha_s(\mu)} d\alpha\ \frac{\gamma(\alpha)}{\beta(\alpha)}
 \label{threedown}
  \,.
\end{align}

The heavy jet mass distribution near the shoulder has the form 
$\frac{d\sigma}{d\rho} = \sum c_{n,m} \alpha_s^n r \ln^m r$. 
Because the integrals over $m_{\ell}$ and $m_{h}$ in Eq.~\eqref{dsdrform} are UV divergent
(as discussed in~\cite{Bhattacharya:2022dtm}),
we 
consider first the third derivative $\frac{d^3 \sigma}{ d \rho^3}$. Taking an extra two derivatives turns the measurement function
$(r + m_h^2 - m_{\ell}^2) \theta (r +
  m_h^2 - m_{\ell}^2)$
  in Eq.~\eqref{dsdrform}
into a simpler $\delta(r  + m_h^2 - m_{\ell}^2)$.
Using additional boundary conditions obtained from fixed-order matching,
this third derivative can be integrated back up to the physical spectrum,
as we will demonstrate in Section \ref{sec:integration}.
The expansion of $\frac{d^3 \sigma}{ d \rho^3}$ near the shoulder
yields a series in distributions of the form
$\left[\frac{\ln^n r}{r}\right]_+$,
analogous to how the expansion of
$\frac{d \sigma}{d\rho}$ in the dijet limit yields a series  of distributions of the form
$\left[\frac{\ln^n \rho}{\rho}\right]_+$.
The resummed form of $\frac{d^3 \sigma}{ d \rho^3}$ is
\begin{align}
    \frac{1}{\sigma_{\text{LO}}}\frac{d^3 \sigma_g}{d \rho^3} 
&= 
    \Pi_g\int_0^\infty d m_\ell^2 \int_0^\infty  d m_h^2\ 
     \sjt_g \cdot
    \left( \frac{Q m_\ell^2}{\mu_{s\ell}} \right)^{\etal}
    \left( \frac{Q m_h^2}{\mu_{sh}} \right)^\etah \frac{1}{m_\ell^2 m_h^2}
    \frac{e^{-\gamma_E (\eta_\ell +\eta_h)}}{\Gamma(\eta_\ell) \Gamma(\eta_h)}
     \nn \\
 &\qquad \times
\delta (r +  m_h^2 - m_{\ell}^2)  \,,
\label{factform3d}
\end{align}
for the gluon channel, and similarly for the quark channel.

\subsection{NNLL resummation \label{sec:nnll}}

For NNLL resummation, we need the three-loop cusp anomalous dimension,
the three-loop $\beta$-function coefficients, the two-loop regular (non-cusp) anomalous dimensions for the hard, jet and soft functions, and the one-loop finite parts of the hard, jet and soft functions.
The two-loop gluon and quark jet functions~\cite{Becher:2006qw,Becher:2010pd} are the same as for other processes, see e.g.~\cite{Becher:2009th, Jouttenus:2011wh}, and we summarize all perturbative ingredients in Appendix~\ref{sec:anom_dim}. 

 The Sudakov shoulder hard function is obtained by matching the QCD vector 
current onto the corresponding SCET current operator. The virtual diagrams in SCET are all scaleless in dimensional regularization, so the hard function can be extracted from the virtual contribution in QCD. For the case of matching onto three hard partons in the final state, the QCD results can be found in~\cite{Ellis:1980wv}. The result is summarized in Appendix~\ref{sec:anom_dim}.  The hard function 
 is the same up to crossing as for any $q\bar{q}g \gamma^\star$ process, such as direct photon production~\cite{Becher:2009th} or 3-jettiness~\cite{Jouttenus:2011wh}, restricted to the trijet configuration. We have checked the hard function against these results as well. 

The soft function for the heavy jet mass Sudakov shoulder is defined as a matrix element of Wilson lines in the trijet configuration. The soft function involves collecting emissions into 6 separate sextant regions shaped like wedges of a 6-slice orange. Then the 
soft gluons in the sextants within the light or heavy hemispheres are combined separately with appropriate projections. This leads to a two-argument soft function, which in momentum space is defined as
\begin{equation}
    S_g(k_\ell, k_h)= \frac{1}{N C_F} \sum_{X_s} \bigg\vert\langle X_s \vert \mathbf{Tr}(Y^\dagger_{n_2}Y_{n_1}T^a Y_{n_1}^\dagger Y_{n_3})\vert 0 \rangle\bigg\vert^2\ \hat{\mathscr{M}}(k_\ell,k_h)\ .
    \label{Skkdef}
\end{equation}
Here, $Y_n$ denotes a path-ordered exponential of gauge fields (a Wilson line) extending from $x=0$ to $x=\infty$ along the $n^\mu$ direction.
As written, all of these Wilson lines are in the fundamental representation of $SU(3)$. We conventionally choose $n_1^\mu$ to point along the thrust axis towards the light hemisphere
so that Eq.~\eqref{Skkdef} corresponds to a soft function with a gluon jet in the light hemisphere. The other channels arise from permuting the Wilson line indices. For heavy jet mass, the trijet measurement function $\hat{\mathscr{M}}(k_\ell,k_h)$ is defined as
\begin{align}
&\hat{\mathscr{M}}(k_\ell,k_h)
 \\
 &\ =\delta\left(k_\ell -\sum_{k_i \in \vert X_s\rangle }\left[\theta_{\mathbf{1}}(k_i)\left(\frac{2}{3}n_1\cdot k_{i}\right) +\theta_{\mathbf{\bar{2}}}(k_i)\left(\frac{2}{3}N_2\cdot k_{i}\right)+\theta_{\mathbf{\bar{3}}}(k_i)\left(\frac{2}{3}N_3\cdot k_{i}\right)\right]\right)
 \nn\\
&\ \ \times \delta\left(k_h -\sum_{k_m \in \vert X_s\rangle }\left[\theta_{\mathbf{\bar{1}}}(k_m)\left(\frac{2}{3}\bar{n}_1\cdot k_{m}\right) +\theta_{\mathbf{2}}(k_m)\left(\frac{2}{3}n_2\cdot k_{m}\right)+\theta_{\mathbf{3}}(k_m)\left(\frac{2}{3}n_3\cdot k_{m}\right)\right]\right)
 \nn .
\end{align}
where the sum runs over the momenta of all particles in the state $|X_s\rangle$. The notation $\theta_\mathbf{1}(k)$ refers to a projection onto the sextant wedge centered on the direction $n_1^\mu$ (see~\cite{Bhattacharya:2022dtm}), i.e. $\theta_\mathbf{1}(k) =\theta(n_2\cdot k- \bar{n}_2 \cdot k )\theta(n_3\cdot k- \bar{n}_3 \cdot k)$, and analogously for $\theta_\mathbf{2}(k)$ and $\theta_\mathbf{3}(k)$. Similarly, $\theta_\mathbf{\bar{1}}$ refers to a projection onto the wedge in the direction $\bar{n}_1^\mu$, which is opposite to the one in the direction $n_1^\mu$, $\theta_{\overline{\mathbf{1}}}(k) = \theta(\bar{n}_2\cdot k - n_2\cdot k) \theta(\bar{n}_3 \cdot k - n_3 \cdot k)$, and similarly for $\theta_\mathbf{\bar{2}}(k)$ 
and $\theta_\mathbf{\bar{3}}(k)$. 

In each region only one component of the momentum $k_i$ is retained. Among the six projections, four of them are $\frac{2}{3} n \cdot k$ with $n^\mu$ a lightlike vector pointing to the center of the sextant. However, for the two sextants adjacent to the light-hemisphere direction $n_1^\mu$, the projections involve two spacelike directions
\begin{equation}\label{eq:capN}
    N_2^\mu=\bn_3^\mu +\frac{1}{2}(n_1^\mu-\bn_1^\mu),\quad  N_3^\mu=\bn_2^\mu+\frac{1}{2}(n_1^\mu-\bn_1^\mu)
  \,.
\end{equation}

Ref.~\cite{Bhattacharya:2022dtm} computed the one-loop soft anomalous dimension needed for NLL resummation. The complete one-loop soft function including the soft constant is calculated in  Appendix \ref{sec:appendix_soft_calc}.
For NNLL resummation, we also need the 2-loop regular soft anomalous dimension $\gamma_s$. 
This can be fixed by RG invariance and is expressible in terms of the hard and jet anomalous dimensions. The result is also summarized in Appendix~\ref{sec:appendix_soft_calc}.

\section{Scale setting} \label{sec:scale_setting}
In this Section, we show that by choosing scales in position space, the Sudakov Landau pole is avoided. We also explore reasons why this should be true.

\subsection{Momentum space vs.\ position space}
One can do the integrals over $m_h^2$ and $m_\ell^2$ exactly before setting scales. The relevant integral has the form
\begin{align}
f (r) & \equiv \frac{1}{\Gamma (a)} \frac{1}{\Gamma (b)} \int_0^{\infty} d x
\int_0^{\infty} d y\ x^{a - 1} y^{b - 1} \delta (r + y - x)
 \nn \\
&= \frac{1}{\Gamma (a + b)} \left[r^{a + b - 1} \frac{\sin (\pi a)}{\sin (\pi
(a + b))} \theta (r) + (-r)^{a + b - 1}\frac{\sin (\pi b)}{\sin (\pi (a + b))} 
\theta (- r) \right] \label{frmom}
 \,,
\end{align}
where we shorten notation by denoting $x\equiv m_\ell^2\ ,\ y\equiv m_h^2$.
This expression has poles at $a+b \in {\mathbb Z}$. The pole at $a+b=1$ is the Sudakov Landau pole frustrating the numerical resummation of the Sudakov shoulder logarithms. It comes from the region where $x,y \gg 1$ and $x-y=r$.

Although the function $f(r)$ in Eq.\eqref{frmom} seems complicated, its Fourier transform is not.
It is easiest to compute the Fourier transform before integrating over $x$ and $y$, doing the integrals in a different order in the $r>0$ and $r<0$ region:
\begin{align}
  \tilde{f} (z) &= \frac{1}{\Gamma (a)} \frac{1}{\Gamma (b)} \left[ \mathcal{}
  \int_0^{\infty} d r\ e^{i z r} \int_0^{\infty} d y\ (r + y)^{a - 1} y^{b - 1}
  + \int_{- \infty}^0 d r\ e^{i z r}  \int d x\ x^{a - 1} (x - r)^{b - 1}
  \right]
  \nn\\
  &= \frac{\Gamma (1 - a - b)}{\pi} \left[ \mathcal{} \sin (\pi a)
  \int_0^{\infty} d r\ e^{i z r} \mathcal{} r^{- 1 + a + b} + \sin (\pi b)
  \int_{- \infty}^0 d r\ e^{i z r}  (- r)^{- 1 + a + b} \right]
 \nn\\
  &= (- i z)^{- a - b} \frac{\sin (\pi a)}{\sin (\pi (a + b))} + (i z)^{- a -b} \frac{\sin (\pi b)}{\sin (\pi (a + b))}
  \nn\\
  &= (-i z)^{- a} (i z)^{- b} \label{eq:fspace_dist}
  \,.
\end{align}
So the Fourier transform is {\it remarkably} simple. 

The  simple form of the Fourier transform belies the complicated
non-analytic behavior of $(-i z)^{- a} ( i z)^{- b}$. In taking the inverse
Fourier transform, one must carefully adjust the contour to avoid the
singularity at $z = 0$. For example, to take the Fourier transform of a single power $z^{-c}$ we
must adjust the contour above or below the branch point. Then closing the
contour either encircles the branch point or gives zero, depending on the sign
of the argument. The result is
\begin{equation}
  \int_{- \infty}^{\infty} d z\ e^{-i z r} (z \pm i \varepsilon)^{- c} =
  \pm 2 \pi i \frac{(i r)^{c - 1}}{\Gamma (c)} \theta (\pm r) 
  \,.
\end{equation}
In a similar manner, the branch point in the position space function $(-i z)^{- a} (i z)^{- b}$ is ultimately responsible for the complicated piecewise form in Eq.~\eqref{frmom}.%
\footnote{
As an aside, it is worth noting that the result of applying the commonly used one-sided Laplace transform to $f(r)$ is not so simple. The Laplace transform is defined with a semi-infinite integration $\int_0^\infty d r e^{-\nu r}$ and so for fixed $\nu$ cannot be convergent for both positive and negative $r$. The best one can do is flip the sign of the exponent by hand between the $r > 0$ and $r < 0$ regions leading to
\begin{equation}
  \mathcal{L} [f](\nu) =  \int_0^{\infty} d r\ e^{- \nu r} f (r) + \int_{-
  \infty}^0 d r\ e^{\nu r} f (r) = \nu^{- a - b} \frac{\sin (\pi a) + \sin (\pi
  b)}{\sin (\pi (a + b))}
  \,.
\end{equation}
This form is not simple at all, and moreover still has the Sudakov Landau pole singularities at $a + b \in \mathbb{Z}$. 
}

The pole at $a + b = 1$ comes from the large $y$ region:
\begin{equation}
  f (r) \approx \int_0^{\Lambda} d y\ y^{a + b - 2} = \frac{\Lambda^{a + b - 1}}{a +
  b - 1}
  \,.
\end{equation}
This is the region of large jet mass shifts $m_\ell^2 = x \gg 1$ and $m_h^2 = y \gg 1$ where their difference $r=x-y$ is small.
If one sets scales in momentum space, so $a\sim b \sim \alpha
_s \ln r$ then the pole at $a+b=1$  corresponds to one value of $r$.
Large $x$ and $y$ in momentum space corresponds to small $|z|$ in position space. For small $|z|$ the integral is approximately
\begin{equation}
  f (r) \approx \int \frac{d z}{2 \pi} (-i z)^{- a} (i z)^{- b} 
  \,.
\end{equation}
This would indeed be singular for $a+b=1$ at small $|z|$. However, when scales are set in position space, so $a \sim b \sim \alpha_s \ln |z|$, then $a$ and $b$ change over the integration region, in contrast to momentum space scale setting where $a$ and $b$ are fixed. Moreover, the region where $a+b=1$ does not correspond to small $z$. In this way, the singular region is simply avoided if one sets
scales in position space.

\subsection{Scale setting}
Above we showed how setting scales in position space should avoid the Sudakov Landau poles. Now we will explore the differences between momentum and position space-setting in more detail. 

The idea behind resummation is to include systematically terms in the perturbation expansion of the form $c_{n,m} \alpha ^n L^m$ for some infinite series of $c_{n,m}$ where $L$ is a relevant logarithm and $\alpha$ the relevant coupling. It can be helpful to distinguish logarithmic counting schemes.
All these schemes agree to some order in resummed perturbation theory and differ only by higher-order terms. However, these higher-order terms can be important, and in this case, are the source of the spurious Sudakov Landau poles as we will see.

A common scheme counts logs in the log of the cumulant~\cite{Catani:1992ua}
\begin{equation}
    R(\alpha,L) = \exp[ L g_1(\alpha_s L) + g_2(\alpha_s L ) + \alpha_s g_3(\alpha_s L) + \cdots] \, .\label{RaLform}
\end{equation}
Here $R(\alpha,L)$ is the cumulant of some observable.
This is in contrast to the differential distribution, 
which is the derivative of
 $R(\alpha,L)$ with respect to the observable and
which generally has an expansion in terms of plus functions and other distributions.
With this organization, $g_1(\alpha L)$ is the leading-logarithmic series, $g_2(\alpha L)$ is the next-to-leading logarithmic series, and so on.

An alternative set of schemes is based on the perturbative expansion of anomalous dimensions. Here, the leading cusp anomalous dimension $\Gamma_0$ and the one-loop $\beta$ function are included for LL resummation, $\Gamma_1$, the two-loop $\beta$ function, and the one-loop regular anomalous dimensions $\gamma_0$ are included for NLL, and one higher order for each to go to NNLL, and so on.
In addition, the finite parts of the hard, jet, and soft functions are required
to one-loop (two-loop) order to achieve NNLL (N$^3$LL) resummation.
Another commonly used combination is that of NNLL RG evolution (i.e., two-loop regular anomalous dimensions)
with two-loop fixed-order finite parts, which we refer to as NNLL$'$ and which typically features improved perturbative uncertainties over NNLL. Within this set of schemes one still has the freedom to set scales in distribution space, cumulant space, or position space. Ultimately, we will be setting scales in position space for our final results. 

 In~\cite{Almeida:2014uva} different log-counting schemes are compared for a number of dijet-observables (we review the case of dijet thrust below) and
 it is shown that results in all spaces coincide up to subleading logarithmic terms.  However, it is precisely these subleading terms which are responsible for the spurious divergences in the Sudakov shoulder case. What we will show is that there is an exact correspondence between the terms associated with the leading cusp anomalous dimension and the double logarithmic series $(\alpha_s L^2)^n$ only if scales are set in position space. In contrast, when scales are set in momentum space, problematic subleading terms are produced.

\subsection*{Dijet thrust}

Before discussing the shoulder case, let us review scale setting in thrust in the dijet region. Let us work in the simplest possible setting, the ``$\Gamma_0$ approximation'' where we take the leading cusp anomalous dimension $\Gamma_0 \ne 0$ but we  set $\beta_j = 0$ and $\gamma_j = 0$ for $j\ge 0$ and set
$\Gamma_j = 0$ for $j > 0$.
We also take $H = 1$ and set $\mu_h=Q$ throughout as the hard scale is not of particular interest. 
This $\Gamma_0$-approximation suffices to show how one obtains
different results depending on in which space scales are set. 
All choices correctly predict the result at the fixed-scale double-logarithmic order where $R\sim \exp(\alpha_s L^2)$, 
but give different results for higher order terms. 
It is exactly this issue of how to treat terms beyond the logarithmic accuracy one it targeting which we wish to address, since for some observables the higher order terms can induce spurious singularities.
In this section, we define $\hat{\alpha} =  \frac{\alpha_s}{4\pi} \Gamma_0$. 

In the $\Gamma_0$-approximation, the thrust cumulant is
\begin{equation}
\label{cumulant}
R(\tau)
 = e^{  - 2 \hat{\alpha} C_F \ln^2 \frac{\mu_h}{\mu_{j}}- 2\hat{\alpha} C_F \ln^2 \frac{\mu_s}{\mu_j}}
\left[\tilde{j} \left( \partial_\eta + \ln \frac{Q \mu_s}{\mu_j^2} \right)\right]^2 
\tilde{s} (\partial_\eta)
 \left( \frac{\tau Q}{\mu_s} \right)^{\eta}
 \frac{e^{-\gamma_E \eta}}{\Gamma(1+\eta)} \theta(\tau)
 \,,
\end{equation}
with $\eta = 4C_F \ahat\ln\frac{\mu_j}{\mu_s}$.
From this, we can read off the  canonical scales $\mu_s=Q\tau$ $\mu_j=\sqrt{Q \mu_s}$. Using these values  and setting $\tilde{j}=\tilde{s}=1$ the resummed cumulant in momentum space takes the form
\begin{equation}
\left[R(\tau)\right]_{\text{mom}}
 = e^{  - \ahat C_F \ln^2 \tau}
 \frac{e^{-\gamma_E \eta}}{\Gamma(1+\eta)} \theta(\tau)
\,,
\end{equation}
with $\eta=-2C_F \ahat  \ln \tau$.
$R(\tau)$ has the form in Eq.~\eqref{RaLform} with $g_1 =- \ahat C_F \ln \tau$ and $g_2 = \ln \frac{e^{-\gamma_E \eta}}{\Gamma(1+\eta)}$. On the other hand if we Laplace transform before setting scales, then
\begin{equation}
{\mathcal L}[R](\nu)
 = \frac{1}{\nu} e^{  - 2 \ahat C_F \ln^2 \frac{\mu_h}{\mu_{j}}- 2\ahat C_F \ln^2 \frac{\mu_s}{\mu_j}}
 \left( \frac{Qe^{-\gamma_E}}{\nu \mu_s} \right)^{\eta}
\,.
\end{equation}
Here we can read off canonical scales as $\mu_s=\frac{Q e^{-\gamma_E}}{\nu}$ and $\mu_j=\sqrt{Q \mu_s}$. With the Laplace-space scale choice
\begin{equation}
{\mathcal L}[R](\nu)
 = \frac{1}{\nu} e^{  -  \ahat C_F \ln^2 (\nu  e^{\gamma_E})}
\,,
\end{equation}
which is a purely LL expansion. The inverse Laplace transform gives 
\begin{equation}
\left[R(\tau)\right]_{\text{Lap.}}
 = 
 e^{-  \ahat C_F \ln^2 \tau}
 e^{-  \ahat C_F \partial_\eta^2}
 \frac{e^{-\gamma_E \eta}}{\Gamma(1+\eta)} \theta(\tau) \,.
\end{equation}
This agrees with the momentum space form up to terms formally of order NNLL and higher,
so the results of setting scales in either space are equivalent. (In practice, setting scales in Laplace space requires some care because one has to numerically integrate over the Landau pole in QCD~\cite{Parisi:1979se, Collins:1981va}.)

In summary, we saw from the example of thrust in the dijet limit that setting scales in position space allows for exact agreement between the logarithmic counting and the solution of the RG equations. In momentum space, some subleading terms are generated even if only $\Gamma_0$ is kept as in this simplified example. It is these subleading terms which are problematic in the Sudakov shoulder case, as we will see next. 

\subsection*{Heavy jet mass shoulder}

Now we return to the Sudakov shoulder of heavy jet mass. 
We work again in the $\Gamma_0$ approximation with $\mu_h=Q$, for which the factorization formula in Eq.~\eqref{factform3d} for the gluon channel becomes
\begin{align}
& \frac{1}{\sigma_{\text{LO}}}\frac{d^3 \sigma_g}{d \rho^3}
 = e^{   
 - 2 \ahat C_F \ln^2 \frac{\mu_h}{\mu_{j h}}
 - 2 \ahat C_F \ln^2 \frac{\mu_{sh}}{\mu_{j h}} 
 - \ahat C_A \ln^2 \frac{\mu_h}{\mu_{j \ell}} 
 - \ahat C_A \ln^2 \frac{\mu_{s \ell}}{\mu_{j \ell}}
}\times
 \frac{e^{-\gamma_E(a+b)}}{\Gamma (a + b)}
\\ 
 &\ \times \left[ 
 \frac{1}{r} \left( \frac{r Q}{\mu_{s\ell}} \right)^{a} \left( \frac{r Q}{\mu_{sh}} \right)^{b}
 \frac{\sin (\pi a)}{\sin (\pi(a + b))} \theta (r) 
 +\frac{1}{(-r)} \left( \frac{-r Q}{\mu_{s\ell}} \right)^{a} \left( \frac{-r Q}{\mu_{sh}} \right)^{b}
  \frac{\sin (\pi b)}{\sin (\pi (a + b))}
\theta (- r) \right] 
. \nonumber
\end{align}
Here, $a= 2 C_A A_{\Gamma} (\mu_j, \mu_{s\ell})$ and $b= 4 C_F A_{\Gamma} (\mu_{jh}, \mu_s)$.
Canonical scales in momentum space are then
\begin{equation}
\mu_{s \ell} = \mu_{sh} = |r| Q, \quad
\mu_{j \ell}^2 =\mu_{j h}^2 = Q \mu_{s \ell} =
 Q \mu_{sh}
\label{canonmomentum}
  \,.
\end{equation}
Note that $|r|$ must be used to obtain canonical scales for both the $r<0$ and $r>0$ regions, and hence the scales are continuous but not smooth functions.
With these choices, the momentum space distribution becomes
\begin{align}
&\left(\frac{1}{\sigma_{\mathrm{LO}}}\frac{d^3\sigma_g}{d \rho^3}\right)_{\text{mom}}
\\
&\ \
= \frac{1}{r}e^{- \Gamma_0 \frac{1}{2} C_A \ahat \ln^2 |r| - \Gamma_0 C_F \ln^2 |r|} 
\frac{e^{-\gamma_E(a+b)}}{\Gamma (a + b)} 
 \left[ \frac{\sin (\pi a)}{\sin (\pi
(a + b))} \theta (r) - \frac{\sin (\pi b)}{\sin (\pi (a + b))} 
\theta (- r) \right] 
\label{sigmomform}
,\nonumber
\end{align}
with $a=-C_A \ahat \Gamma_0\ln |r|$ and $b = -2 C_F \ahat \Gamma_0 \ln |r|$.
This has the form of a NLL distribution $\exp[ L g_1(\ahat L) + g_2(\ahat L)]$. The Sudakov Landau pole is in the NLL contribution, $g_2(\ahat L)$. 

Now let us look at the distribution in position space. Before setting scales, it has the form
\begin{align}
   \widetilde{\sigma}_g(z) &= \int_{-\infty}^\infty dr\  e^{i z r}\ \frac{1}{\sigma_{\mathrm{LO}}}\frac{d^3\sigma_g}{d \rho^3}  
 \\
 &= e^{   
 - 2 \ahat C_F \ln^2 \frac{\mu_h}{\mu_{j h}}
 - 2 \ahat C_F \ln^2 \frac{\mu_{sh}}{\mu_{j h}} 
 - \ahat C_A \ln^2 \frac{\mu_h}{\mu_{j \ell}} 
 - \ahat C_A \ln^2 \frac{\mu_{s \ell}}{\mu_{j \ell}}
}
\left(-iz\frac{\mu_{s\ell} e^{\gamma_E}}{Q}\right)^{-a}
\left(iz\frac{\mu_{sh} e^{\gamma_E}}{Q}\right)^{-b}
.\nonumber
\end{align}
We can read off that the canonical scales are 
\begin{equation}
\mu_{s \ell} =i \frac{Q e^{- \gamma_E}}{z}, \quad 
\mu_{s h} = -i\frac{Q e^{- \gamma_E}}{z},  \quad
\mu_{j \ell}^2 = Q \mu_{s \ell}  \quad\text{and}\quad
\mu_{j h}^2 = Q \mu_{sh}
\label{complexcanon}
 \,.
\end{equation}
With these scale choices,
\begin{equation}
\widetilde{\sigma}_g(z) =\exp \left[ - \frac{1}{2} \ahat C_A
  \Gamma_0 \ln^2 (-i z e^{\gamma_E}) - \ahat C_F \Gamma_0 \ln^2 ( i z
  e^{\gamma_E})\right]
  \label{stzform}
 \,.
\end{equation}
Thus in position space this is exactly an LL expansion with no higher terms.

\begin{figure}
    \centering
    \includegraphics[scale=0.7]{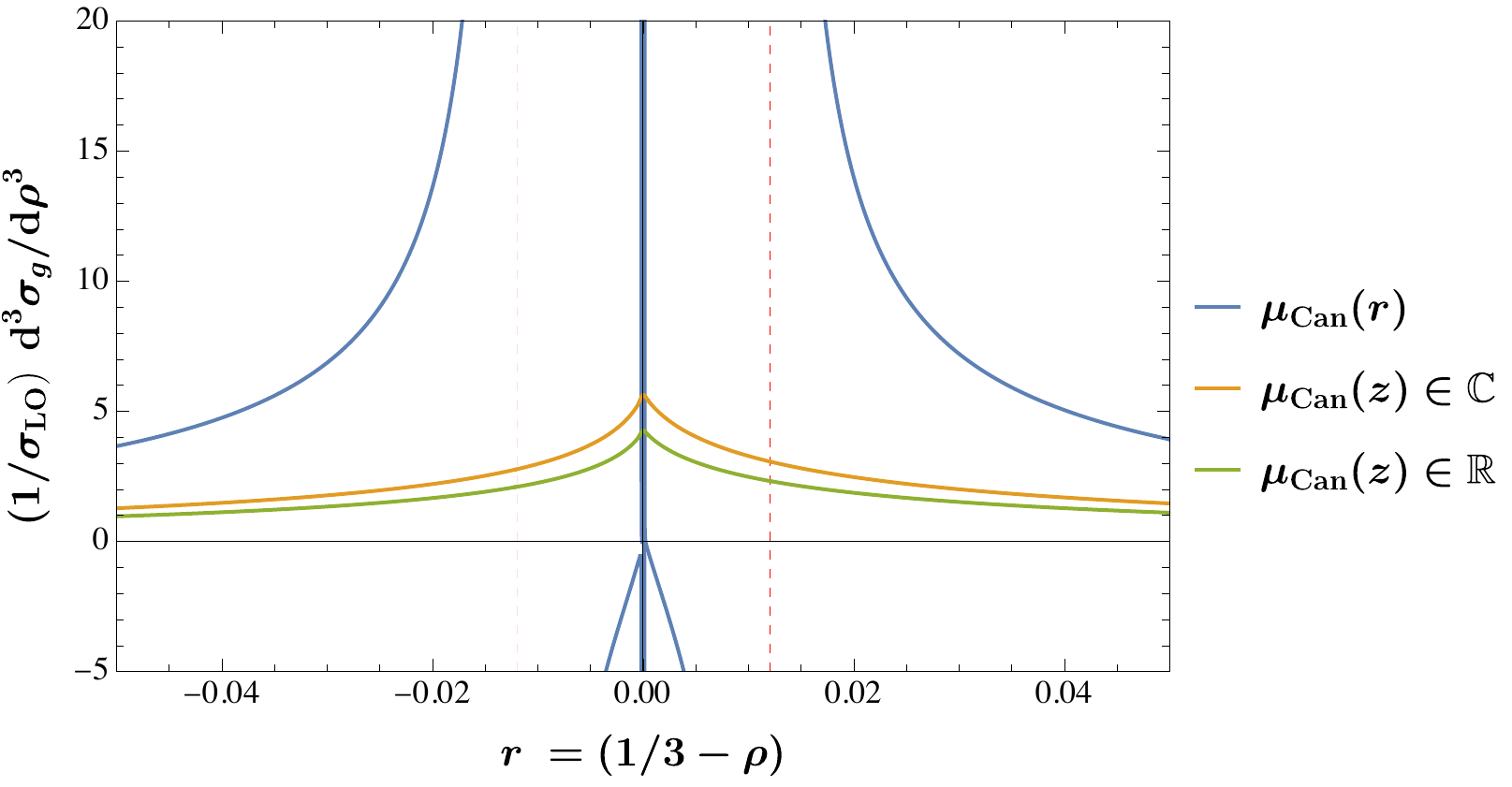}
    \caption{Comparison of the resummed $\frac{d^3 \sigma_g}{d \rho^3}$ distribution in the $\Gamma_0$ approximation when scales are set in position or momentum space for $\ahat=\frac{\alpha_s}{4\pi} = 0.01$. The blue curve shows the result with canonical scales in momentum space from Eq.~\eqref{canonmomentum}, the orange curve with canonical complex scales from Eq.~\eqref{complexcanon} and the green curve with real scales from Eq.~\eqref{realcanon}. 
    The Sudakov Landau poles  are divergences in blue curve at $a+b=1$ ($r=0.012$) and $a+b=2$ ($r=0.00015$) are seen as an artifact of momentum-space scale setting.}
    \label{fig:posmomscale}
\end{figure}

Although we cannot Fourier transform back to momentum space analytically, we can do so numerically. A comparison between the inverse Fourier transform of $\widetilde{\sigma}_g(z)$ with $\mu_{\rm Can}(z)$, and $\left(\frac{d^3\sigma_g}{d \rho^3}\right)_{\text{mom}}$ with $\mu_{\rm Can}(r)$, is shown in Fig.~\ref{fig:posmomscale}. We see that when canonical scales are chosen in position space, $\mu_{\rm Can}(z)$, the Sudakov Landau pole is gone.

When specifying the canonical scales, we are always free to make alternate choices that agree up to ${\cal O}(1)$ factors, since such choices still minimize large logs.
An alternative is to use real scales, such as
\begin{equation}
\mu_{s \ell} = \mu_{sh} =\frac{Qe^{-\gamma_E}}{|z|}, \quad
\mu_{j \ell}^2 =\mu_{j h}^2 = Q \mu_{s \ell} =
 Q \mu_{sh}
 \label{realcanon} 
  \,.
\end{equation}
Choosing real scales like this in position space will still remove the Sudakov Landau pole. Real scales are additionally easier to implement since they avoid analytically continuing $\alpha_s(\mu)$ to complex scales, as e.g. done for the resummation of so-called timelike logarithms in gluon-fusion Higgs production~\cite{Ahrens:2009cxz, Berger:2010xi, Becher:2012yn, Becher:2013xia, Stewart:2013faa, Ebert:2017uel}. 
Moreover, real scales are also preferable since they guarantee that the distribution will be real when Fourier transformed back into momentum space. For generic complex scales 
the higher order terms, beyond the logarithmic order one is working, can in general be complex.
With the scales in Eq.~\eqref{realcanon} the position space distribution is
\begin{equation}
\widetilde{\sigma}_g(z) =\exp\left[ -\frac{C_A+2C_F}{2} \ahat 
  \Gamma_0 \ln^2 (|z|e^{\gamma_E})\right]
  \left(-i\frac{z}{|z|} \right)^{C_A \ahat \Gamma_0 \ln (|z|e^{\gamma_E})}
  \left(i\frac{z}{|z|} \right)^{2C_F \ahat \Gamma_0 \ln (|z|e^{\gamma_E})} \, .
  \label{ab_gamma0_exp_realscales}
\end{equation}
The numerical inverse Fourier transform of this function is also shown in Fig.~\ref{fig:posmomscale}.

When setting scales in position space and transforming back, one might be concerned that an entirely new expression is generated, with no relation to the one with scales set in momentum space. First of all, we know that if the computations and matching are done to common fixed order $\alpha_s^n$, then the momentum and position space expressions must both be scale-independent to that order. Thus when we Fourier transform the position space expression back to momentum space it must agree to that order with the momentum space expression. Although it is harder to show directly, the series of large logarithms must also agree. In the $\Gamma_0$ approximation, we can derive an analytic relation between the resummed expressions that also gives insight into how the Sudakov Landau pole is removed.
To Fourier transform the distribution using canonical position space scales, we can write formally that 
\begin{align}
&  \left( \frac{1}{\sigma_{\mathrm{LO}}}\frac{d^3 \sigma_g}{d \rho^3} \right)_{\text{pos}} =
  \int_{-\infty}^{\infty} \frac{dz}{2\pi} e^{-i z r} \tilde{\sigma}_g(z)\nonumber\\
 & =  e^{- \ahat
  \Gamma_0 \frac{1}{2} C_A \partial_a^2 - \ahat \Gamma_0 C_F
  \partial_b^2} 
r^{a+b-1}
 \frac{e^{-\gamma_E(a+b)}}{\Gamma (a + b)} \left[ \frac{\sin (\pi a)}{\sin (\pi
(a + b))} \theta (r) - \frac{\sin (\pi b)}{\sin (\pi (a + b))}
\theta (- r) \right]_{a = b = 0}
\nonumber\\
  & = e^{- \ahat
  \Gamma_0 \frac{1}{2} C_A \partial_a^2 - \ahat \Gamma_0 C_F
  \partial_b^2} 
  \left(\frac{1}{\sigma_{\mathrm{LO}}}\frac{d^3 \sigma_g}{d \rho^3} \right)_{\text{mom}} 
 \,.
\end{align}
To get to the last line, we used $f (\partial_a) r^a = r^a f (\partial_a + \ln r)$ and then expanded $(\partial_a+L)^2$ and used $e^{x\partial_a} f(a) = f(a+x)$.  In this final form, we must set $a=-C_A \ahat \Gamma_0\ln |r|$ and $b = -2 C_F \ahat \Gamma_0 \ln |r|$  after taking the derivatives as in Eq.~\eqref{sigmomform}. Since $a\sim b \sim \ahat \ln r$ each application of $\ahat \partial_a^2$ or $\ahat \partial_b^2$ gives a contribution suppressed by $\ahat$ which makes the position and momentum space distributions differ at the NNLL level and beyond. Nevertheless, these derivatives cure the Sudakov Landau pole: near the pole $\partial_a^2 \sim \partial_r^2 \to \infty$ and the divergence gets exponentially suppressed.

\subsection{Comparison to \texorpdfstring{$q_T$}{qT} resummation \label{sec:qtresummation}}

It has been known for a long time
that color-singlet transverse-momentum ($q_T$) spectra in hadron collisions
feature poles at finite values of $q_T$
when strictly truncating the logarithmic expansion in momentum space~\cite{Frixione:1998dw},
or equivalently when naively setting scales~\cite{Ebert:2016gcn} in momentum space.%
\footnote{
We refer the reader to ref.~\cite{Ebert:2016gcn} for a detailed discussion of the issues of naive momentum-space scale setting,
as well as for a formal momentum-space construction that avoids them, essentially by employing cumulant-space scales
for every individual boundary term in the factorization theorem.
See also ref.~\cite{Monni:2016ktx}, which avoids the issues of momentum-space scale setting by approximating logarithms of $q_T$
by logarithms of the hardest emission's $p_T$, and ref.~\cite{Becher:2010tm, Becher:2011xn, Neill:2015roa, Kang:2017cjk},
where only a subset of scales is set in momentum space.}
In both $q_T$ and the Sudakov shoulder, 
these spurious poles are due to cancellations
between energetic emissions that contribute to the observable:
through momenta with opposite direction in the transverse plane for $q_T = \abs{\vec{q}_T}$,
and through mass shifts with opposite sign for the Sudakov shoulder, as in Eq.~\eqref{massshift}.
For the $q_T$ distribution, it is in fact standard practice
to perform the resummation by setting scales (or truncating the logarithmic expansion)
as a function of the Fourier-conjugate variable $b_T = \abs{\vec{b}_T}$ of $q_T$~\cite{Parisi:1979se, Collins:1981va, Collins:1984kg},
and then performing a numerical~\cite{Balazs:1997xd}
or approximate analytic~\cite{Kulesza:1999gm} inverse Fourier integral.
This procedure is well known to lift the spurious poles
by retaining an appropriate set of terms that would be subleading when counting logarithms directly in momentum space,
in close analogy to our discussion in Section~\ref{sec:scale_setting}.
It is thus interesting to ask in what ways our results here differ from the $q_T$ case. Two intriguing differences are
\begin{itemize}
   \item
   In the $\vec{q}_T$ case, each individual beam and soft function (or TMD PDF)
   is an azimuthally symmetric function of the magnitude $k_T = \vec{k}_T$
   of the total transverse momentum of soft or collinear radiation.
   Accordingly, the resulting spectrum is azimuthally symmetric and only a function of $\abs{\vec{q}_T}$.
   This is distinct from the case of the heavy jet mass shoulder,
   where individual jet functions always contribute with a given fixed sign
   and the soft function itself is not symmetric under $m_h \leftrightarrow m_\ell$.
   In position space, this physical asymmetry between hemispheres
   is encoded in an imaginary part of the Fourier transform of the cross section,
   cf.~Fig.~\ref{fig:sigmaz}, which is an interesting novel feature of the Sudakov shoulder case.
   
   \item
   To our knowledge, the heavy-jet mass distribution in the shoulder region is the first invariant mass-like
   observable which features these spurious poles,
   and for which position-space resummation is required to arrive at physical results.
   Examples of transverse momentum-like observables exist
   whose resummation can be carried out in momentum space,
   either because they are strictly additive scalar quantities (transverse energy $E_T$)
   or because they involve an ordering that vetoes harder emissions (leading jet $p_T$).
   Our analysis shows that an observable being transverse
   is neither sufficient \emph{nor} necessary
   for the appearance of spurious poles,
   and that position-space resummation is a crucial tool beyond the case of $q_T$. In other language~\cite{Bauer:2002aj}, we can say that this is the first SCET$_\mathrm{I}$ observable where position space scale-setting is critical, while earlier examples are in SCET$_\mathrm{II}$.
\end{itemize}

\subsection{Complex scales and non-global logarithms \label{sec:complexscales}}
In the $\Gamma_0$ approximation, the HJM shoulder distribution receives distinct contributions from the light and heavy hemispheres. These separate contributions are well-defined since the leading Sudakov double logarithms come from the soft-collinear limit where the soft emissions are always near a jet and therefore uniquely associated with a hemisphere. This means we can choose distinct complex soft scales as in Eq.~\eqref{complexcanon}.

Beyond the leading-log approximation, when single logs come into play, it is no longer possible to cleanly separate left and right hemisphere contributions. In the dijet limit, the hemisphere soft function factorizes as~\cite{Fleming:2007xt,Hoang:2008fs,Becher:2008cf,Chien:2010kc,Kelley:2011ng,Hornig:2011iu}
\begin{equation}
    \tilde{s}_{\text{hemi}}(\nu_1,\nu_2,\mu) = \tilde{s}_\mu(\nu_1,\mu), \tilde{s}_\mu(\nu_2,\mu) \, \tilde{s}_f(\nu_1,\nu_2) 
\,,\end{equation}
where $\nu_1$ and $\nu_2$ are the Laplace conjugates of the left and right hemisphere masses. This factorization is non-trivial. It follows from RG invariance and symmetry of the hemisphere soft function under the interchange of the two hemisphere masses. It is also not uniquely defined -- one can shuffle constants among the three factors. We conventionally define $\tilde{s}_\mu(\nu,\mu)$ to be all the terms determined by RG invariance. The factor $\tilde{s}_f(\nu_1,\nu_2)$ describes the non-global structure of the hemisphere soft function. This function is known to 2-loops~\cite{Kelley:2011ng,Hornig:2011iu} and contains the leading non-global logarithm~\cite{Dasgupta:2001sh}. 

The position-space trijet hemisphere soft function $S(z_\ell,z_h,\mu)$ relevant for the heavy jet mass Sudakov shoulder resummation is superficially similar. We might like to write
\begin{equation}
    \tilde{s}(z_\ell,z_h,\mu) =\tilde{s}_\ell(z_\ell,\mu) \,, \tilde{s}_h(z_h,\mu) \, \tilde{s}_{\text{ng}}(z_\ell,z_h)  
\label{trijetfact}
\,.\end{equation}
Unfortunately, although we can still use RG invariance to determine its $\mu$-dependence we no longer have a symmetry in the interchange of the two hemispheres to give a precise meaning to two factors $\tilde{s}_\ell(z_\ell,\mu)$ and $\tilde{s}_h(z_h,\mu)$ (beyond the cusp anomalous dimension). 
To be explicit, in the gluon channel, 
the coefficient of the overall soft anomalous dimension at ${\cal O}(\alpha_s)$
can be written as 
$\gamma_0^s = \gamma_0^{sg} + \gamma_0^{sqq}$
where
\begin{equation}
\gamma_0^{sg} = -2C_A \ln 3 + 4 C_F \ln 2,\quad \gamma_0^{sqq} = -4 C_F \ln 6
  \,.
\end{equation}
This separation is motivated by the contributions to the soft function from radiation into the light hemisphere containing the single gluon jet, or the heavy hemisphere containing the quark and antiquark jets. This suggests that we might identify 
$\gamma_0^\ell = \gamma_{0}^{sg}$ 
as the anomalous dimension governing $\tilde{s}_\ell(z_\ell,\mu)$ and 
$\gamma_0^h = \gamma_{0}^{sqq}$ 
as the anomalous dimension governing $\tilde{s}_h(z_h,\mu)$. Also, there is no guarantee that resummation achieved by assuming separate anomalous dimensions will predict any correct logarithms at higher orders.

\begin{figure}
    \centering
    \includegraphics[scale=1.0]{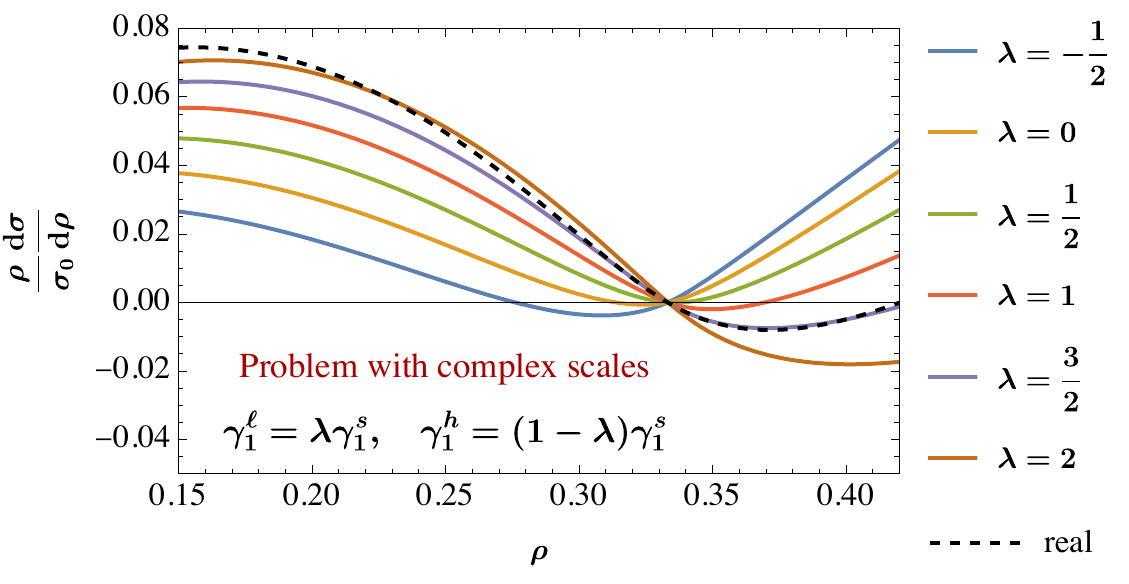}
    \caption{Problem with complex scales, using NNLL resummation and no matching. This figure shows the effect on the resummed distribution if the two-loop soft anomalous dimension is artificially separated into light and heavy hemisphere components so that separate complex scales can be chosen. Different choices of the split introduce unphysical logarithms which distort the distribution. We see that complex scales introduce spurious uncertainty. }
    \label{fig:complexscalesnnll}
\end{figure}
If we cannot define single-scale objects $\tilde{s}_\ell(z,\mu)$ and $\tilde{s}_h(z,\mu)$ there is no sense to choosing different scales for the heavy and light hemispheres. Moreover, it can be problematic to choose different {\it complex} scales for the two hemispheres as in Eq.~\eqref{complexcanon}. 
In Fig.~\ref{fig:complexscalesnnll} we show the prediction (using the full NNLL setup we describe below) with canonical complex scales for different choices of how the two-loop soft anomalous dimension splits into heavy and light components:
\begin{equation}
    \gamma_1^\ell = \lambda \gamma_1^s,\quad
    \gamma_1^h = (1-\lambda) \gamma_1^s  
   \,.
\end{equation}
Curves for $\lambda=-\frac{1}{2} \ldots 2$ are shown in Fig.~\ref{fig:complexscalesnnll} and compared with the real scale choice in Eq.~\eqref{realcanon}. We see that artificially separating the anomalous dimension can generate large distortions in the spectrum.
 It effectively induces large logarithms of the ratio of these scales $\ln \frac{\mu_\ell}{\mu_h} \sim \pi$. These logarithms are formally of the same order as non-global logarithms of $k_\ell/k_h$ in the $\tilde{s}_\text{ng}$ part of the trijet soft function in Eq.~\eqref{trijetfact}, which we necessarily treat at fixed order when evaluating $\tilde{s}_\text{ng}$ at any given individual scale.
We thus conclude that there is no
sensible way to split the soft anomalous dimension into light and heavy components, and that a common scale should be used for all parts of the soft function.

\subsection{Summary}
In this section, we examined scale setting in position, momentum, and Laplace space for the Sudakov shoulder region of heavy jet mass in the $\Gamma_0$ approximation. From the momentum-space point of view, the regions with $r<0$ and $r>0$ are distinct, and the canonical scale choice $\mu_s \sim |r| Q$ is not smooth across $r=0$. However, $r=0$ is not special from the point of QCD: the shoulder at $r=0$ is an artifact of perturbation theory. In position space, the natural complex scales are $\mu_{s\ell} \sim -\frac{i}{z}Q$ and $\mu_{sh} \sim \frac{i}{z}Q$. 
Choosing scales in position space makes the distribution smooth across $r=0$ and removes the spurious Sudakov Landau singularity caused by momentum-space scale setting. 

To choose canonical complex scales requires a separation into contributions from heavy or light hemispheres, which is a sensible criterion only in the LL approximation. Beyond that, artificially separating scales in the heavy and light hemispheres is problematic. Instead, one should choose equal soft scales such as $\mu_{s\ell}= \mu_{sh} \sim \frac{Q}{|z|}$. 
The Sudakov Landau pole is still removed as long as scales are chosen in position rather than momentum space.

\section{Integration and matching \label{sec:integration_and_matching}}

With the factorization formula and the insights from  Section~\ref{sec:fact} and~\ref{sec:scale_setting} in hand, we can now turn to full NNLL resummation.
As discussed, choosing scales in momentum space is problematic: it generates spurious Sudakov Landau poles from the region where the heavy and light jet mass shifts are both large, but there difference is still small. 

In position space this region corresponds to small $z$. When we set scales in position space the distribution is heuristically $z^{-\ahat \ln z} = e^{-\ahat \ln^2 z}$ so the small $z$ region becomes Sudakov suppressed.
However, despite this suppression,
 one must still be careful when performing the inverse Fourier transform back to momentum space in actual QCD because of the Landau pole in $\alpha_s(\mu)$. 
In addition, the factorization formula we have been using so far holds for $\frac{d^3 \sigma}{d\rho^3}$, so one must integrate over $\rho$ twice to get back the physical spectrum $\frac{d \sigma}{d\rho}$. This integration requires boundary conditions generating a linear function $c_0 + c_1 \rho$. These two constants $c_0$ and $c_1$ correspond to finite remainders after adding counterterms to remove the logarithmic and linear UV divergences in the original $\frac{d\sigma}{d\rho}$ distribution in Eq.~\eqref{dsdrform}. More physically, they correspond to the fact that the offset and slope of the distribution at $r=0$ are not predicted by the shoulder factorization formula alone: there can be contributions from four-parton and
higher-order configurations which have $\rho \sim \frac{1}{3}$ but do not arise from configurations close to the trijet configuration. The integration constants $c_0$ and $c_1$ must be determined by some boundary conditions for the double integral from $\frac{d^3 \sigma}{d\rho^3}\to \frac{d \sigma}{d\rho}$  as we discuss in this section.

\subsection{Integration} \label{sec:integration}
In terms of the position space distribution, the factorization formula has the form
\begin{equation}
    \frac{1}{\sigma_{\text{LO}}}\frac{d^3 \sigma_g}{d\rho^3}=  
    \int_{-\infty}^\infty \frac{dz}{2\pi}\, e^{-i z r}\, \widetilde\sigma_g(z)
  \,,
\end{equation}
where $r=\frac{1}{3}-\rho$ and
\begin{equation}
\label{eq:resummed_sigmaz}
    \widetilde\sigma_g(z) = \Pi_g\ \cdot 
    \sjt_g \cdot
\left( -i z\frac{e^{\gamma_E} \mu_{s\ell} }{Q} \right)^{-a}
\left( i z\frac{e^{\gamma_E} \mu_{sh} }{Q} \right)^{-b}
  \,,
\end{equation}
with $\Pi_g$ the same as in Eq.~\eqref{Pig}, but now $\sjt_g$ is a function rather than an operator
\begin{equation}
\sjt_g =     
\widetilde{j}_q
\left( 
\ln \frac{Q^2 e^{-\gamma_E}}{i z \mu_{jh}^2} 
\right) 
\widetilde{j}_{\bar{q}}
\left( 
\ln \frac{Q^2 e^{-\gamma_E}}{i z \mu_{jh}^2}
\right) 
\widetilde{j}_g
\left( 
\ln \frac{Q^2 e^{-\gamma_E}}{-i z \mu_{jl}^2} 
\right)
\widetilde{s}_{q \bar{q};g} \left(
\ln \frac{Q e^{-\gamma_E}}{-i z \mu_{s\ell}},
\ln \frac{Q e^{-\gamma_E}}{i z \mu_{sh}}
\right) 
.
\end{equation}
For NLL${}^\prime$ or NNLL resummation we expand $\sjt_g$ to order $\alpha_s$. The quark channel is analogous.

To obtain the physical spectrum, we need to inverse Fourier transform and integrate over $r$ twice. A sufficient condition for $\sigma(\rho)$ to be real is that  $\widetilde{\sigma}^{\star}(-z)=\widetilde{\sigma}(z)$. This condition on $\widetilde{\sigma}$ holds automatically for real scale choices, since $(i (-z))^\star = iz$. We can thus write the inverse Fourier integral as
\begin{equation}
 \frac{1}{\sigma_{\text{LO}}}\frac{d^3\sigma_i}{d\rho^3}
 =\int_{-\infty}^{+\infty}\frac{d z}{2\pi}\, e^{-i z r}\,
   \widetilde{\sigma_i}(z)
 =2\ \re\left[\int_0^{\infty} \frac{d z}{2\pi} e^{-i z r}
   \widetilde{\sigma_i}(z)\right]
\,.\end{equation}
Integrating twice with respect to $r$ generates a factor of $\frac{1}{z^2}$ in the integrand along with two integration constants. We can then write the integral as\footnote{In practice, we also introduce $u\equiv \frac{1}{1+z}$ to map the Fourier integration region into a finite region $u\in[0,1]$. And to improve the convergence, we set a cutoff $\delta=10^{-12}$ in the upper limit of $u$, and by varying $\delta$ from $10^{-8}$ to $10^{-20}$, we find that the error is very tiny.}
\begin{align}
	\frac{1}{\sigma_{\text{LO}}}\frac{d\sigma_i}{d\rho}= 2\ \re\left[\int_0^{\infty} \frac{d z}{2\pi} { K(z,r)} \widetilde{\sigma_i}(z)\right]
 \label{rint} 
 \,,
\end{align}
with a kernel function
\begin{equation}
	 K(z,r) =  - \frac{e^{-i r z}}{z^2} + K_0(z) +K_1(z) \, r
\,,\end{equation}
where $K_0$ and $K_1$  serve to fix the constants in the boundary conditions. 
For example, if we performed the integrals from a lower cutoff $r_L$ so that
\begin{equation}
\frac{1}{\sigma_{\text{LO}}}\frac{d\sigma_i}{d\rho}=\int_{r_L}^{r} d r^{\prime} \int_{r_L}^{r^{\prime}} d r^{\prime\prime} \frac{1}{\sigma_{\text{LO}}}\frac{d^3\sigma}{d\rho^3}
\,,\end{equation}
then we would find
\begin{equation}
	 K^{r_L}(z,r) = \frac{e^{-ir_L z}-e^{-i r z}}{z^2}+i\frac{e^{-ir_L z}(r_L-r)}{z}
  \label{kerneldef}
\,.\end{equation}
Unlike for dijet observables, where the limit of integration $\rho=0$ or $\tau=0$ is natural, there is no natural choice for $r_L$ for the Sudakov shoulder. Thus, this boundary condition forcing the distribution to vanish at $r=r_L$ is not ideal.

A more natural boundary condition is to ask that the unmatched distribution when expanded to fixed order reproduces the leading behavior of the full theory distribution near $r=0$. It does this automatically for all the $r \ln^m r$ terms, which are true predictions of the effective theory factorization formula. It also reproduces the difference in slopes in the left and right shoulders. This difference comes from integrating over the $\delta$-function terms in the triple-derivative. For example,
\begin{equation}
    \int_{r_L}^r dr' \int_{r_L}^{r'}dr'' \delta(r'') = r \theta(r)  - r \theta(r_L)= \frac{1}{2}r\Big[\theta(r) -\theta(-r)\Big] +\left[\frac{1}{2} - \theta(r_L)\right] r
  \,.
\end{equation}
The first term on the right is independent of $r_L$ and therefore a prediction of the theory. 
The value for the second term is determined by one of the boundary conditions.

More generally,  consider the unmatched resummed distribution, expanded in $\alpha_s$. The distribution
$g(r) = \frac{1}{\sigma_0} \frac{d\sigma}{d\rho}$ has the generic form
\begin{align}
\hspace{1cm}
g^\text{res}(r) &= 
  \theta(r)\sum_{m\geq 1} d_m^+  r \ln^m r+
 \theta(r)\sum_{m\geq 1} d_m^- r  \ln^m (-r)
 \nn \\
  &\ \  
  + b r [\theta(r)-\theta(-r)] + c_0 + c_1 \,  r
 \,,\hspace{1cm}
\end{align}
with $c_1,b,d_m^\pm$ all functions of $\alpha_s$, but only $c_0$ and $c_1$ depend on the choice of boundary conditions. That is, $d^\pm$ and $b$ are predicted by the factorized expression while $c_0$ and $c_1$ are not. We can then demand that the offset at $r=0$ and the sum of the slopes on the left and right shoulders at $r=0$ agree with a fixed-order calculation to fix $c_0$ and $c_1$. 
Moreover, using perturbation theory to determine the boundary conditions implies that the uncertainty on these constants will be included in the uncertainty associated with the fixed-order contribution. Thus we will not have to account for it separately.

At leading order, the distribution in the two mass shifts is
\begin{equation}
\frac{1}{\sigma_0}  \frac{d^2 \sigma}{d m_{\ell}^2 d m_h^2} =\delta(m_\ell^2) \,\delta(m_h^2)
\,,\end{equation}
so the integrals over $m_\ell^2$ and $m_h^2$ in Eq.~\eqref{massshift} are finite and
the unmatched distribution is
\begin{equation}
  g(r) = \int_0^{\infty} d m_h^2  \int_0^{\infty} d m_{\ell}^2 \,
  \frac{d^2 \sigma}{d m_{\ell}^2 d m_h^2} (r + m_h^2 - m_{\ell}^2) \, \theta (r +  m_h^2 - m_{\ell}^2) =
  r\theta(r)
 \,.
\end{equation}
We read off $d_m^\pm=0$, $b=c_1=\frac{1}{2}$ and $c_0=0$. Note that at leading order there is no UV divergence in the integrals over $m_\ell^2$ and $m_h^2$ so the integration constants are calculable without using the full theory.
Since $g''(r) = \delta(r)$ we can implement these integration constants with an integral kernel
\begin{equation}
    K(z,r)=\frac{1}{z^2}\left[1-e^{-i z r}+ r(1-e^{iz})\right] \label{eq:LOkernel}
  \,.
\end{equation}
This kernel when integrated against $\tilde{\sigma}(z) =1$ gives $r\theta(r)$ as desired. 

At higher orders we need to modify the boundary conditions by the inclusion of $\alpha_s$ corrections from fixed-order perturbation theory. Furthermore, we also want to add any fixed order contributions that are not enhanced by large logarithms, and hence not part of the resummed cross-section (so called ``non-singular'' terms).
This can be accomplished by writing
\begin{align}
\label{eq:matching}
	\frac{d\sigma^{\text{match,sh}}}{d\rho}
&=\frac{d\sigma^{\text{FO}}(\mu_{\text{FO}})}{d\rho}
+2\re\biggl\{\int_0^{\infty} \frac{d z}{2\pi} K(z,r) \left[\tilde{\sigma}^\text{sh}(z,\mu_{\text{sh}})-\widetilde{\sigma}^\text{sh}(z,\mu^{\text{FO}})\right]\biggr\}
\nonumber \\
&= \overbrace{2\re\biggl\{\int_0^{\infty} \frac{d z}{2\pi} K(z,r) \, \tilde{\sigma}^\text{sh}(z,\mu_{\text{sh}})\biggr\}}^{\text{resummed}}
+ \underbrace{\biggl[ \frac{d\sigma^\text{FO}(\mu_{\text{FO}})}{d\rho} - \overbrace{\frac{d\sigma^\text{sh}(\mu_{\text{FO}})}{d\rho}}^{\text{singular}}\biggr]}_{\text{non-singular}}
\,,\end{align}
where $\mu^{\text{res}}$ stands for the scale choices leading to resummation and $\mu^{\text{FO}}$ refers to setting all the scales equal to a common fixed-order scale.
When writing the matched cross section as on the first line, the integration constants will contribute only beyond the order to which the resummed cross section is matched, since the latter is scale-invariant to that order. Thus the uncertainty on the constants will be within the fixed-order (hard scale) uncertainty. We therefore use the leading-order kernel in Eq.~\eqref{eq:LOkernel} for simplicity.
On the second line, we have rearranged terms and performed the analytic 
$z$ integral over the fixed-order singular cross section 
to isolate the ``non-singular" term in square brackets, which is analytic near the shoulder.

\subsection{Profile functions}
\label{sec:profiles}

Although choosing canonical real scales resums the correct logarithms to NNLL order, in practice it is important to choose scales which 
avoid 
the Landau pole at $\mu=\Lambda_{\rm QCD}$, and which allow for resummation to turn off in regions that are more accurately described by an alternate description. 
These features can be achieved with profile functions~\cite{Ligeti:2008ac,Abbate:2010xh}. In our case,  we want to turn off the resummation for the shoulder region as we approach the dijet region, where conventional dijet resummation applies, and to switch to fixed-order perturbation theory when we are far enough above the shoulder threshold.

To construct profile functions, we follow the procedure used for $q_T$ resummation in~\cite{Lustermans:2019plv}. We start with the canonical (real) scales, as justified in Section~\ref{sec:fact}. To ensure that the soft and jet scales do not hit the Landau pole of QCD, we adjust these canonical scales to be
\begin{align} \label{canonhjs}
\mu_h^{\can} =Q
\,, \qquad
\mu_s^{\can}(z)
=\sqrt{
   \Big(\frac{Q e^{-\gamma_E}}{|z|}\Big)^2
   +\Big(\mu_{s}^{\text{min}}\Big)^2
}
\,, \qquad
\mu_j^{\can}(z) =\sqrt{\mu_h^{\can} \mu_s^{\can} }
\,,\end{align}
and take $\mu_s^{\text{min}}=2$ GeV. 
We would like the canonical scales to be used for a certain range of $\rho$ and then smoothly turn off away from the resummation region. 
To do so, we choose central jet and soft profile scales based on the canonical scales as
\begin{align} 
    \mu_{j,s}^{\pro}(z,\rho) = \mu_h^{1-g(\rho)} [\mu^{\mathrm{can}}_{j,s}(z)]^{g(\rho)}
    \label{cantopro}
\,,\end{align}
where $0 \leq g(\rho) \leq 1$ in the exponent determines whether the resummation is fully on ($g = 1$)
or turned off to recover the fixed-order singular result ($g = 0$),
and smoothly interpolates between the two cases through a weighted geometric mean.
Since the RG evolution resums logarithms of $\mu_s/\mu_j$ and $\mu_j/\mu_h$,
this effectively amounts to having $g(\rho)$ multiply logarithms of $z$ in the resummation exponent. 

To achieve a matched result, the shoulder resummation needs to turn off outside of the resummation region.
To implement this, we take $g(\rho)$ to be a product of profile functions
governing the turn-off towards the left and the right of the shoulder,
\begin{align}
g(\rho) = g_L(\rho) \, g_R(\rho)
\,.\end{align}
We use two different functional forms for the individual profile functions,
each with their own advantages.
For the left profile, we use a sigmoid function:
\begin{align} \label{eq:baseprofile}
   g_L(\rho) =
   1-\frac{1}{1+e^{(\rho - \rho_L)/\sigma_L}}
      \ =\ \raisebox{-0.5\height}{\includegraphics[width=4cm]{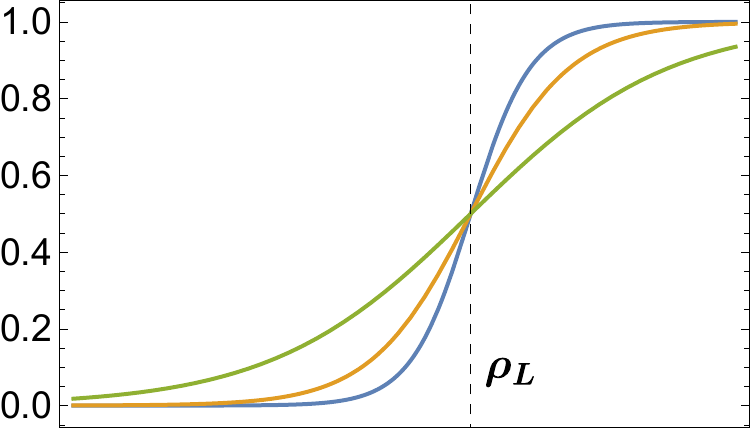}}
\,.\end{align}
The advantage of using an infinitely differentiable function like the sigmoid is that it does not introduce unphysical kinks in the matched distribution which may bias comparisons to data.
A disadvantage is that the sigmoid function turns the resummation exactly on or off
only in the asymptotic limits $\rho \to \pm \infty$. (In practice, $g_L(\rho)\geq 0.9$ for $\rho > \rho_L + 2\sigma_L$,
which leads to scales that are equivalent to canonical ones
within the variations we consider.
Similarly, we find that $g_L(\rho) \leq 0.1$
for $\rho < \rho_L - 2\sigma_L$ is close enough to the strict fixed-order limit
to the left of the shoulder for our purposes.)
For the right profile, we use a piecewise quadratic function~\cite{Lustermans:2019plv}:
\begin{align} \label{eq:profile_quadratic}
   g_R(\rho)
   = \begin{cases}
      1
      \,,\quad &
      \rho<\rho_{R1}
      \,, \\
      1 - \frac{2(\rho-\rho_{R1})^2}{(\rho_{R1}-\rho_{R 2})^2}
      \,,\quad &
      \rho_{R1}<\rho<\tfrac{\rho_{R1} + \rho_{R2}}{2}
      \,, \\
      \frac{2(\rho-\rho_{R 2})^2}{(\rho_{R1}-\rho_{R 2})^2}
      \,,\quad &
      \tfrac{\rho_{R1} + \rho_{R2}}{2}<\rho<\rho_{R 2}
      \,, \\
      0
      \,,\quad &
      \rho>\rho_{R 2}
   \,,\end{cases}
 \quad=\ \raisebox{-0.5\height}{\includegraphics[width=4cm]{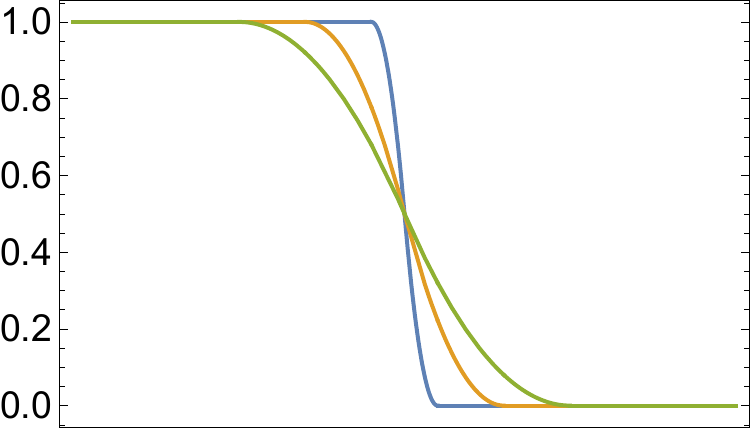}}
  \,.
\end{align}
This form has the advantage that the resummation is exactly off at $\rho \geq \rho_{R 2}$. Since the cross section on the right shoulder falls rapidly, one must turn off resummation rather quickly to avoid unphysical negative values from the resummed distribution~\cite{Abbate:2010xh}. This is easiest to achieve with a quadratic profile. 

\begin{figure}
    \centering
    \includegraphics[height=0.28\textwidth]{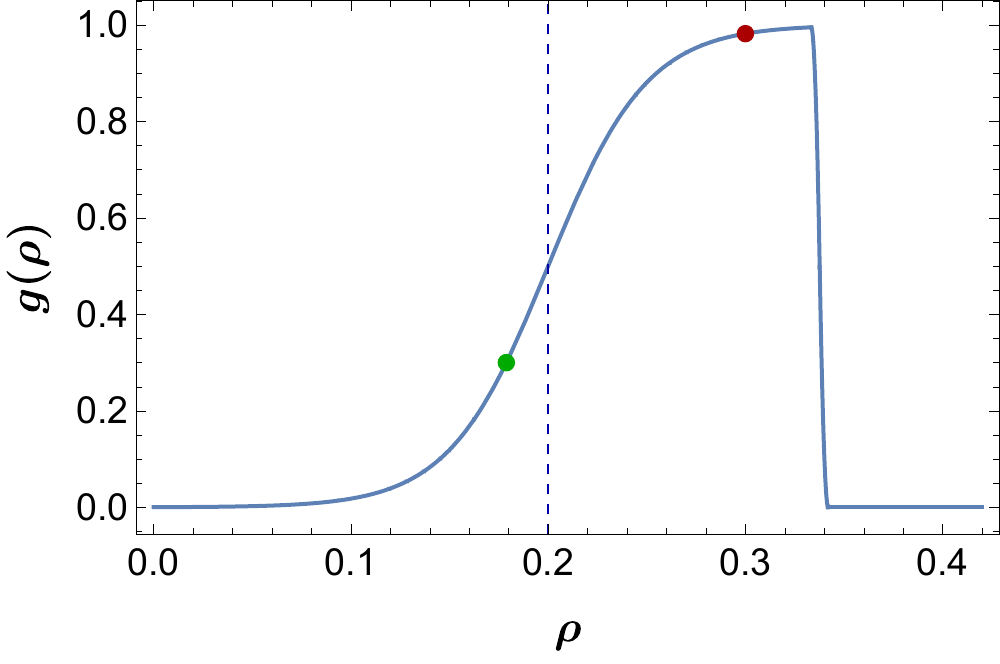}
    \includegraphics[height=0.28\textwidth]{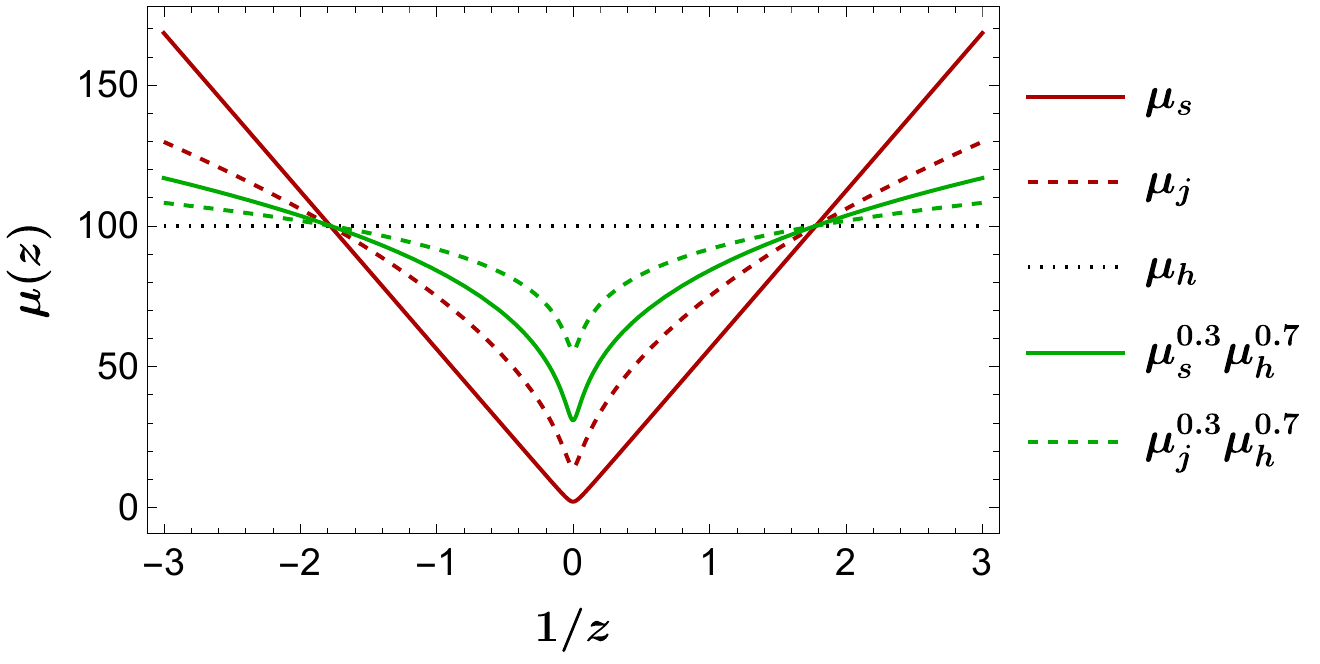}
    \caption{The left panel shows the weight function used in the shoulder profiles to interpolate between canonical scales ($g=1$) and fixed scales ($g=0$). The dashed line indicates the central transition point for the left-shoulder profile, at $\rho_L = 0.2$.
    The right panel shows the position space scales we use for  canonical scales ($g \approx 1$; red dot on left, red curves on right) and for scales which are part way on the transition to fixed scales  ($g=0.3$; green dot on left, green curves on right). We take $Q=100~\text{GeV}$ and $\mu_s^{\min}=2~\text{GeV}$, $\rho_L = 0.2, \sigma_L = 0.025, \rho_{R1} = \frac{1}{3}$ and $\rho_{R2} = 0.342$. The region where $|z|<e^{-\gamma_E}= 0.56$ has $\mu_j,\mu_s>Q$ but gives a negligible contribution to the momentum space distribution.}
    \label{fig:scales}
\end{figure}

\begin{figure}
    \centering
    \includegraphics[width=0.49\textwidth]{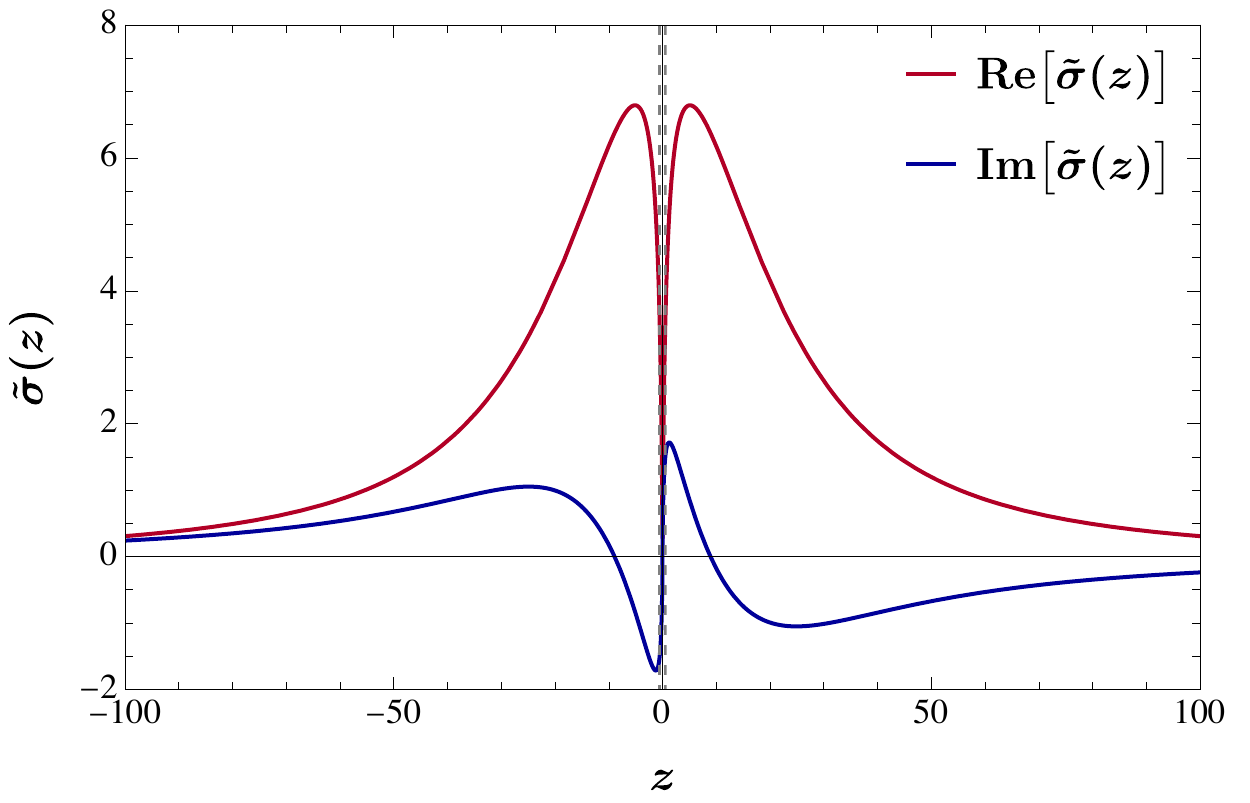}
    \hfill
    \includegraphics[width=0.49\textwidth]{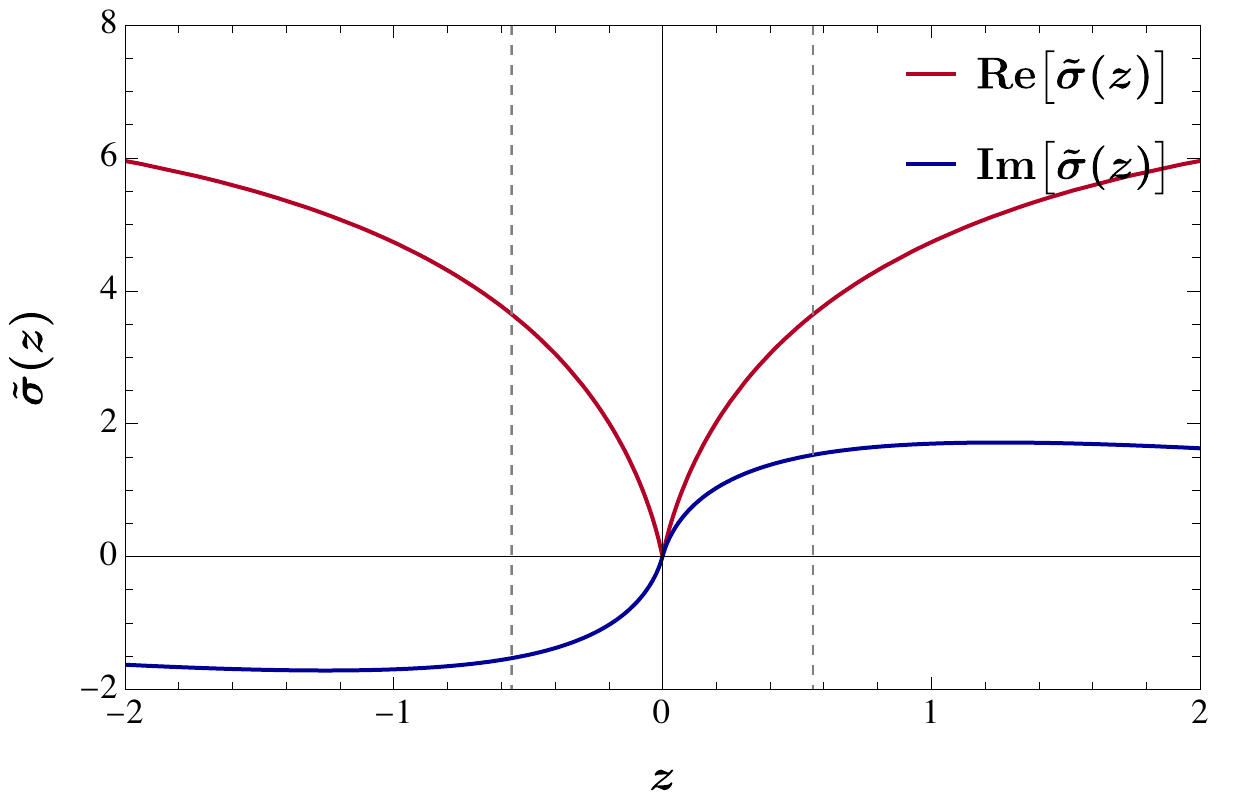}
    \caption{The Fourier transform $\widetilde{\sigma}(z)$ of the cross section with NNLL resummation and real scale choices. Right panel zooms in for the small $z$ region.
    The region within the two grey dashed lines have canonical scales above $Q$ but contributes negligibly to the momentum space distribution.
    }
    \label{fig:sigmaz}
\end{figure}

The left panel of Fig.~\ref{fig:scales} shows the combined interpolating function $g(\rho)$ we use for the full shoulder region. The right panel of this figure shows the scales as a function of $z$ for a point in the canonical region where $g=1$, and a point with $g=0.3$ in the transition region, illustrating the weighted geometric mean of scales.
Although the region with $|z| \lesssim 0.56$  where the jet and soft scales are larger than $Q$ is not excluded from the Fourier integral, this region contributes negligibly.\footnote{In the
$|z|\lesssim 0.56$ region, since $r$ is  small, we can approximate $e^{i z r} = 1$ in the inverse Fourier transform. Thus not only does the region $|z| \lesssim 0.56$ not contribute much overall (because the integrand is bounded), but the amount it contributes is also independent of $r$ and therefore its leading contribution is fixed by the $c_0+c_1 r$ boundary condition.} We can understand this from looking directly at $\tilde{\sigma}(z)$ as shown in Fig.~\ref{fig:sigmaz}.

To determine the profile midpoint $\rho_L$
and the transition width $\sigma_L$ on the left shoulder,
we compare the size of the singular cross section
predicted by the shoulder factorization at fixed order
to the total cross section 
in the top left panel of Fig.~\ref{fig:sing_vs_non_sing2}.
The non-singular remainder, which is given by the difference of the two,
has the same size as the singular cross section around $\rho \approx 0.15$,
suggesting that the resummation should be turned off
around this point. Taking $\rho \lesssim 0.15$
to be the region where the resummation is (nearly) off,
this suggests $\rho_L - 2 \sigma_L = 0.15$.
On the other hand, the singular cross section dominates over the remainder by a factor $\geq 5$ down to $\rho \approx 0.25$,
which motivates keeping the resummation on
until $\rho_L + 2 \sigma_L \approx 0.25$.
Thus the $g_L(\rho)$ sigmoid profile is configured to make the bulk of the transition  between $\rho=0.25$ and $\rho=0.15$, as evident in Fig.~\ref{fig:scales}.

For the right shoulder, a comparison of singular and non-singular contributions is not a useful tool
to determine the resummation region. The problem is that the cross section dies off very fast for $\rho \gtrsim \frac{1}{3}$ and moreover there is a second shoulder determined by the 4-parton threshold at $ \rho = 0.42$. 
In fact, the resummed distribution with canonical scale choices, although smooth at $\rho = \frac{1}{3}$, falls rapidly for larger $\rho$ and becomes negative by $\rho = 0.345$ (see Fig.~\ref{fig:shcen} below). To obtain a physical, positive cross section, the resummation must be turned off at least by then, so we take $\rho_{R2}=0.342$.
We take $\rho_{R1} = \frac{1}{3}$ to at least extend the right shoulder resummation as far as possible. 
 Part of the reason for using a quadratic profile in this region, which has $g_R(\rho)=0$ exactly for $\rho \ge \rho_{R2}$ is that even the very small difference between the sigmoid and quadratic profiles has big effects relative to the value of the cross section.
When we match the resummed distribution to fixed order, we add the singular and nonsingular contributions (see Eq.~\eqref{eq:matching}).
In the fully fixed order region the singular and nonsingular terms are individually large with opposite sign, and must sum to give a cross section which is close to zero. 
Even a small amount of leftover resummation from not turning off higher order logarithms can spoil this sensitive cancellation.
The cancellation in the fixed order region is guaranteed by the quadratic profile in Eq.~\eqref{eq:profile_quadratic}, which imposes $g(\rho\ge \rho_{R2})=0$ exactly, as can be seen in Fig.~\ref{fig:scales}.
The same delicate cancellations do not take place in the region controlled by $g_L(\rho)$, allowing us to use the smoother sigmoid for the left shoulder transition.

\subsection{Dijet matching \label{sec:dijetmatching}}

\begin{figure}
    \centering
    \includegraphics[width=0.49\textwidth]{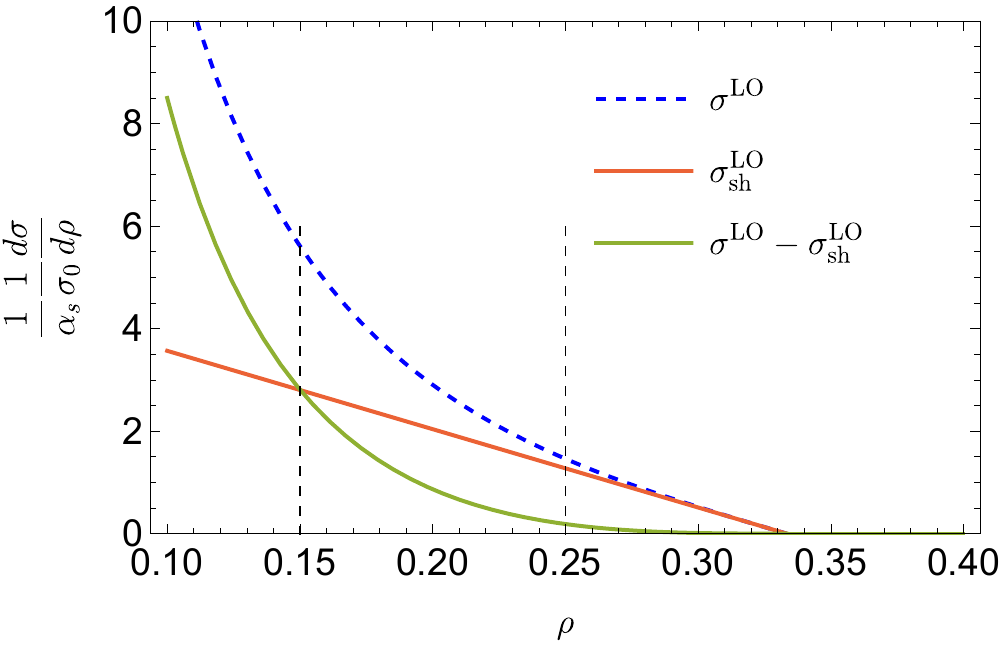}
    \includegraphics[width=0.49\textwidth]{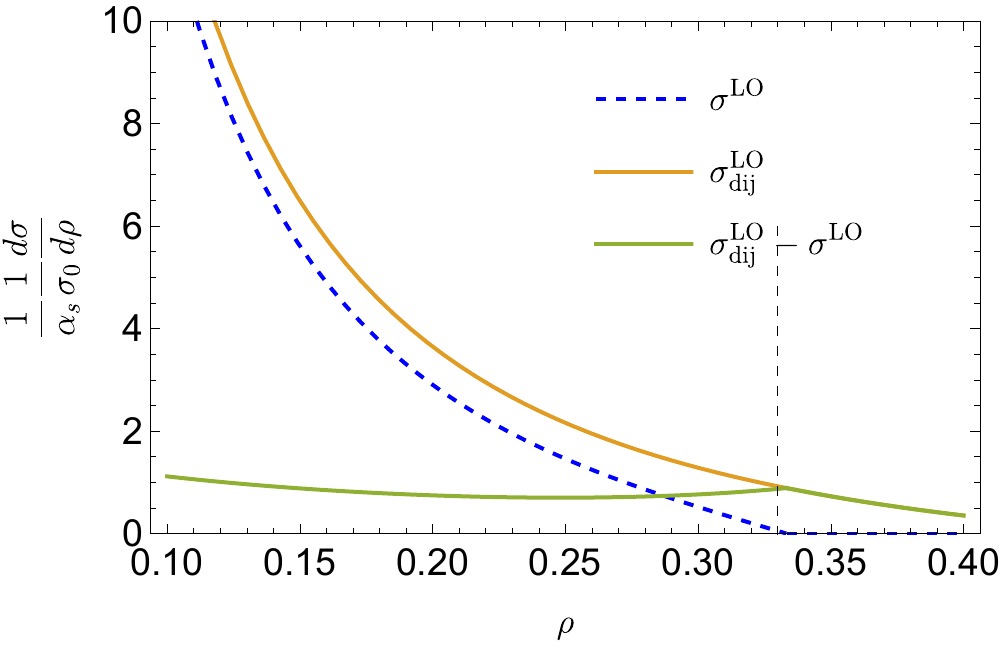}\\
    \includegraphics[width=0.49\textwidth]{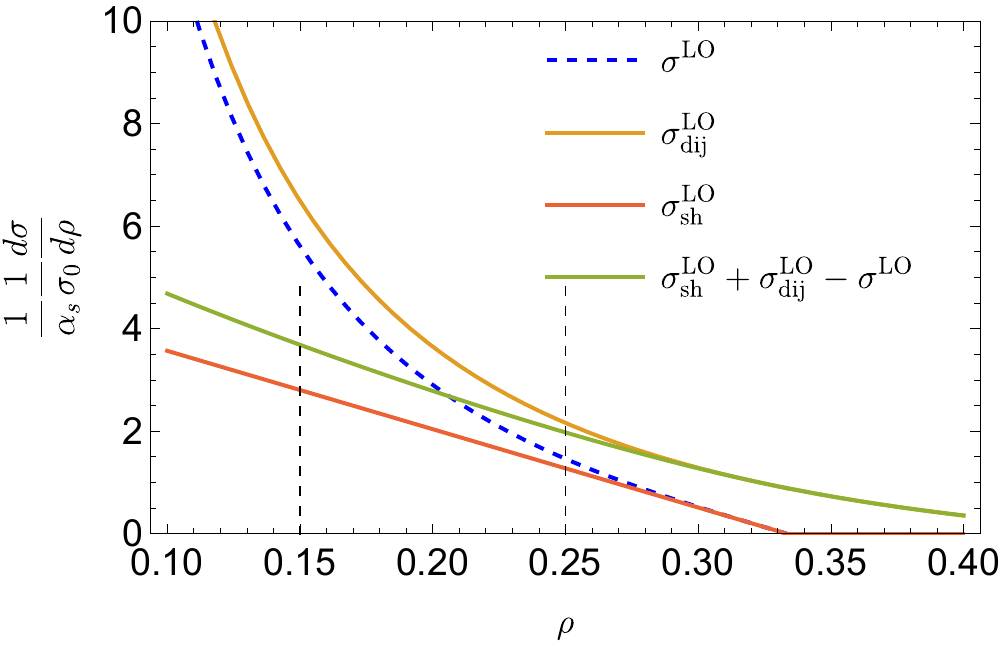}
    \includegraphics[width=0.49\textwidth]{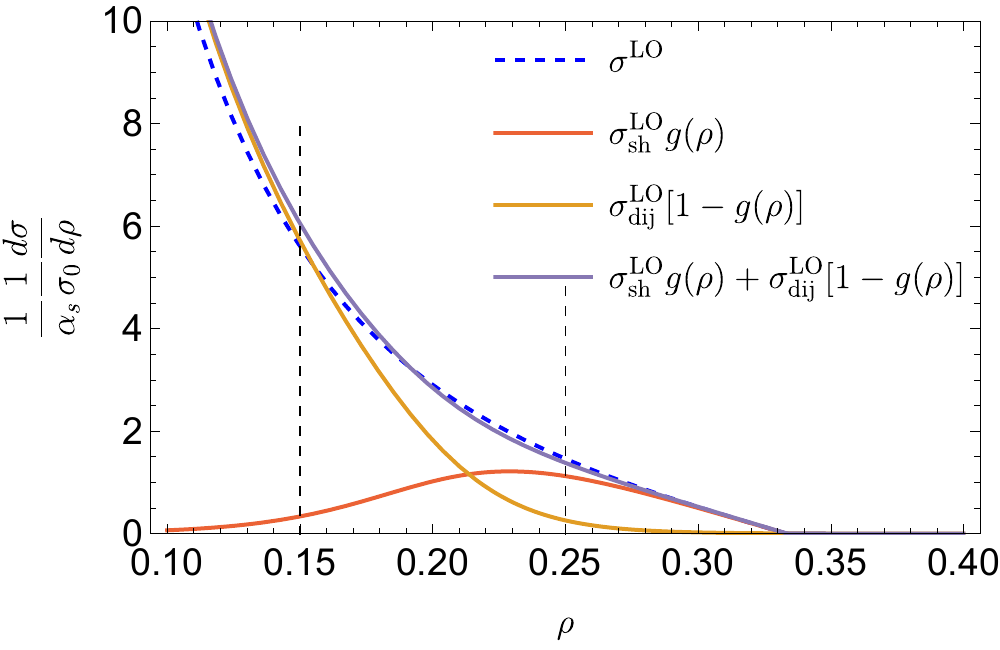}
  \caption{
Comparison of fixed order contributions from various sources to determine which terms dominate in various regions. In the top-left panel we compare the should singular terms $\sigma_\text{sh}^\text{LO}$ with the full LO and their difference. In the top right panel we compare the dijet singular terms $\sigma_\text{dij}^\text{LO}$ with the full LO and their difference. 
The bottom left shows the negative of the remainder when both expansions are subtracted from LO. The bottom right panel shows that taking a linear combination of the two LO singular contributions weighted by the function $g(\rho)$ used for the profile functions can nearly completely reproduce the full LO distribution.
In our actual matched results, the transition between different regions is implemented through scale choices, not a simple linear combination. 
}
    \label{fig:sing_vs_non_sing2}
\end{figure}

To obtain the best possible prediction valid across phase space,
we should match the resummed shoulder distribution to both the fixed-order result
and the resummed dijet distribution.
The resummation for heavy jet mass in the dijet limit has been done up to the N${}^3$LL level~\cite{Chien:2010kc}.
The general form for the resummed distribution in the dijet region is
\begin{align}
\frac{1}{\sigma_0} \frac{d \sigma^{\text{dij}}}{d \rho}
&=\exp\left[4C_F S(\mu_h,\mu_j)+4C_F S(\mu_s,\mu_j)-2A_H(\mu_h,\mu_s)+4A_J (\mu_j,\mu_s)\right]
 \nn\\
&\quad \times \left(\frac{Q^2}{\mu_h^2}\right)^{-2 C_F A_\Gamma (\mu_h,\mu_j)}
  H^{\text{dij}}(Q^2,\mu_h^2)\
  \tilde j_{q}\left(\partial_{\eta_1}+\ln\frac{Q\mu_s}{\mu_j^2}\right) \tilde j_q\left(\partial_{\eta_2}+\ln\frac{Q\mu_s}{\mu_j^2}\right)
 \nn \\
&\quad \times\tilde s_{\mu}(\partial_{\eta_1}) \tilde s_{\mu}(\partial_{\eta_2}) \tilde s_f\left(\partial_{\eta_1}-\partial_{\eta_2}\right)
 \frac{1}{\rho}
 \left(\frac{\rho Q}{\mu_s}\right)^{\eta_1+\eta_2} \frac{e^{-\gamma_E \eta_1}}{\Gamma(\eta_1)} \frac{e^{-\gamma_E \eta_2}}{\Gamma(\eta_2)}
\,,
\end{align}
where after taking derivatives one sets
\begin{align}
    \eta_1 =\eta_2 = 2C_F A_{\Gamma}(\mu_j,\mu_s) \,.
\end{align}
As discussed in Section~\ref{sec:scale_setting}, for the dijet region it is not necessary to set scales in position space (although one could). For simplicity, for the canonical scales we use conventional canonical scales in momentum space,
\begin{align} \label{eq:dijet_canonical_scales}
\mu_h = Q
\,, \qquad
\mu_{j,{\rm dij}}^{\rm can} = Q\sqrt{\rho}
\,, \qquad
\mu_{s,{\rm dij}}^{\rm can} = Q \rho
\,.\end{align}

Before we present the complete matched result including shoulder resummation,
we first discuss how to match the resummed dijet distribution to fixed order using standard procedures~\cite{Abbate:2010xh}
in a setup where the shoulder is not accounted for as a special region.
Here one adds a non-singular remainder with a common scale $\mu_{\rm FO}$ to the resummed dijet cross section evaluated with soft and jet profile scales (denoted together by $\mu^{\pro}_{\rm dij}$) 
\vspace{-5pt}
\begin{align}
	\frac{d\sigma^{\text{match,dij}}}{d\rho}
   = \overbrace{\frac{d\sigma^{\text{dij}}(\mu^{\pro}_{\rm dij})}{d\rho}}^{\text{resummed}}
   +\underbrace{ \biggl[ \frac{d\sigma^{\text{FO}}(\mu_{\rm FO})}{d\rho}
   - \overbrace{ \frac{d\sigma^{\text{dij}}(\mu_{\rm FO})}{d\rho}}^{\text{singular}} \biggr]}_{\text{non-singular}} \label{eq:prof_eqn}
\,.\end{align}
To ensure that the plain fixed-order result is recovered in the tail,
it is common to turn off the dijet resummation with profile functions
that interpolate between 
canonical scales
in Eq.~\eqref{eq:dijet_canonical_scales}
and equal scales $\mu_s = \mu_j = \mu_h$ directly as a function of $\rho$.
When matching to the plain fixed-order result,
appropriate parameters for the dijet profile can be inferred by the top right panel of Fig.~\ref{fig:sing_vs_non_sing2} in analogy to the discussion for the shoulder.
When the LO cross section vanishes at $\rho=\frac{1}{3}$, the dijet singular cross section exactly cancels the non-singular. Any small mismatch between the singular and non-singular, resulting for example from scales that are close but not equal in the profiles, can lead to large relative errors on the prediction.
This suggests that the dijet resummation should be 
turned off before reaching this point.
In practice, when comparing to partially matched results based on Eq.~\eqref{eq:prof_eqn},
we will make use of the profile functional form
developed in \refcite{Hoang:2014wka} for the $C$ parameter distribution,
adjusting the point $t_s$ (defined in that reference)
where the resummation is exactly turned off to be $t_s = \frac{1}{3}$.

We now turn to the complete matched result
incorporating both dijet and shoulder resummation.
The matched cross section takes the following form,
\begin{align}
	\frac{d\sigma^{\text{match}}}{d\rho}
	= \frac{d\sigma^{\text{dij}}(\mu_{\rm dij}^\pro)}{d\rho}
   + \frac{d\sigma^{\text{sh}}(\mu_{\rm sh}^\pro)}{d\rho}
	+\left[ \frac{d\sigma^{\text{FO}}(\mu_{\rm FO})}{d\rho}
	- \frac{d\sigma^{\text{dij}}(\mu_{\rm FO})}{d\rho}
	- \frac{d\sigma^{\text{sh}}(\mu_{\rm FO})}{d\rho}
	\right] \label{eq:dj_and_sh_match}
\,.\end{align}
This matched distribution has the formal properties that far from the shoulder, when the shoulder scales are equal to the fixed-order scale, it reduces to the dijet distribution matched to fixed order as in Eq.~\eqref{eq:prof_eqn}. Similarly, parametrically far from the dijet region, where the dijet scales reduce to fixed order scales, the dijet resummation turns off and it reduces to a matched distribution of shoulder and fixed order only.
We note that matching different regions of a one-dimensional distribution is in a sense easier than the joint resummation required for double-differential distributions as in~\cite{Lustermans:2019plv}, where generically there is a region where two separate types of logarithms can be simultaneously large.

For Eq.~\eqref{eq:dj_and_sh_match} to have the right limits, we moderate the dijet resummation 
through profile functions similar to Eq.~\eqref{cantopro}. We define the profile scales with a function $g_\text{dij}(\rho)$ playing the role that $g(\rho)$ plays for shoulder resummation:
\begin{align}
\mu_{j,\text{dij}}^{\pro}(\rho) &= \mu_h^{1-g_\text{dij}(\rho)} \big[\mu^{\mathrm{can}}_{j,\text{dij}}(\rho)\big]^{g_\text{dij}(\rho)}
\,, \nn \\
\mu_{s,\text{dij}}^{\pro}(\rho) &= \mu_h^{1-g_\text{dij}(\rho)} \big[\mu^{\mathrm{can}}_{s,\text{dij}}(\rho)\big]^{g_\text{dij}(\rho)}
\label{dijcantopro}
\,.\end{align}
Note that for, dijet resummation, the profile scales only depend on $\rho$, in contrast to Eq.~\eqref{cantopro} where they also depend on $z$, because we perform the resummation directly in momentum space.

The weight function $g_\text{dij}(\rho)$ should interpolate
between the region where the dijet resummation is fully on ($g_\text{dij} = 1$) and the region where it is off ($g_\text{dij} = 0$). To determine the parameters for $g_\text{dij}$, we return to the comparison of the singular and non-singular contributions, but now look at the non-singular with both singular shoulder and singular dijet subtracted. This overall non-singular contribution is given by the expression in square brackets in Eq.~\eqref{eq:dj_and_sh_match}.
At leading order it is shown (multiplied by $-1$) in the bottom-left panel of Fig.~\ref{fig:sing_vs_non_sing2}. 
From this plot, we conclude there is no reason to adjust the shoulder profiles when matching to dijet -- the shoulder singular still accounts for nearly the complete LO distribution for $\rho \gtrsim 0.25$ and the shoulder singular is the same size as the overall non-singular at $\rho \approx 0.15$. However, for the dijet, when matching to the shoulder resummation it is no longer sensible to maintain dijet resummation up to $\rho \approx 0.33$. If dijet resummation is allowed past the point where the dijet singular and the full nonsingular curves are nearly equal in magnitude, $\rho \approx 0.25$, the resummation would upset the large cancellations between them,
biasing the result. If dijet resummation is turned off at $\rho \gtrsim 0.25$, these cancellations are guaranteed.

A profile function for the matched dijet resummation consistent with these constraints is simply the same profile as for the shoulder, but growing in the opposite direction. That is, we take
\begin{align} \label{eq:def_g_dij}
g_\mathrm{dij}(\rho)
= \frac{1}{1+e^{(\rho - \rho_\mathrm{dij})/\sigma_\mathrm{dij}}} \, g_R(\rho)
 \,,
\end{align}
with 
$\rho_\mathrm{dij} = \rho_L$
and $\sigma_\mathrm{dij} = \sigma_L$ equal to those of $g_L$ in Eq.~\eqref{eq:baseprofile}.
The additional factor of $g_R(\rho)$ in Eq.~\eqref{eq:def_g_dij}
 ensures that all resummation (dijet and shoulder) is turned off in the far tail $\rho \gtrsim 0.342$,
maintaining consistency with our choice of profiles
for the right shoulder.

\subsection{Uncertainties}
A common way to assess theoretical uncertainties due to missing higher orders is to vary the factorization scale. We consider three scale variations in the shoulder region. A {\bf hard} variation involves changing $Q\to 2^{v_h} Q$ in all the scales in Eq.~\eqref{canonhjs}. A {\bf soft} variation involves varying the soft scale in the same was as for the hard variation, but not varying the hard and maintaining $\mu_j=\sqrt{\mu_s \mu_h}$.
For the {\bf jet} scale variation, we take $\mu_j =(\mu_h \mu_s)^{v_j}$ and allow $v_j$ to differ from $\frac{1}{2}$.
That is, these variations correspond to
\begin{align} 
	\mu_{h} &= 2^{v_h}Q  \notag \,, \\
 \label{varcanonhjs}
	\mu_{s,\text{sh}}(z) &= \sqrt{
 \Big(2^{v_h}2^{v_s} \frac{Q e^{-\gamma_E}}{|z|}\Big)^2
 +\Big(\mu_{s}^{\text{min}}\Big)^2}\notag \,,\\
	\mu_{j,\text{sh}}(z) &= \bigl[ \mu_{s,\text{sh}(z)} \bigr]^{v_j} \mu_{h}^{1- v_j}
  \,.
\end{align}
These varied scales are then inserted into the profiles using Eq.~\eqref{cantopro}. To vary the scales by a factor of 2 we separately vary $v_h$ and $v_s$ from $-1$ to $1$. This variation is inherited by the jet scale, maintaining its canonical relation with the soft and hard jet scale. In addition, we vary the jet scale away from its canonical value by varying the power $v_j= \{ 0.4, 0.6\}$ in the exponent.

For the dijet resummed distribution we consider scale variations of the form:
\begin{align} \label{dj_var}
    \mu_{h}&=2^{v_h}Q
    \,, \nn \\
    \mu_{s,\text{dij}}(\rho)&=2^{v_h} 2^{v^{\prime}_s} Q\rho
    \,, \nn \\
    \mu_{j,\text{dij}}(\rho)&=
      \bigl[ \mu_{s,\text{dij}(\rho)} \bigr]^{v_j'} \mu_{h}^{1- v_j'}
\,,\end{align} 
i.e., we vary the dijet scales in a fashion that is similar to the shoulder scales. We use a single hard scale variation $v_h$, which is the same for both dijet and shoulder logs and appears in the nonsingular cross section in Eq.~\eqref{eq:dj_and_sh_match} as well, where we take $\mu_\mathrm{FO} = \mu_h$.

We independently vary the profile parameters $\rho_L$, $\sigma_L$, $\rho_\mathrm{dij}$, and $\sigma_\mathrm{dij}$.
We do not consider variations of the profile functions in the right shoulder
since resummation must turn off quite quickly on the right shoulder, so any reasonable variations which do not make the cross section negative are in any case smaller than the hard scale variation.

A summary of the profile parameters along with their variation ranges is shown in Table~\ref{tab:varytab}. As a preliminary estimate of the overall perturbative uncertainty,
we consider an envelope of all the scale variations and the profile parameter variations.

\begin{table}
\centering
\begingroup
\renewcommand{\arraystretch}{1.5}
\begin{tabular}{ |c|c|c| }	
\hline
	\text{Parameter} & \text{Central value} & \text{Variation Range} \\[2pt]
	\hline
	$v_h$ & $0$ & $-1$\text{ to }$1$\\[-5pt]
	$v_s, v_s'$ & $0$ & $-1$\text{ to }$1$\\[-5pt]
	$v_j, v_j'$ & $0.5$ & $0.4$\text{ to }$0.6$\\[-5pt]
	$\rho_L, \rho_{\rm dij}$ & 0.2 & $0.18$\text{ to }$0.22$\\[-5pt]
	$\sigma_L, \sigma_{\rm dij}$ & $0.025$ & $0.02$\text{ to }$0.03$\\
 \hline
\end{tabular}
\endgroup
\caption{Parameter ranges used for the uncertainty estimates in both the shoulder profile function and dijet profile function. We fix the profile function for the right shoulder
to have $\rho_{R 1}=\frac{1}{3}$ and $\rho_{R 2} = 0.342$, and fix $\mu_s^{\text{min}}= 2 \GeV$. Varying these parameters has negligible effect on the distribution.
For the dijet resummation (when matching to the fixed-order result only),
we use the profile function from \cite{Hoang:2014wka}, with the following set of parameters defined in that reference:
$n_0=2$, $n_1=10$, $t_2=0.2\pm 0.025$, $t_s=\frac{1}{3}\pm 0.025$, $r_s=1^{\times 2}_{\div 2}$, $e_j=0\pm 1$, $e_h=1^{\times 2}_{\div 2}$.
}

\label{tab:varytab}
\end{table}

\section{Numerical results \label{sec:numerics}} 
\begin{figure}[t]
	\centering
	\includegraphics[width=0.49\textwidth]{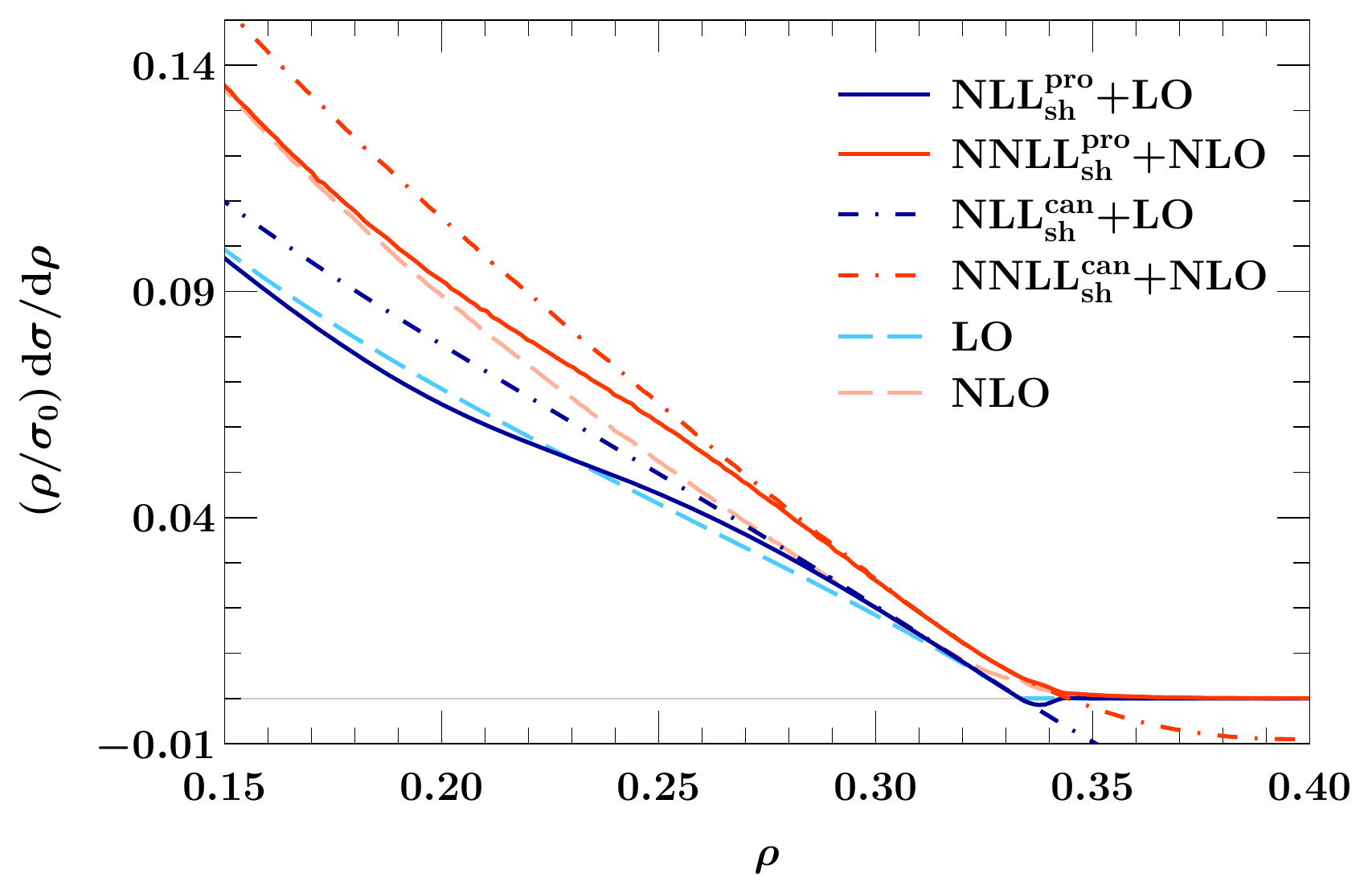}
	\hfill
	\includegraphics[width=0.49\textwidth]{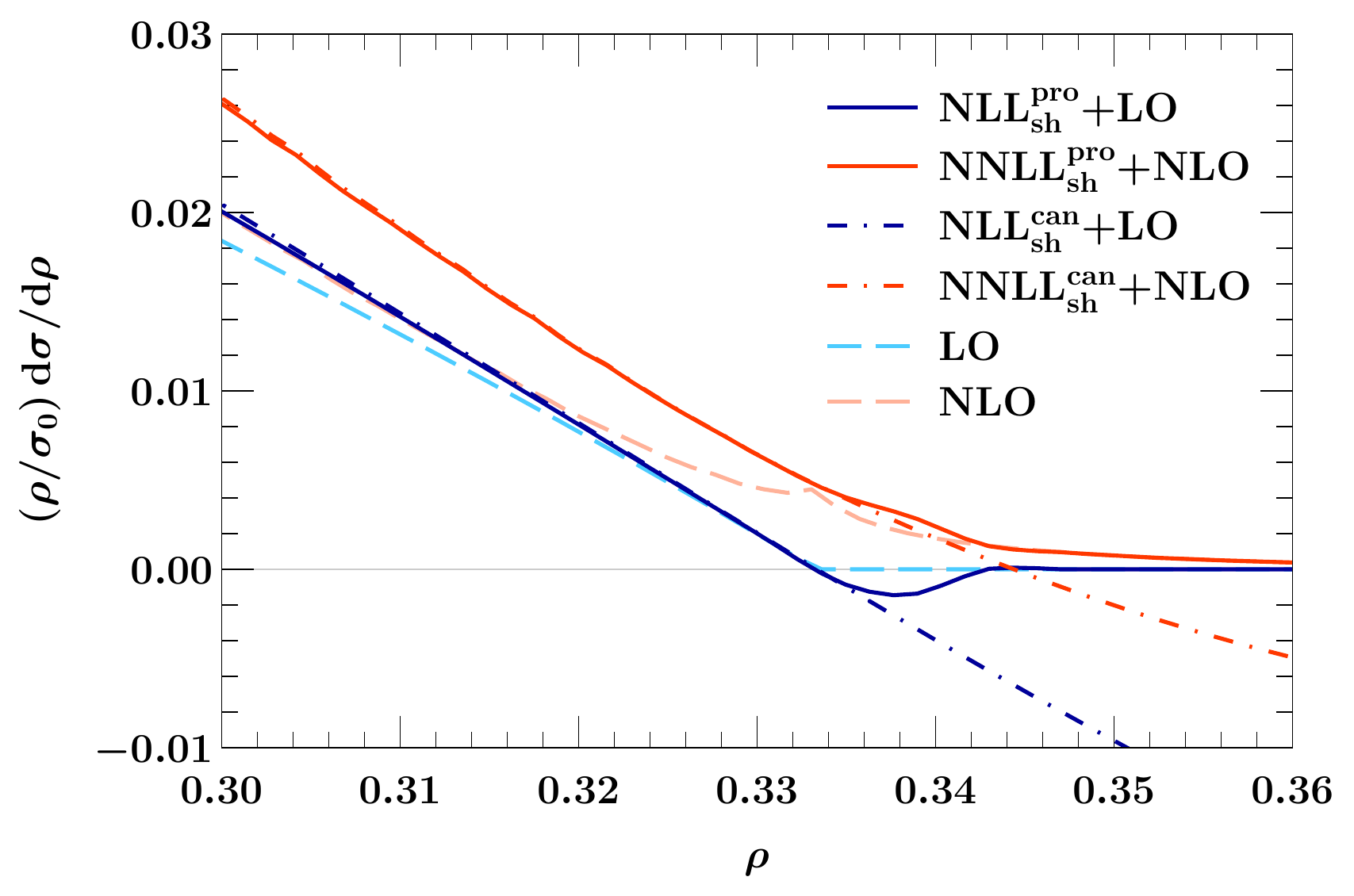}
	\caption{
    Plots of Sudakov shoulder resummation in heavy jet mass distribution, with matching to FO distributions. Profiled scales are indicated by solid curves, and interpolate between canonical scales near $\rho=\frac{1}{3}$ (dot-dashed curves) and  fixed order scales  ($\mu_h =\mu_{j,\text{sh}}=\mu_{s,\text{sh}}$, dashed curves). The right panel indicates a zoomed in view of the distribution near $\rho=\frac{1}{3}$. 
 }
	\label{fig:shcen}
\end{figure}
With all the ingredients in place, we can now look at numerical predictions. All our results are given for $e^+e^-$ hadron events at $Q=91.2$ GeV with $\alpha_s = 0.1179$~\cite{ParticleDataGroup:2022pth}.

We first examine the prediction using Sudakov shoulder resummation without dijet matching in Fig.~\ref{fig:shcen}. We compare canonical scales choices, as given in Eq.~\eqref{canonhjs}, to profile scales, as given in Eq.~\eqref{cantopro}, and compare to fixed order distributions at LO and NLO~\cite{Catani:1996jh,Catani:1996vz}. 
One can also see from this figure that the profile functions smoothly transition out of the region where shoulder logarithms are large ($\rho \sim \frac{1}{3}$) into the fixed order region at smaller $\rho$.
This figure also confirms that setting scales in position space eliminates the Sudakov Landau poles encountered in Ref.~\cite{Bhattacharya:2022dtm}.
The right panel in Fig.~\ref{fig:shcen} zooms in on the region around $\rho = \frac{1}{3}$.
It shows that in the matched curves, the non-analytic behavior (kink) of the fixed-order distributions is completely resolved by the resummation of shoulder logarithms,
leaving behind a distribution with continuous first derivative. 
Although the NLL+LO matched curve goes negative for $\rho > \frac{1}{3}$, this is due to our boundary conditions on integrating the third derivative which makes it match exactly to LO at $\rho=\frac{1}{3}$. For the NNLL+NLO curve, the offset is positive and the profile functions of the right shoulder are chosen to turn off the shoulder resummation quickly enough to avoid a negative cross section.
Comparing the NLO curve to those at NNLL+NLO, we see that the shoulder resummation has a considerable impact for values of $\rho \gtrsim 0.25$.

\begin{figure}[t]
	\centering
        \includegraphics[width=0.49\textwidth]{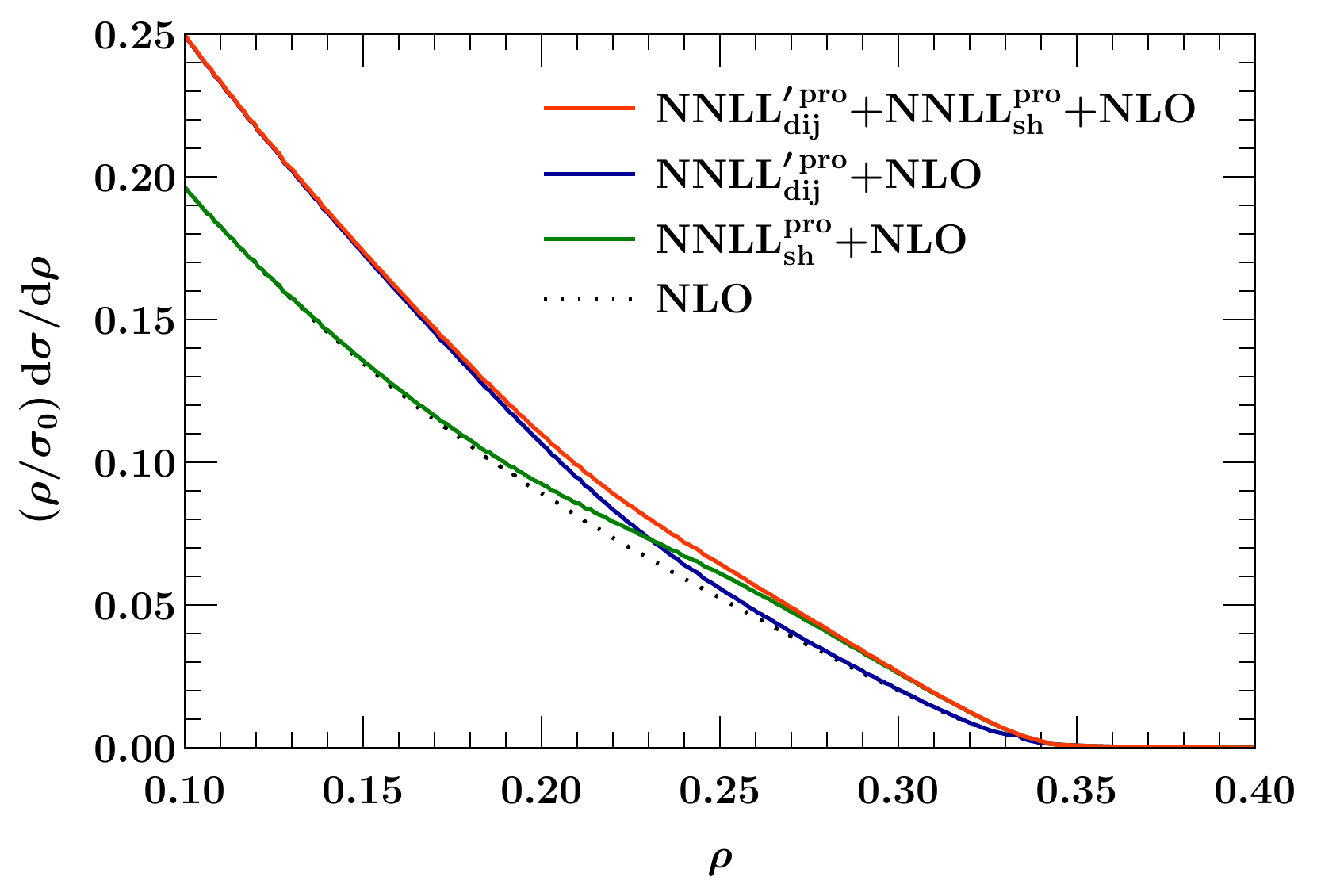}
        \hfill
	\includegraphics[width=0.49\textwidth]{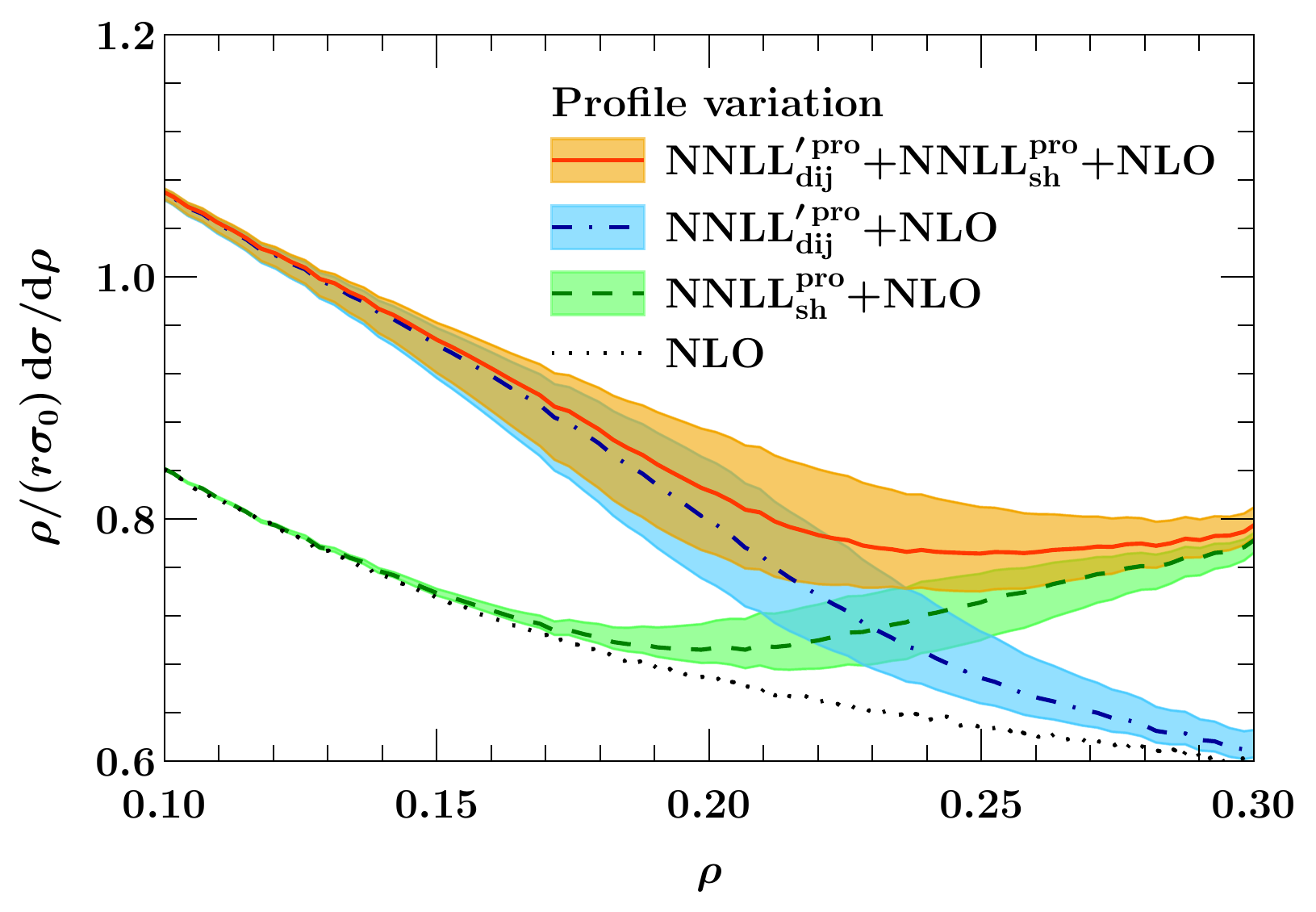}\\
	\caption{Relative contributions dijet and shoulder resummation to the matched distribution. Right panel shows the ratio with respect to $r=\frac{1}{3}-\rho$, i.e. $\frac{\rho}{r} \frac{1}{\sigma_0}\frac{d\sigma}{d\rho}$.
This is done to highlight and expand the transition region without distorting the curves.
For the green (blue) curve, we vary the shoulder (dijet) profile only, and for the red curve, we present the envelope of both profiles. 
}	\label{fig:joincen}
\end{figure}

\begin{figure}[!htp]
	\centering
    \includegraphics[width=0.49\textwidth]{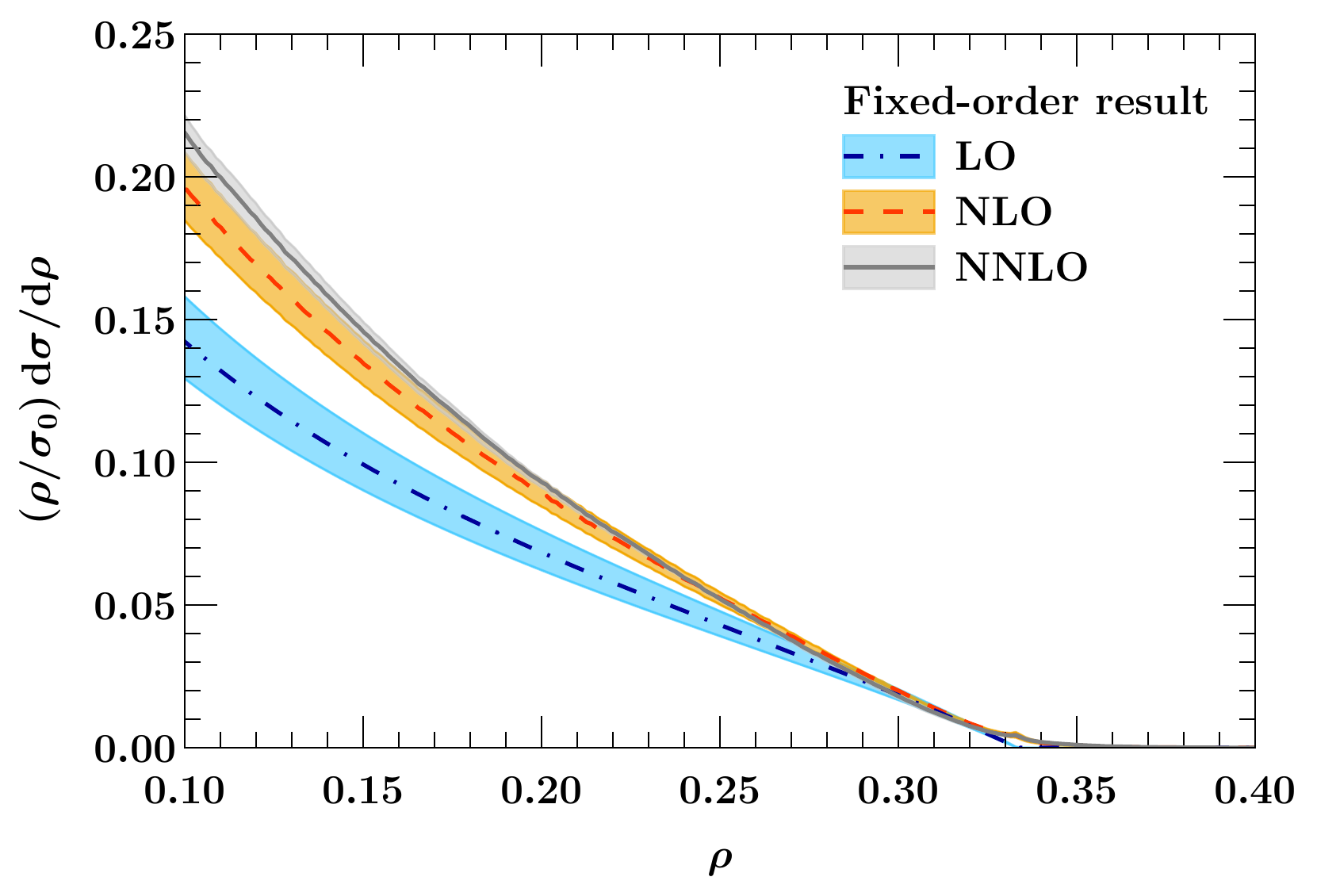}
	\hfill
	\includegraphics[width=0.49\textwidth]{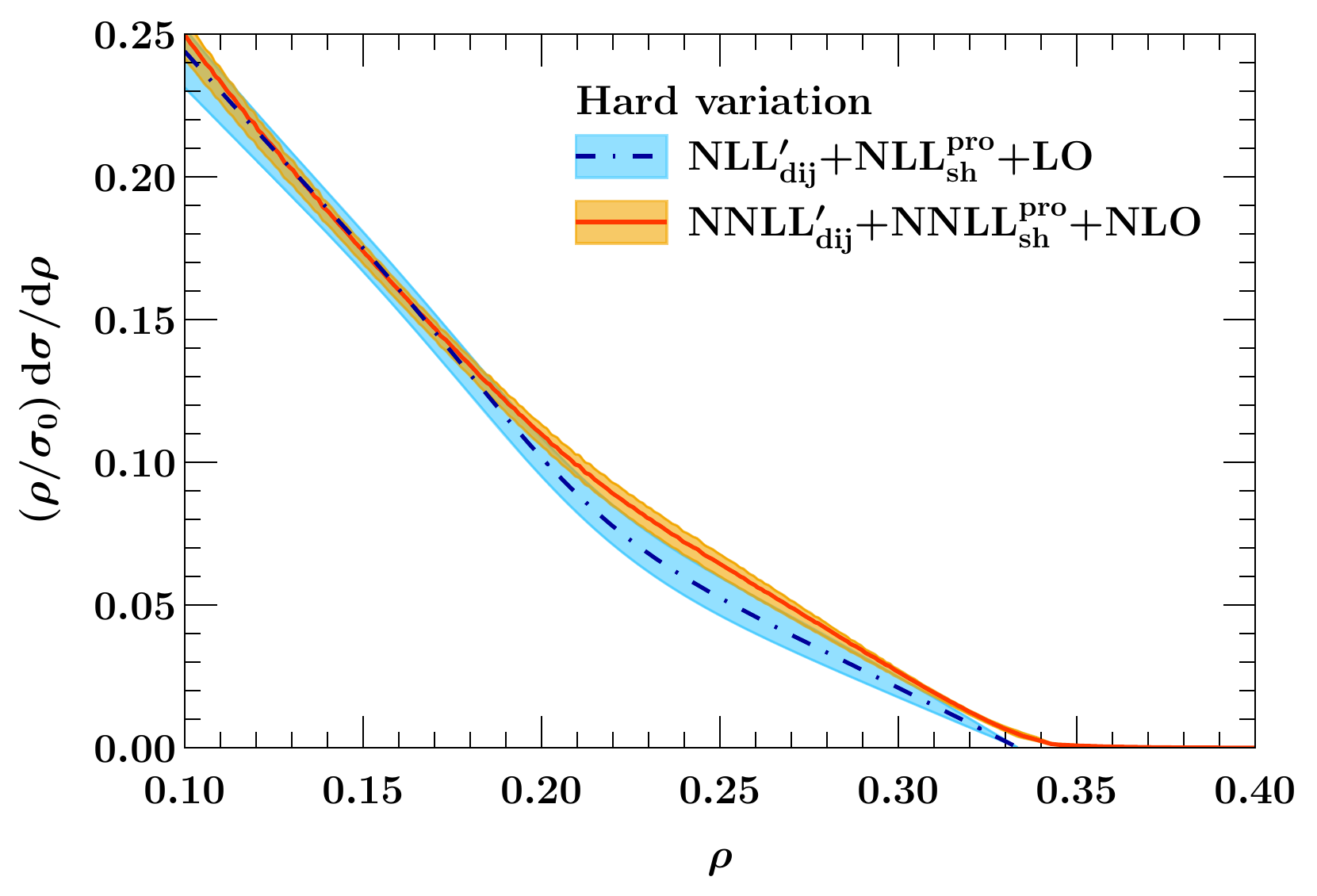}\\
	\includegraphics[width=0.49\textwidth]{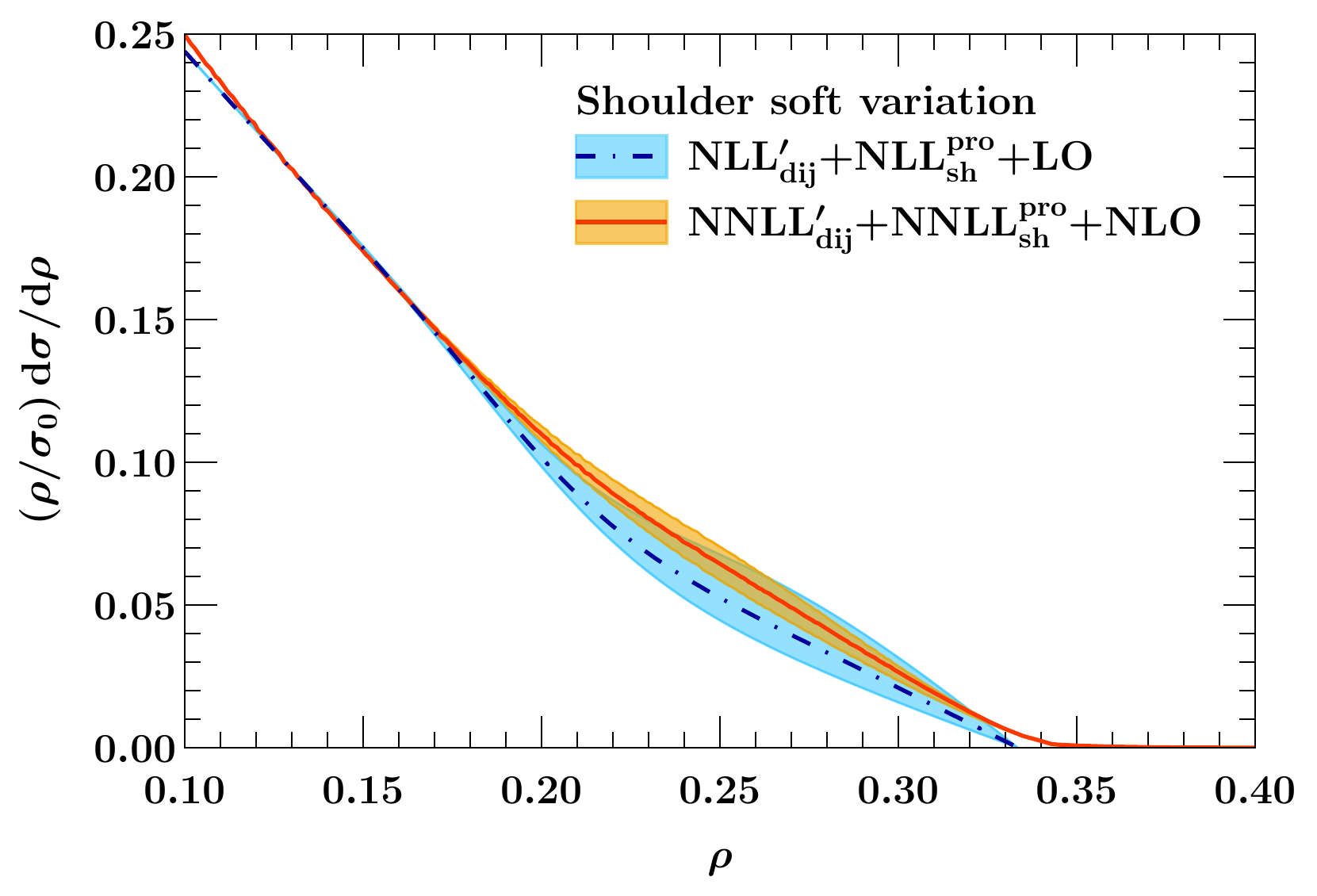}
	\hfill
    \includegraphics[width=0.49\textwidth]{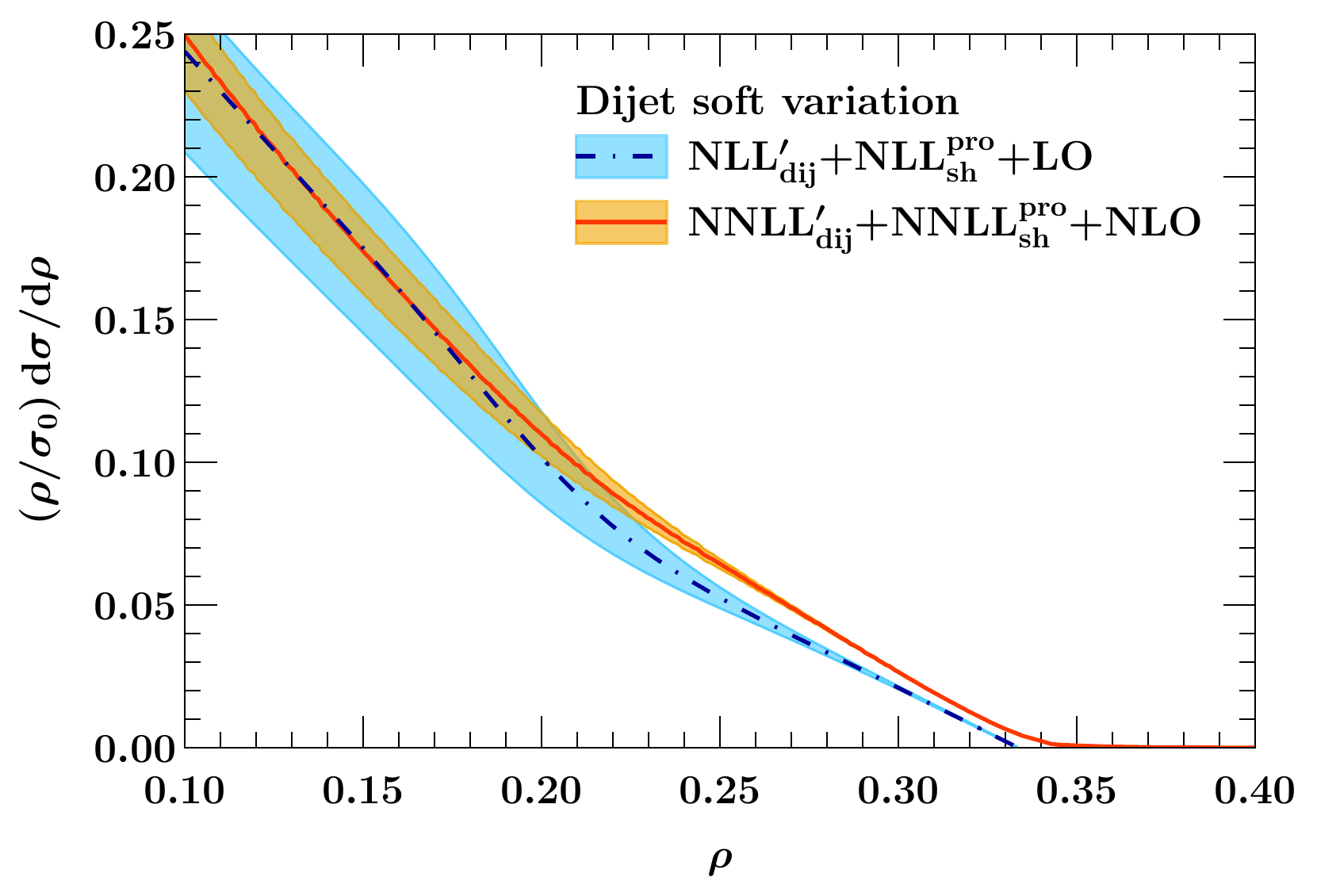}\\
	\includegraphics[width=0.49\textwidth]{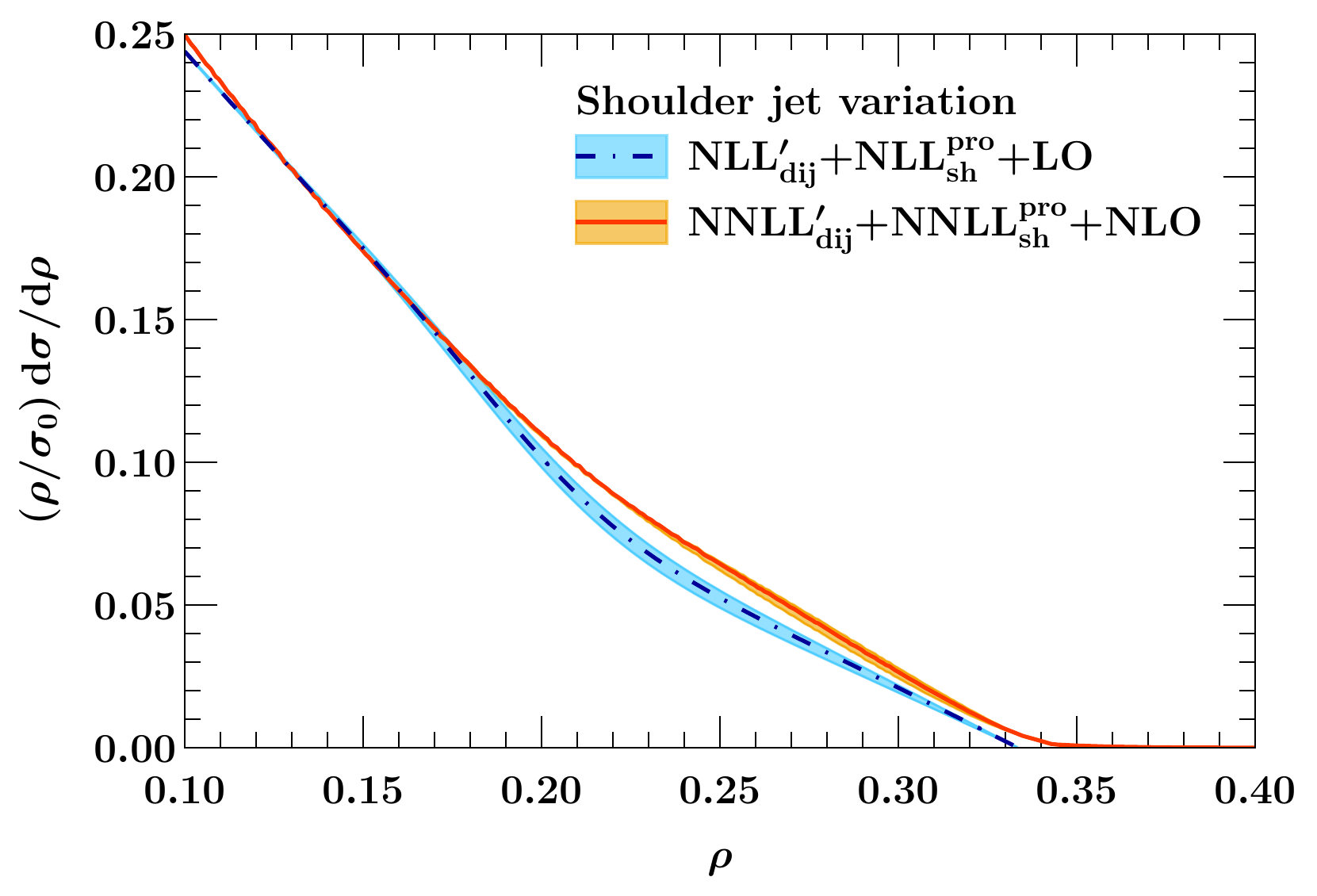}
    \hfill
    \includegraphics[width=0.49\textwidth]{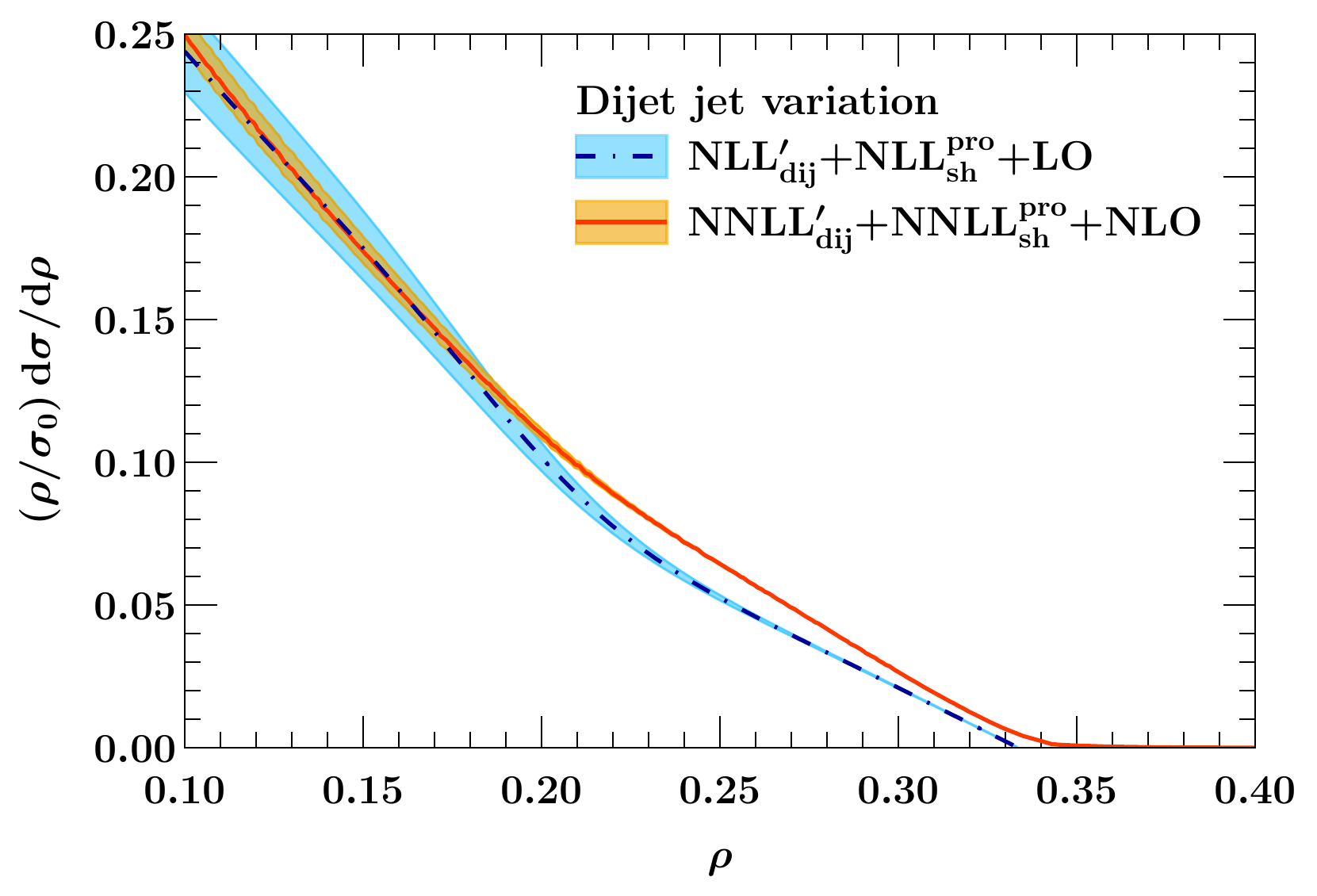}\\
	\includegraphics[width=0.49\textwidth]{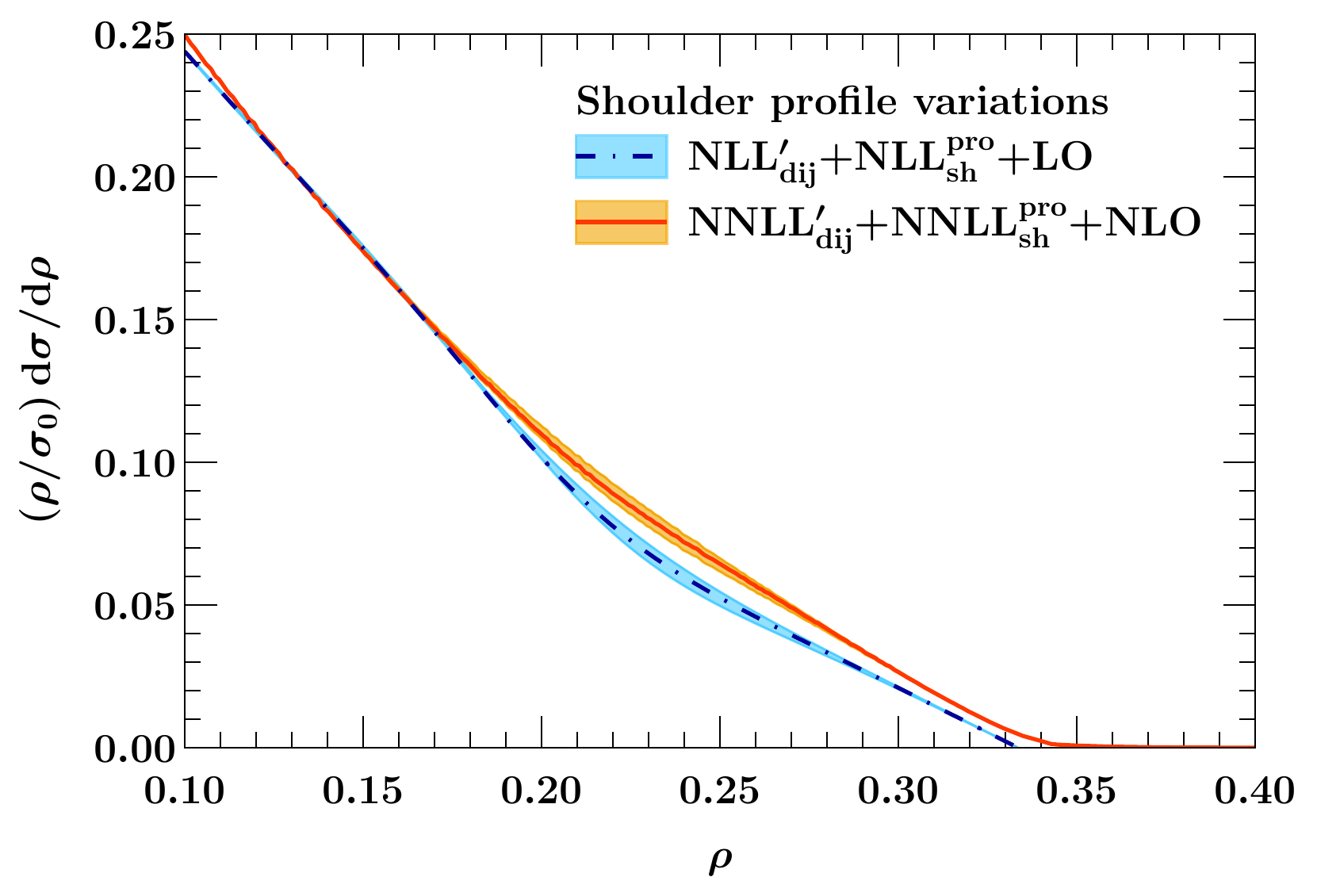}
	\hfill
    \includegraphics[width=0.49\textwidth]{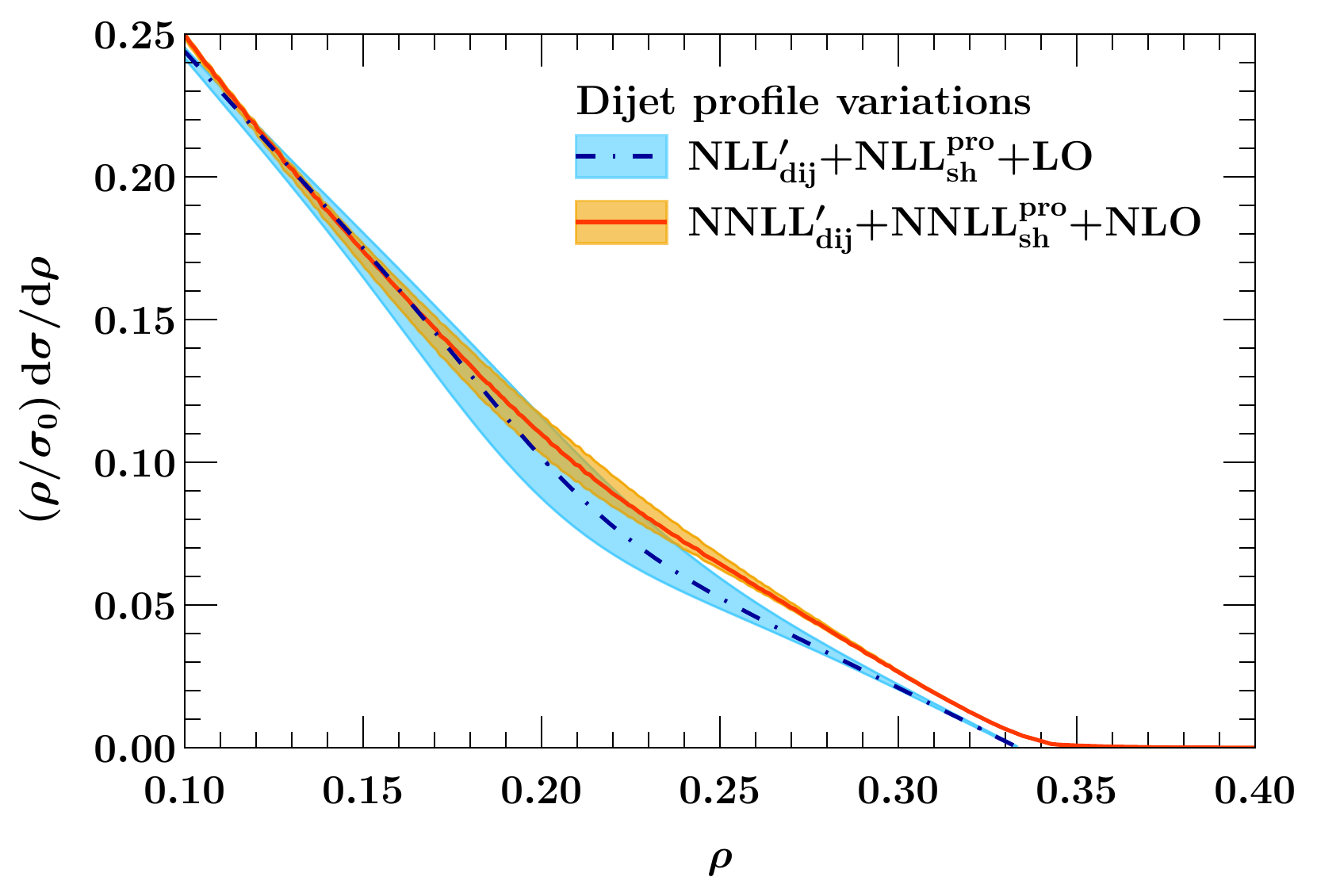}\\
	\caption{Plots showing various uncertainties associated with the scale variations. Top left panel shows the fixed order distributions for reference only (these uncertainties are obtained by varying the renormalization scale by a factor of 2, which are part of our hard uncertainties and not included in the overall envelope). 
 The top-right panel shows the hard variation, where $\mu_h$ is varied in the dijet, shoulder and matched contributions simultaneously. The remaining panels show soft, jet and profile variations for the shoulder and dijet contributions separately.
}
	\label{fig:shvar}
\end{figure}

Next, we consider results with both shoulder and dijet resummation, and show in Fig.~\ref{fig:joincen} how the profile functions interpolate between the regions where shoulder resummation and dijet resummation are important.
As discussed in Section~\ref{sec:dijetmatching},  we use the same profile functions for both dijet resummation and shoulder resummation, such that the dijet resummation is active in the range $0\leq \rho \lesssim 0.2$, and the shoulder resummation is active in the range $\rho\gtrsim 0.25$,  with a smooth transition between them. The parameters used and their variations are given in Table~\ref{tab:varytab}.
We use NNLL resummation for the shoulder, NNLL$^{\prime}$ resummation for the dijet,
and match to the fixed NLO ($\ord{\as^2}$) distribution.
These form a consistent set of resummed orders since they both contain the complete
$\ord{\as^2}$ information in the respective singular limit.
We see that the matched resummed distribution smoothly merges into the dijet resummation near $\rho\sim 0.2$, and into the shoulder resummation near $\rho \sim 0.25$, as desired. In the right panel of Fig.~\ref{fig:joincen}, we highlight the transition region by dividing the distribution by $r=\rho-\frac{1}{3}$. In this panel we include profile 
uncertainty estimates obtained  by varying the profile parameters within ranges given in Table~\ref{tab:varytab}.
In all cases the band shown is the envelope of the relevant individual variations.

\begin{figure}[t!]
	\centering
	\includegraphics[width=0.7\textwidth]{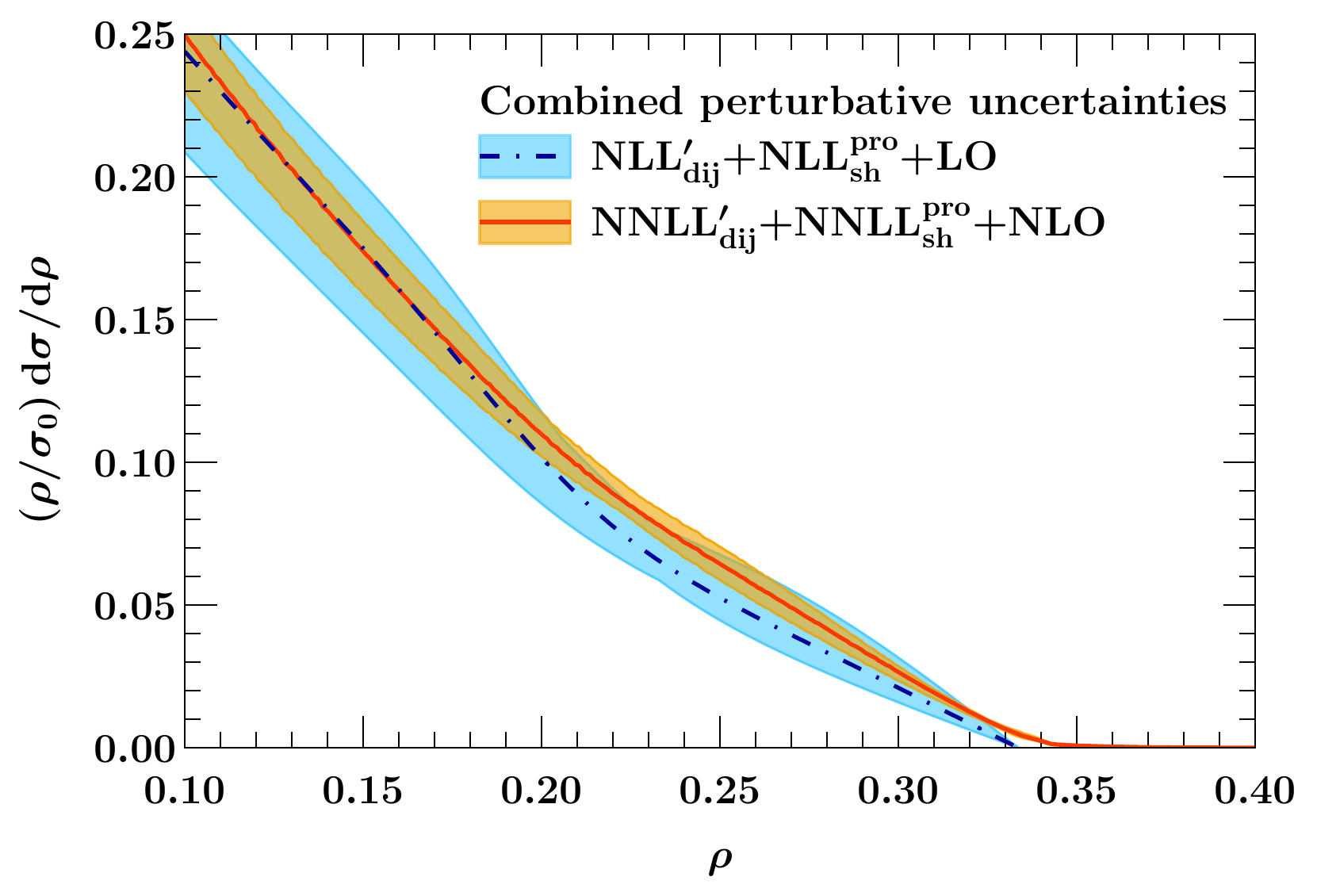}
	\caption{Envelope of all perturbative scale variations
	in the matched predictions shown in Fig~\ref{fig:shvar}.
} 
	\label{fig:allvarplot}
\end{figure}

We next plot the various uncertainties from individual hard, jet, and soft scale variations and profile variations in Fig.~\ref{fig:shvar}. The first panel shows the fixed-order predictions. We use \textsc{Event2}~\cite{Catani:1996jh,Catani:1996vz} 
for NLO  and \textsc{CoLoRFulNNLO} \cite{DelDuca:2016csb,DelDuca:2016ily} for  NNLO.
The hard scale variation for the shoulder and dijet resummation is correlated, since there is only a single common hard scale for both factorizations. We vary the hard scale by a factor of 2 above and below its central value at 91.2 GeV. The soft scales ($\mu_{s,\text{sh}}, \mu_{s,\text{dij}}$) are both independently varied for the shoulder and the dijet. The range of variation is approximately a factor of 2 relative to their pointwise central value, while simultaneously varying the jet scale using the relation $\mu_j=\sqrt{\mu_h \mu_s}$. The jet scale variation is done as per the prescription in Eqs.~\eqref{varcanonhjs} and \eqref{dj_var}. Note that all variations maintain the monotonicity condition $\mu_s < \mu_j <\mu_h$ (up to the numerically negligible region where $|z|<0.56$), and are thus sensibly defined. 
Finally, the profile parameter variations are performed
as summarized in Table~\ref{tab:varytab}.
We recall that the shoulder profile variation is performed while fixing the profile for dijet scales and varying only the shoulder profiles, and vice-versa for the dijet profile variation.

Our final prediction, with uncertainties given by an overall envelope of all variations, is shown in Fig.~\ref{fig:allvarplot}. Increasing the resummation order leads to a visible decrease in perturbative uncertainty.
At the same time, the NNLL result is fully covered by the larger uncertainty estimate at NLL.
The dominant source of perturbative uncertainty in both the shoulder and dijet region is the variation of the respective soft scale.

\begin{figure}[t!]
	\centering
	\includegraphics[width=0.49\textwidth]{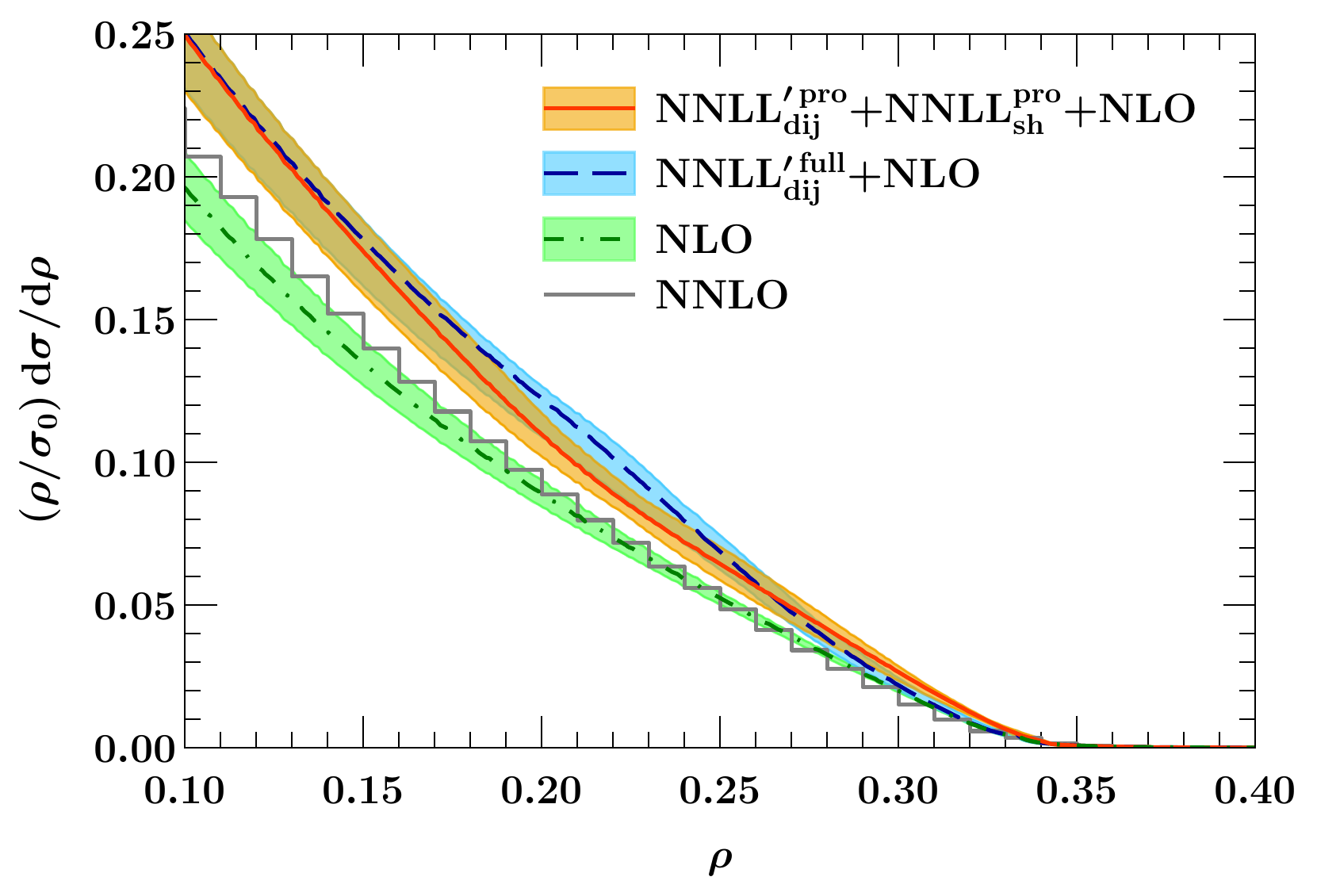}
	\hfill
	\includegraphics[width=0.49\textwidth]{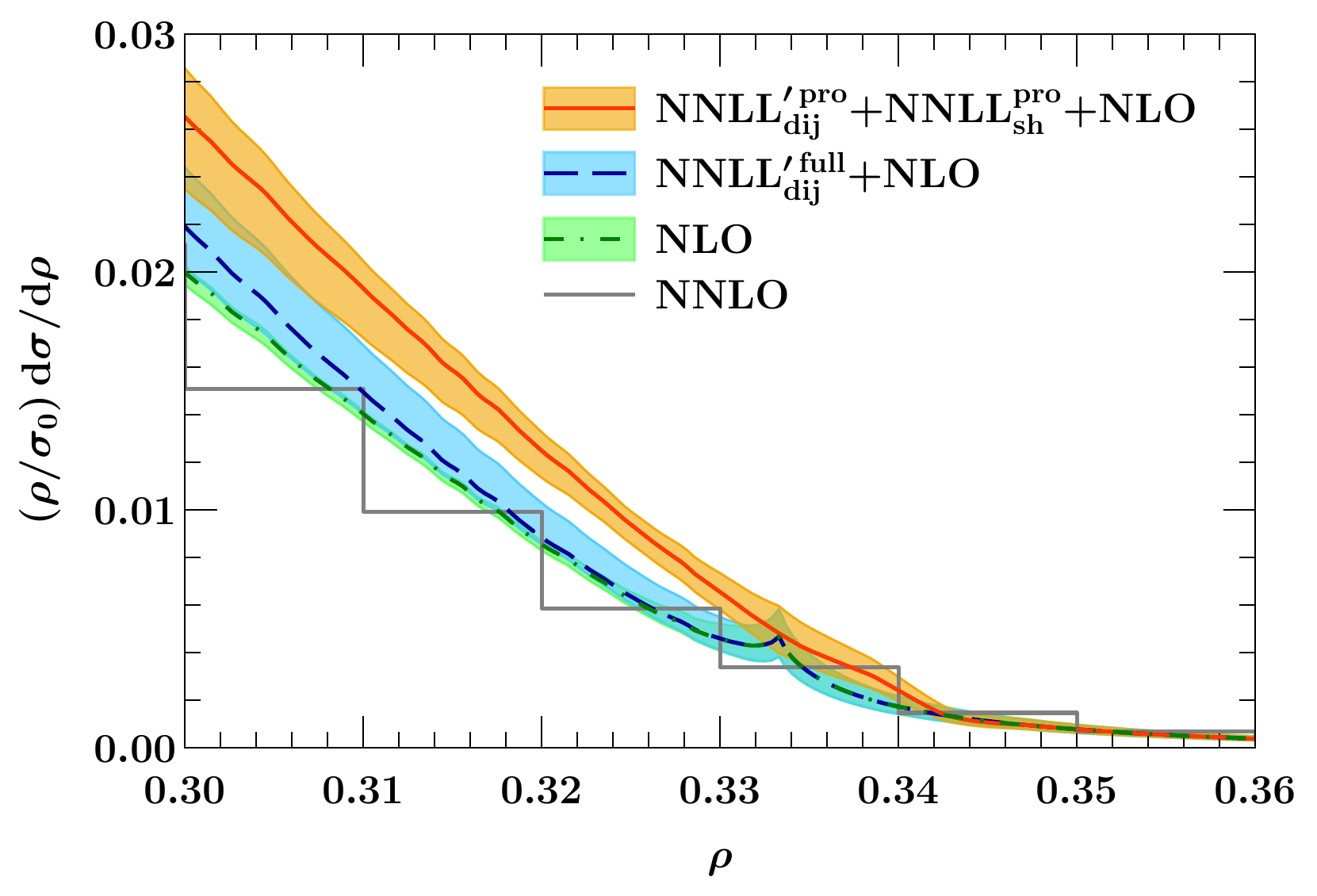}
 \\
        \includegraphics[width=0.49\textwidth]{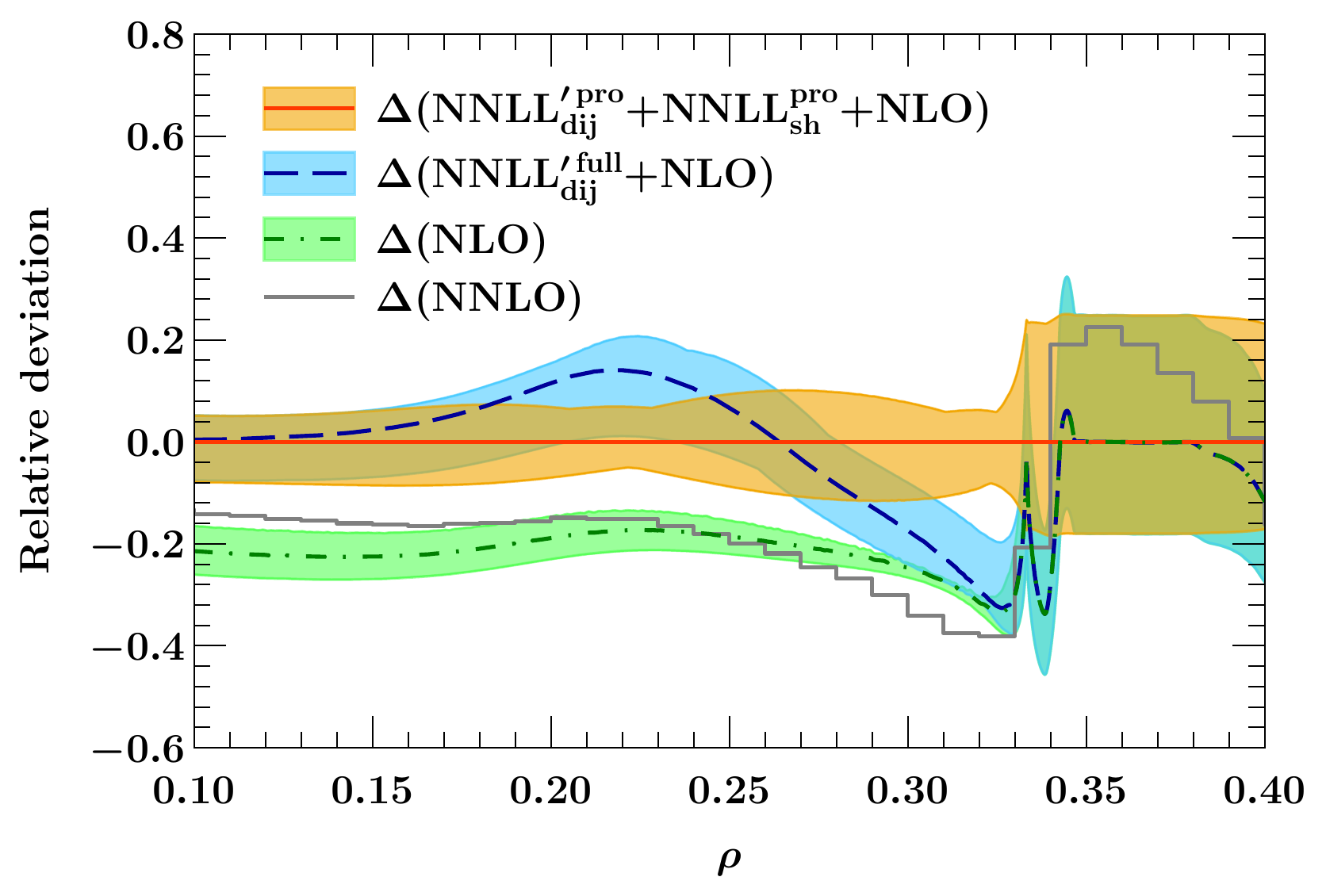}
	\hfill
	\includegraphics[width=0.49\textwidth]{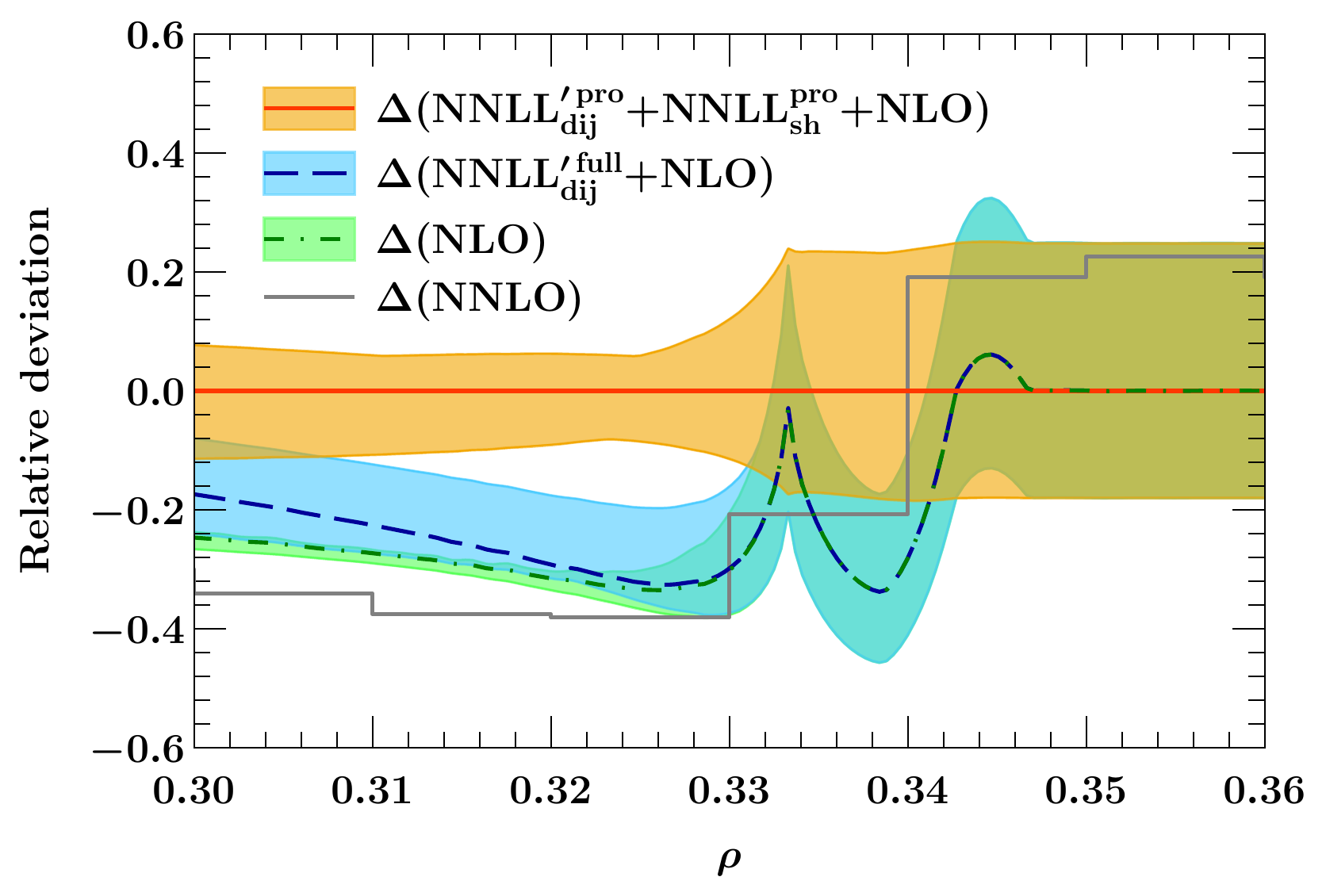}
	\caption{
Plots comparing our dijet + shoulder resummation to both ``full" dijet resummation (matched to NLO, but no shoulder) and fixed order results at NLO and NNLO.  
In the left panels we show results over a wider range of $\rho$ where both shoulder and dijet resummation are active, while in the right panels we zoom in on values of $\rho$ near the Sudakov shoulder. 
In the upper panels we show absolute cross sections, while in the lower panels we show variations relative to the central curve for our highest order shoulder+dijet resummed result, namely
$\Delta(\text{resum})\equiv \frac{\text{resum}}{\text{NNLL}^{\prime\text{pro}}_{\text{dij}}+\text{NNLL}^{\text{pro}}_{\text{sh}}+\text{NLO}}-1$.}
    \label{fig:money_plot}
\end{figure}

Finally, we compare our prediction combining shoulder and dijet resummation to the previous approach where only dijet resummation was included~\cite{Chien:2010kc} in Fig.~\ref{fig:money_plot}. For reference, we also plot the NLO and NNLO distributions in the range $0.10\lesssim\rho\lesssim 0.4$.
Fig.~\ref{fig:money_plot} also shows the associated perturbative uncertainty with each resummation, which we obtain by considering an envelope of all scale and profile variations. (Note that the resummed results with and without shoulder resummation use different profile functions, as described in Section \ref{sec:profiles}.) The uncertainty associated with the FO curves is given by a scale variation by a standard factor of 2 around the central value $\mu_{\text{FO}}=m_Z$.
The sharp spikes in the dijet resummation and FO distributions are indicative of the presence of large shoulder logs, which are exponentiated by shoulder resummation, leaving behind a smooth result.
The addition of shoulder resummation has a numerically significant impact on the value of the cross section in the region $\rho \gtrsim 0.2$, which includes both the transition region between dijet and shoulder resummation $0.2\lesssim\rho\lesssim0.25$ and the region where shoulder resummation dominates, for $\rho\gtrsim 0.25$.
These differences will be relevant for $\alpha_s(m_Z)$ fits using the heavy jet mass distribution.
Next, turning to the region of very large $\rho > \frac{1}{3}$ to the right of the shoulder, we can see that the fully matched prediction merges onto the fixed NLO result past $\rho \gtrsim 0.34$, as desired for both the central value and uncertainty band. Since the NLO is the first non-zero order to contribute in this region, its $\ord{20 \%}$ uncertainty is substantially larger than any of the predictions below the shoulder.  
Finally, we observe that the fully matched prediction with shoulder resummation provides a realistic uncertainty estimate across the shoulder,
smoothly connecting the NNLL uncertainties on the left with the NLO (effectively LO) uncertainties on the right.
In contrast, both fixed order and dijet-resummation-only predictions are discontinuous at the shoulder and their  
uncertainty bands can be seen to be unreliable, since they do not capture the impact of the large logarithmic contributions. From the NNLO curve in Fig~\ref{fig:money_plot} we see that this higher order contribution does not improve the ability of fixed order results to capture the large resummation effects in the $0.1<\rho<0.34$ region. Including NNLO results can be expected to reduce the perturbative uncertainty for $\rho>0.34$, where we note that the NNLO central value is within the NLO uncertainties.

\section{Conclusions \label{sec:conclusions}}
In this paper we have considered the resummation of Sudakov shoulder logarithms in the heavy jet mass distribution in $e^+e^-$ events. These logarithms were identified by Catani and Webber~\cite{Catani:1997xc} in 1997, and a systematically improvable method to compute and resum these logarithms was developed in 2022~\cite{Bhattacharya:2022dtm}. When the factorized distribution is resummed in momentum space, that is the space where the heavy jet mass itself lives, spurious ``Sudakov Landau poles" result. These poles are divergences in the distribution arising from scale setting and have no physical significance. Similar poles have been found in factorized expressions at hadron colliders, such as the $q_T$ spectrum in Drell-Yan. This is the first example we know of for Sudakov Landau poles in a non-transverse momentum distribution.

The Sudakov Landau poles in the heavy jet mass distribution were found in~\cite{Bhattacharya:2022dtm} and prohibited the quantitative application of the resummation formula. In this paper we have identified the origin of these poles as arising from an inconsistent solution of the evolution equations: when scales are set in momentum space, subleading terms in logarithmic power counting are included which give rise to these singularities. In contrast, when the resummation is performed in position space and the momentum-space distribution is only generated by a final Fourier transform after scales have been set, then the Sudakov Landau poles are removed. We verified this analytically and numerically. 

In addition to the Sudakov Landau pole, another complication in the HJM shoulder is the existence of a linear UV divergence in $\frac{d\sigma}{d\rho}$ associated with the region where the light and jet mass are both very large, but their difference is finite. We regulate this divergence by considering the third-derivative $\frac{d^3\sigma}{d\rho^3}$, and specifying appropriate boundary conditions to carry out the double integral, which fix terms of the form $c_0 + c_1 \rho$.
In particular we demanded that the resummed and matched distribution and its first derivative agree with the fixed order distribution at $r=\frac{1}{3}-\rho=0$, where perturbative uncertainties provide natural variations in this constraint.

The trijet hemisphere soft function for the Sudakov shoulder is a two-scale function, like the one for heavy jet mass in the dijet region. Thus it has non-global structure which is worth exploring. Unlike the hemisphere soft function, the trijet hemisphere soft function is not symmetric in its arguments, and there is not a natural way to separate its global and non-global parts. While one might like to try to choose different complex scales $\mu_{s\ell}$ and $\mu_{sh}$ for the light and heavy hemispheres, we find such complex scales are problematic,
since it is difficult to ensure logs of the form $\ln\frac{\mu_{s\ell}}{\mu_{sh}}$ do not induce spurious effects. If instead, one uses a single real soft scale, then the only missing logarithms are non-global in nature and integrate to a fixed-order contribution to the trijet soft function starting at order $\alpha_s^3$, similar to the leading non-global logarithmic contribution to the dijet heavy jet mass distribution. Such contributions are formally N${}^3$LL, the same order as the unknown 2-loop soft function constant. Thus our NNLL calculation is self-consistent with real scales and dropping the non-global contributions.

To make a quantitative prediction, we needed to match between the dijet and shoulder regions. This requires us to modulate the hard, jet and soft scales with profile functions, where as $\rho$ grows we turn off the dijet resummation and turn on the shoulder resummation. We included such profiles and provided the first quantitative predictions that include both types of resummation. We find that at the NNLL level, the Sudakov shoulder resummation provides as much as a 20\% increase over the NLO cross section for values of $\rho \gtrsim 0.25$, and captures effects that may be numerically important for $\alpha_s(m_Z)$ fits for $\rho \gtrsim 0.2$.

For a full phenomenological analysis and comparison to data, the main remaining ingredient needed is the incorporation of non-perturbative power corrections due to hadronization. In the dijet region, much is understood about how the power corrections affect the distribution. In the trijet region, much less is known. It will be important to understand how power corrections coming from factorization formula in the trijet region can be combined with power corrections from the dijet region in a manner that goes beyond hadronization models. Combining the NNLL resummation of Sudakov shoulder logarithms that we have developed in this paper with a rigorous field theoretic treatment of
power corrections could be critical in resolving long-standing discrepancies between the values of $\alpha_s$ extracted from the heavy jet mass distribution and the world average. 

 \acknowledgments

The authors would like to thank T. Becher, P. Monni, A. Gao and Y. Li for useful conversations. 
M.D.S., XY.Z. and A.B. are supported in part  by the U.S. Department of Energy under contract DE-SC0013607. I.S. and J.M. were supported in part by the U.S. Department of Energy contract DE-SC0011090. 
I.S. was also supported in part by the Simons Foundation through the Investigator grant 327942.


\appendix
\section{One-loop trijet hemisphere soft function}
\label{sec:appendix_soft_calc}

In this section, we review the calculation of trijet hemisphere soft function needed for both thrust and hemisphere. Based on the analysis of trijet kinematics in~\cite{Bhattacharya:2022dtm}, we define the six-directional differential soft function via 
\begin{equation}
\label{eq:def_diffsoft}
	S_{6,g/q}(q_i)=2g_s^2\mu^{2\epsilon} \iota^\epsilon \int \frac{d^d k}{(2\pi)^{d-1}}\delta^{+}(k^2)\hat{\mathscr{M}}(k,q_i)|\mathcal{M}_{g/q}|^2,\quad i=1,2,3,\bar 1,\bar 2,\bar 3
\end{equation}
with the conventional factor $\iota=\frac{e^{\gamma_E}}{4\pi}$ and  the eikonal matrix elements
\begin{align}
	|\mathcal{M}_{g}|^2=\left[\left(C_F-\frac{1}{2}C_A\right)\frac{n_2\cdot n_3}{(n_2\cdot k)(n_3\cdot k)}+\frac{1}{2}C_A\frac{n_1\cdot n_2}{(n_1\cdot k)(n_2\cdot k)}+\frac{1}{2}C_A\frac{n_1\cdot n_3}{(n_1\cdot k)(n_3\cdot k)}\right] 
  ,\notag\\
	|\mathcal{M}_{q}|^2=\left[\left(C_F-\frac{1}{2}C_A\right)\frac{n_1\cdot n_3}{(n_1\cdot k)(n_3\cdot k)}+\frac{1}{2}C_A\frac{n_1\cdot n_2}{(n_1\cdot k)(n_2\cdot k)}+\frac{1}{2}C_A\frac{n_2\cdot n_3}{(n_2\cdot k)(n_3\cdot k)}\right]
  ,
\end{align}
and the soft measurement function ($\hat{\mathscr{M}}$) at one-loop
\begin{align}
\label{eq:soft_measure}
	\hat{\mathscr{M}}(k,q_i)&=\theta(n_2\cdot k-\bar n_2\cdot k) \theta(n_3\cdot k-\bar n_3\cdot k)\delta\left(q_{j\neq 1}\right)\delta\left(q_1-\frac{2}{3}n_1\cdot k\right)\notag\\
	&+\theta(n_1\cdot k-\bar n_1\cdot k) \theta(n_3\cdot k-\bar n_3\cdot k)\delta\left(q_{j\neq 2}\right)\delta\left(q_2-\frac{2}{3}n_2\cdot k\right)\notag\\
	&+\theta(n_1\cdot k-\bar n_1\cdot k) \theta(n_2\cdot k-\bar n_2\cdot k)\delta\left(q_{j\neq 3}\right)\delta\left(q_3-\frac{2}{3}n_3\cdot k\right)\notag\\
	&+\theta(\bar n_3\cdot k-n_3\cdot k)\theta(\bar n_2\cdot k-n_2\cdot k)\delta\left(q_{j\neq \bar 1}\right)\delta\left(q_{\bar 1}-\frac{2}{3}\bar n_1\cdot k\right)\notag\\
	&+\theta(\bar n_1\cdot k-n_1\cdot k)\theta(\bar n_3\cdot k-n_3\cdot k)\delta\left(q_{j\neq \bar 2}\right)\delta\left(q_{\bar 2}-\frac{2}{3} N_2\cdot k\right)\notag\\
	&+\theta(\bar n_1\cdot k-n_1\cdot k)\theta(\bar n_2\cdot k-n_2\cdot k)\delta\left(q_{j\neq \bar 3}\right)\delta\left(q_{\bar 3}-\frac{2}{3} N_3\cdot k\right)
\,.\end{align}
Here $n_1$, $n_2$ and $n_3$ are three lightlike vectors that describe the trijet configuration. For concreteness, we align
\begin{align}
      n_1&=\left(1,0,0,1\right), & n_{2}&=\left(1,0,\frac{\sqrt{3}}{2},-\frac{1}{2}\right), & n_{3}&=\left(1,0,-\frac{\sqrt{3}}{2},-\frac{1}{2}\right)\ .
\end{align}For thrust, the last two projections $N_{2,3}$ are lightlike while for HJM, it is spacelike. Explicitly,
\begin{align}
	\text{Thrust}:&\quad  N_2=\bar n_2,\quad N_3=\bar n_3\notag\\
	\text{HJM}:&\quad  N_2=\left(1,0,\frac{\sqrt{3}}{2},\frac{3}{2}\right),\quad N_3=\left(1,0,-\frac{\sqrt{3}}{2},\frac{3}{2}\right)
  \,.
\end{align}
The short-hand notation $\delta(q_{j\neq 1})$ stands for  $\delta(q_2)\delta(q_3)\delta(q_{\bar 1})\delta(q_{\bar 2})\delta(q_{\bar 3})$.

In the following, we will focus on the HJM case. To simplify the calculation, we perform the standard topology identification. First of all, let us introduce the notation 
\begin{equation}
	I_{n_a,n_b,n_c,n_d,N}=\int d^d k \frac{n_a\cdot n_b}{(n_a\cdot k)(n_b \cdot k)} \delta^{+}(k^2)\delta\left(q-\frac{2}{3}N\cdot k\right)\theta\left(n_c\cdot k-\bar n_c\cdot k\right)\theta\left(n_d\cdot k-\bar n_d\cdot k\right)
\end{equation}
then our soft function can be written as a linear combination of $I_{n_a,n_b,n_c,n_d,N}$. For example, the first term in Eq.~\eqref{eq:soft_measure} can be mapped to
\begin{equation}
	S_{6,g}(q_i)\supset \left(C_F-\frac{1}{2}C_A\right) I_{n_2,n_3,n_2,n_3,n_1}+\frac{1}{2}C_A I_{n_1,n_2,n_2,n_3,n_1}+\frac{1}{2}C_A I_{n_1,n_3,n_2,n_3,n_1}
\,.\end{equation}

Using rotations and reflection, we can rename the projection vectors and reduce the soft function into 7 master integrals:
\begin{align}
	&I_0=I_{n_1,n_2,n_2,n_3,n_1},\quad I_1=I_{n_2,n_3,n_2,n_3,n_1},\quad I_2=I_{n_1,n_3,\bar n_2,\bar n_3,\bar n_1}, \quad I_3=I_{n_2,n_3,\bar n_2,\bar n_3, \bar n_1}, \quad\notag\\
	& I_4=I_{n_2,n_3,\bar n_1,\bar n_2, N_3}, \quad I_5=I_{n_1,n_2,\bar n_1,\bar n_2, N_3}, \quad I_6=I_{n_1,n_3,\bar n_1,\bar n_2,N_3}
\,.\end{align}
To evaluate the integrals, we use the lightcone parameterization, appropriately expand to the first order in $d=4-2\epsilon$ and compute each piece numerically. It turns out some of the numbers can be reconstructed with the pentagon table~\cite{Chicherin:2020oor} via PSLQ~\cite{PSLQref,Bailey:1999nv}. $I_0$ is the only divergent integral, which reads
\begin{align}
	I_0&=\mathcal{N}\bigg[\frac{4\pi}{3\epsilon}+\frac{1}{3}\pi\ln\frac{81}{64}-2\kappa+\epsilon\bigg(\frac{24 c_1}{5}+\frac{32 c_2}{3}+\frac{10}{9} \pi  \text{Li}_2\left(-\frac{1}{2}\right)-\frac{91 \pi ^3}{270}+\frac{17}{30}
   \pi  \ln ^2 3\notag\\
   &\qquad\qquad\qquad\qquad\qquad\qquad\qquad-\frac{25}{9} \pi  \ln ^2 2-2 \pi  \ln 2 \ln 3\bigg)\bigg]
\,.\end{align}
For others, we get
\begin{align}
   I_1&=\mathcal{N}\bigg[4\kappa-\frac{4}{3}\pi\ln 2+\epsilon c_3\bigg],\quad 
   I_2=\mathcal{N}\bigg[-2\kappa+2\pi\ln 2+\epsilon c_4\bigg]
   \,,\notag\\
   I_3&=\mathcal{N}\bigg[4\kappa+\frac{4}{3}\pi\ln 2+\epsilon c_5\bigg], \quad 
   I_4=\mathcal{N}\bigg[-2\kappa+2\pi\ln 2+\epsilon c_6\bigg]
   \,, \notag\\
   I_5&=\mathcal{N}\bigg[4\kappa+\frac{4}{3}\pi\ln 2+\epsilon c_7\bigg],\quad 
   I_6=\mathcal{N}\bigg[-2\kappa+2\pi\ln 2+\epsilon c_8\bigg]
\,,\end{align}
where the constants $\kappa$ and $c_i$ are
\begin{align}
	\kappa=\Im\left[\text{Li}_2\left(e^{\frac{i\pi}{3}}\right)\right],\quad c_1=\Im\left[\text{Li}_3\left(\frac{i}{\sqrt{3}}\right)\right],\quad c_2=\Im\left[\text{Li}_3\left(1+i\sqrt{3}\right)\right],\quad c_3=-1.89958 ,\notag\\
	 c_4=-3.83452,\quad c_5=-12.3488,\quad c_6=5.56704,\quad c_7=15.9482,\quad c_8=3.52263 ,
\end{align}
and the overall normalization is
\begin{equation}
	\mathcal{N}=\frac{3}{8}\Omega_{d-3} \frac{1}{q^{1+2\epsilon}} \left(\frac{2}{3}\right)^{2\epsilon},\quad \Omega_{d}\equiv \frac{2\pi^{d/2}}{\Gamma(d/2)}
\,.\end{equation}

Putting back the color factors and renormalizing in $\overline{\text{MS}}$, we obtain the differential soft function defined in Eq.~\eqref{eq:def_diffsoft}. To get the trijet hemisphere soft function, we need to integrate the soft momenta over the hemispheres, i.e.
\begin{align}
	S_i(q_L,q_H,\mu)
  &=\int d^6 q_i S_{6i}(q_i,\mu)\delta\left(q_L-q_1-q_{\bar 2}-q_{\bar 3}\right)\delta\left(q_H-q_{\bar 1}-q_{2}-q_{3}\right) 
  \nn\\
  & \equiv S_{iL}(q_L,\mu) S_{iH}(q_H,\mu) S_{f}(q_L/q_H) 
\,.\end{align}
Note that for convenience, we split the soft constant into $S_{iL}(q_L,\mu)$ and $S_{iH}(q_H,\mu)$ and leave only non-global logarithmic structure in $S_{f}(q_L/q_H)$. Since non-global logarithms are the same order as the finite part of the 2-loop soft function, which is relevant only for N${}^3$LL resummation, we simply set
$S_{f}(q_L/q_H)= 1$. The final result is
\begin{align}
	S_{gL}(q_L,\mu)&=\delta(q_L)\left(1+\frac{\alpha_s}{4\pi} c_{gL}\right)+\frac{\alpha_s}{4\pi}  \bigg[\frac{-2C_A\Gamma_0 \ln\frac{q_L}{\mu}+2\gamma_{sg}}{q_L}\bigg]_{\star} 
  \,, \notag\\
	S_{gH}(q_H,\mu)&=\delta(q_H)\left(1+\frac{\alpha_s}{4\pi} c_{gH}\right)+\frac{\alpha_s}{4\pi}\bigg[\frac{-4C_F\Gamma_0 \ln\frac{q_H}{\mu}+2\gamma_{sqq}}{q_H}\bigg]_{\star}
  \,, \notag\\
	S_{qL}(q_L,\mu)&=\delta(q_L)\left(1+\frac{\alpha_s}{4\pi} c_{qL}\right)+\frac{\alpha_s}{4\pi}\bigg[\frac{-2C_F\Gamma_0 \ln\frac{q_L}{\mu}+2\gamma_{sq}}{q_L}\bigg]_{\star}
  \,, \notag\\
	S_{qH}(q_H,\mu)&=\delta(q_H)\left(1+\frac{\alpha_s}{4\pi} c_{qH}\right)+\frac{\alpha_s}{4\pi}\bigg[\frac{-2(C_F+C_A)\Gamma_0 \ln\frac{q_H}{\mu}+2\gamma_{sqg}}{q_H}\bigg]_{\star}
  \,,
\end{align}
with the non-cusp anomalous dimensions
\begin{align}
	\gamma_{sg}&=8C_F\ln2-4C_A\ln3,\quad 
     \gamma_{sqq}=-8C_F \ln 6
   \,, \notag\\
  \gamma_{sq}&=4C_A\ln 2+4C_F\ln\frac{2}{3},\quad 
     \gamma_{sqg}=-4(C_A+C_F)\ln 6
    \,,
\end{align}
and the soft constants
\begin{align} \label{eq:hjm_soft_constants}
	c_{gL}&=-4.44002C_A+1.68285C_F,
& c_{gH} & =-0.210218C_A+5.10882C_F  ,\notag \\
c_{qL}&=0.841426C_A-3.59860C_F ,
&c_{qH}& =2.55411C_A+2.34389C_F
\,.\end{align}

For completeness, we can also extract the exact 1-loop soft function for thrust in the trijet region. The only difference between the thrust and HJM trijet soft functions is that the soft constants in the light hemisphere are different:
\begin{align}
	c_{gL}^\text{thrust}&=8.09467C_A+10.6607C_F,
 \quad
c_{gH}^\text{thrust}=-0.210218C_A+5.10882C_F ,\notag \\
c_{qL}^\text{thrust}&=5.33034C_A+13.4250C_F ,\quad
c_{qH}^\text{thrust}=2.55411C_A+2.34389C_F
\,.\end{align}

\section{Hard, jet and soft functions}\label{sec:anom_dim}
In this Appendix, we summarize the hard and jet functions. We write the cusp anomalous dimension as
\begin{align} \Gamma_{\text{cusp}}&=\left(\frac{\alpha_s}{4\pi}\right)\Gamma_0+\left(\frac{\alpha_s}{4\pi}\right)^2 \Gamma_1+\left(\frac{\alpha_s}{4\pi}\right)^3 \Gamma_2+\cdots\notag 
 \,,
\end{align}
where \cite{Korchemsky:1987wg,Moch:2004pa}
\begin{align}
    \Gamma_0&=4 
      \,, \notag\\
	\Gamma_1&=4\left[C_A\left(\frac{67}{9}-\frac{\pi^2}{3}\right)-\frac{20}{9}T_F n_F\right]
      \,, \notag\\
	\Gamma_2&=4 \bigg[\left(-\frac{56 \zeta_3}{3}-\frac{418}{27}+\frac{40 \pi ^2}{27}\right)
   C_A n_f T_F+\left(\frac{22 \zeta_3}{3}+\frac{245}{6}-\frac{134 \pi ^2}{27}+\frac{11
   \pi ^4}{45}\right) C_A^2
      \notag\\
   &\qquad+\left(16 \zeta_3-\frac{55}{3}\right) C_F n_f T_F-\frac{16}{27}
   n_f^2 T_F^2\bigg]
\,.\end{align}

\subsubsection*{Hard Functions }
Hard functions for event shape factorizations in SCET are obtained by matching QCD form factors onto corresponding SCET operators. In general, the hard function RGE takes the form
\begin{align}
    \frac{d\ln H(Q^2,\mu^2)}{d\ln \mu}=2\Gamma^{H} \ln \frac{Q^2}{\mu^2}+2\gamma^H
   \,,
\end{align}
where $\Gamma^H$ is the anomalous dimension proportional to the cusp anomalous dimension, and $\gamma_H$ depends on the relevant factorization and kinematics. The solution to the above RGE is given by
\begin{align}
    H(\mu^2,Q^2)=\exp\left[4S(\mu_h, \mu)-2A_{\gamma_H}(\mu_h, \mu)\right]\left(\frac{Q^2}{\mu_h^2}\right)^{-2A_{\Gamma}(\mu_h,\mu)} \times h(L)
   \,,
\end{align}
where $L\equiv \ln\left(\frac{Q^2}{\mu_h^2}\right)$ is the fixed order hard function at the scale $\mu_h$. Perturbatively, we have
\begin{align}
    h(L)&=1+\left(\frac{\alpha_s}{4\pi}\right)\left[-\Gamma_0^H\frac{L^2}{2}-\gamma_0^H L+c_1^H\right]+\left(\frac{\alpha_s}{4\pi}\right)^2 \bigg[(\Gamma_0^H)^2 \frac{L^4}{8}+(\beta_0+3\gamma_0^H)\Gamma_0^H \frac{L^3}{6}\notag\\
    &+\left(-\Gamma_1^H+(\gamma_0^H)^2+\beta_0\gamma_0^H-c_1^H \Gamma_0^H\right)\frac{L^2}{2}+(-\gamma_1^H-\gamma_0^H c_1^H-\beta_0 c_1^H)L+c_2^H\bigg]
   \,.
\end{align}
The constants $c_i^H$ encode the fixed order, non-log pieces in the hard function. At our working order, the hard anomalous dimension has the form~\cite{Becher:2009qa}
\begin{align}
    \frac{d}{d\ln{\mu}} \ln{H(Q^2,\mu^2)}\ =-\Gamma_{\cusp} \sum_{i<j} \mathbf{T}_i \cdot \mathbf{T}_{j} \ln{\left[\frac{2q_i \cdot q_j}{\mu^2}\right]} +2 \gamma^{\prime}_H \label{eq:neubert_becher_ansatz}
\,,\end{align} where $\mathbf{T}_j^a$ denotes the color charge of the $j^{\mathrm{th}}$ external hard parton, written using color space formalism~\cite{Catani:1998bh}. Here $q_i$ is the momentum of the outgoing $i^{\mathrm{th}}$ hard parton. For the shoulder resummation, the hard kinematics is a trijet one, equally spaced in angle given by
\begin{align} \label{eq:kin_shoulder}
    q_i^{\mu} &= \frac{Q}{3} n^{i}_\mu\ , & 2q_{i}\cdot q_{j}&= \frac{Q^2}{3}\ . 
\end{align}with the relevant partons being $q,\bar{q},g$. The relevant color sums are
\begin{align}\label{eq:T_eq_color}
    \mathbf{T}_q\cdot \mathbf{T}_{\bar{q}}&= \frac{C_A}{2}-C_F
    \,, \nn \\
    \mathbf{T}_q\cdot \mathbf{T}_{g}&=\mathbf{T}_{\bar{q}}\cdot \mathbf{T}_{g}= -\frac{C_A}{2}
\,.\end{align}
The non-cusp hard anomalous dimension $\gamma^{\prime}_H$ is also color diagonal up~to two loop order, and is thus a sum of quark and gluon pieces, (i.e)
\begin{align}\label{eq:gamma_h_small}
    \gamma^{\prime}_H=2\gamma^q +\gamma^g
\,.\end{align}
These are extracted from the expressions in~\cite{Becher:2007ty, Ahrens:2009cxz} and are given by
\begin{align}
    \gamma^{q}=\left(\frac{\as}{4\pi}\right)\gamma_{0}^{q} + \left(\frac{\as}{4\pi}\right)^2\gamma^{q}_{1}+\cdots
    \,,  \qquad
    \gamma^{g}=\left(\frac{\as}{4\pi}\right)\gamma_{0}^{g} + \left(\frac{\as}{4\pi}\right)^2\gamma^{g}_{1}+\cdots
\,,\end{align}
where~\cite{Moch:2005id, Moch:2005tm, Idilbi:2005ni, Idilbi:2006dg, Becher:2006mr}
\begin{align}
    \gamma^{q}_0&=-6 C_F
    \,,  \\
    \gamma^{g}_0&= -2\beta_0
    \,, \nn \\
    \gamma^{q}_{1}&=C_F^2 \left(4\pi^2-3-48\zeta_3\right) +C_F C_A \left(-\frac{11\pi^2}{3}-\frac{961}{27}+52 \zeta_3\right)+ C_F T_F n_f \left(\frac{4\pi^2}{3}+\frac{260}{27}\right)
    \,, \nn \\
    \gamma^g_{1}&=C_A^2 \left(\frac{11\pi^2}{9}-\frac{1384}{27}+4\zeta_3\right) +C_A T_F n_f\left(-\frac{4\pi^2}{9}+\frac{512}{27}\right)+8 C_F  T_F n_f
\,.\nn
\end{align}
Using Eqns.~\eqref{eq:kin_shoulder}, \eqref{eq:T_eq_color}, \eqref{eq:gamma_h_small} in~\eqref{eq:neubert_becher_ansatz}, we find that the anomalous dimension $\Gamma^H$ for the shoulder trijet hard function is
\begin{align}
    \Gamma^H&=\left(C_F+\frac{1}{2}C_A\right)\Gamma_{\text{cusp}}
   \,,\\
    \gamma^{H}&=\left(\frac{\as}{4\pi}\right)\gamma_0^{H} + \left(\frac{\as}{4\pi}\right)^2\gamma_1^{H} +\cdots\ ,
  \nn
\end{align}
where 
\begin{align}
    \gamma_0^H&=-\beta_0-6C_F-2(C_A+2C_F)\ln 3
     \,,\nn \\
    \gamma_1^H&=C_A T_F n_f \left(\frac{256}{27}-\frac{2}{9}\pi^2+\frac{40}{9}\ln 3\right)+C_A^2\left(2\zeta_3+\frac{11}{18}\pi^2-\frac{692}{27}+\frac{2\pi^2}{3}\ln 3-\frac{134}{9}\ln 3\right)\notag\\
    &\quad
    +C_F T_F n_f\left(\frac{368}{27}+\frac{4\pi^2}{3}+\frac{80}{9}\ln 3\right)+C_F^2\left(-48\zeta_3+4\pi^2-3\right)\notag\\
    &\quad
    +C_F C_A\left(52\zeta_3-\frac{11}{3}\pi^2-\frac{961}{27}+\frac{4\pi^2}{3}\ln 3-\frac{268}{9}\ln 3\right)
     \,.
\end{align}
For NNLL resummation, we also need the $c_i^H$ constants to one-loop, namely $c_1^H$. It can be extracted from the real-virtual correction of $\gamma^{\star} \rightarrow q\bar q$ squared matrix elements~\cite{Ellis:1980wv,Garland:2001tf}, 
\begin{align}
    c_1^H&=C_F\left[-\frac{65}{4}+\frac{3\pi^2}{2}-\frac{21}{8}\ln 3-10 \ln 2\ln 3+3\ln^2 3+10\text{Li}_2\left(\frac{1}{3}\right)\right]\notag\\
    &\quad
    +C_A\left[\frac{3}{4}+\frac{5\pi^2}{4}+\frac{3}{8}\ln 3+\ln 2\ln 3-\frac{3}{2}\ln^2 3-\text{Li}_2\left(\frac{1}{3}\right)\right]\ .
\end{align}
The hard anomalous dimensions are also common to other hard functions that have appeared earlier in the SCET literature~\cite{Jouttenus:2011wh,Becher:2009th}, and we have verified ours against the same. 

\subsubsection*{Jet functions}
The jet function RGE in the Laplace space is 
\begin{equation}
	\frac{d \ln \tilde j(\nu)}{d\ln \mu}=-2 \Gamma^J\ln\frac{1}{\nu \mu^2 e^{\gamma_E}}-2\gamma^J 
   \,,
\end{equation}
with perturbative solution 
\begin{align}
	\tilde j(L)&=1+\left(\frac{\alpha_s}{4\pi}\right) \left[\Gamma_0^J\frac{L^2}{2}+\gamma_0^J L+c_1^J \right]+\left(\frac{\alpha_s}{4\pi}\right)^2\bigg[(\Gamma_0^J)^2\frac{L^4}{8}+(-\beta_0+3\gamma_0^J)\Gamma_0^J\frac{L^3}{6}\notag\\
	&+(\Gamma_1^J+(\gamma_0^J)^2-\beta_0\gamma_0^J+c_1^J\Gamma_0^J)\frac{L^2}{2}+(\gamma_1^J+\gamma_0^J c_1^J-\beta_0 c_1^J)+c_2^J\bigg]
\,.\end{align}
We will need the $\mathcal{O}(\alpha_s)$ piece for NNLL shoulder resummation and $\mathcal{O}(\alpha_s^2)$ piece for NNLL${}^\prime$ dijet resummation.
For quark jets~\cite{Lunghi:2002ju,Bosch:2004th,Becher:2006qw}, the anomalous dimension and constant are
\begin{align}
	\Gamma^{J_q}&=C_F \Gamma_{\text{cusp}} 
  \,, \notag\\ 
     \gamma_0^{J_q}&=-3C_F 
  \,,\notag\\
	\gamma_1^{J_q}&=C_F^2\left(-\frac{3}{2}+2\pi^2-24\zeta_3\right)+C_F C_A\left(-\frac{1759}{54}-\frac{11}{9}\pi^2+40\zeta_3\right)+C_F T_F n_f\left(\frac{242}{27}+\frac{4}{9}\pi^2\right)
  ,\notag\\
	c_1^{J_q}&=C_F\left(7-\frac{2}{3}\pi^2\right)
\,.\end{align}
For gluon jets, we have~\cite{Becher:2009th,Becher:2010pd}
\begin{align}
  \Gamma^{J_g}&=C_A \Gamma_{\text{cusp}} 
    \,, \notag\\
  \gamma_0^{J_q}& =-\beta_0
    \,,\notag\\
  \gamma_1^{J_g}&=C_A^2\left(-\frac{1096}{27}+\frac{11}{9}\pi^2+16\zeta_3\right)+C_A T_F n_f\left(\frac{368}{27}-\frac{4}{9}\pi^2\right)+4 C_F T_F n_f
    ,\notag\\
   c_1^{J_g}&=C_A\left(\frac{67}{9}-\frac{2}{3}\pi^2\right)-\frac{20}{9}T_F n_f
\,.\end{align}

\subsubsection*{Soft functions}

The RGE for the trijet hemisphere soft function is
\begin{align}
    \frac{d\ln \tilde{s}(\nu)}{d\ln \mu}=-2\Gamma^S \ln\frac{1}{\nu \mu e^{\gamma_E}}-2\gamma^S
    \,,
\end{align}
and its solution is
\begin{align}
    \tilde s(L)&=1+\left(\frac{\alpha_s}{4\pi}\right) \left[\Gamma_0^S\frac{L^2}{2}+2\gamma_0^S L+c_1^S \right]+\left(\frac{\alpha_s}{4\pi}\right)^2\bigg[(\Gamma_0^S)^2\frac{L^4}{8}+(-\beta_0+3\gamma_0^S)\Gamma_0^S\frac{L^3}{3}\notag\\
	&+(\Gamma_1^S+4(\gamma_0^S)^2-4\beta_0\gamma_0^S+c_1^S\Gamma_0^S)\frac{L^2}{2}+2(\gamma_1^S+\gamma_0^S c_1^S-\beta_0 c_1^S)L+c_2^S\bigg]
\,.\end{align}
The soft anomalous dimension for each hemisphere is
\begin{alignat}{3}
	\Gamma^{sg} &= -2C_A \Gamma_{\text{cusp}}
	\,, &\quad
	\gamma_0^{sg} &= -2C_A\ln 3+4C_F\ln 2
	\,, \nonumber \\
	\Gamma^{sq} &= -2C_F \Gamma_{\text{cusp}}
	\,, &\quad
	\gamma_0^{sq} &= -2C_F\ln\frac{3}{2}+2C_A\ln 2
	\,, \nonumber \\
	\Gamma^{sqg} &= -2(C_F+C_A)\Gamma_{\text{cusp}}
	\,, & \quad
	\gamma_0^{sqg} &= -2(C_A+C_F)\ln 6
	\,, \nonumber \\
	\Gamma^{sqq} &= -4C_F \Gamma_{\text{cusp}}
	\,, &\quad
	\gamma_0^{sqq} &= -4C_F\ln 6
\,.\end{alignat}
The two-loop non-cusp anomalous dimension can be obtained from RG invariance
\begin{align}
	\gamma_1^{sg}+\gamma_1^{sqq}=\gamma_1^{sq}+\gamma_1^{sqg}&=C_A T_F n_f \left(-\frac{112}{27}+\frac{2\pi^2}{9}+\frac{40}{9}\ln 3\right)+C_F T_F n_f\left(-\frac{224}{27}+\frac{4\pi^2}{9}+\frac{80}{9}\ln 3\right)\notag\\
	&\quad
	+C_F C_A \left(\frac{808}{27}-\frac{11}{9}\pi^2-\frac{268}{9}\ln 3+\frac{4\pi^2}{3}\ln 3-28\zeta_3\right)\notag\\
	&\quad
	+C_A^2\left(\frac{404}{27}-\frac{11}{18}\pi^2-\frac{134}{9}\ln 3+\frac{2\pi^2}{3}\ln 3-14\zeta_3 \right)
\,.\end{align}
The soft constants were computed in  Appendix~\ref{sec:appendix_soft_calc}. In Laplace space and in the above notation they translate to
\begin{align}
	c_1^{sg}=\Gamma_0^{sg}\frac{\pi^2}{12}+c_{gL},\quad c_1^{sqq}=\Gamma_0^{sqq}\frac{\pi^2}{12}+c_{gH},\quad c_1^{sq}=\Gamma_0^{sq}\frac{\pi^2}{12}+c_{qL},\quad c_1^{sqg}=\Gamma_0^{sqg}\frac{\pi^2}{12}+c_{qH}
\,.\end{align}

\section{Running coupling and Sudakov kernels}
\label{sec:appendix_kernels}

We expand the QCD beta-function as
\begin{equation}
\label{eq:beta}
  \frac{\mathrm{d}\alpha_s(\mu)}{\mathrm{d}\ln \mu}= \beta (\alpha_s(\mu)),\quad 
  \beta (\alpha)=- 2 \alpha\, \left[ \left( \frac{\alpha}{4 \pi} \right) \beta_0 + \left(
  \frac{\alpha}{4 \pi} \right)^2 \beta_1 + \left( \frac{\alpha}{4 \pi}
  \right)^3 \beta_2 + \cdots \right]\,,
\end{equation}
where the coefficient up to three loops are given by~\cite{Tarasov:1980au,Larin:1993tp,vanRitbergen:1997va,Czakon:2004bu}
\begin{align}
\beta_0 &=\frac{11}{3} C_A - \frac{4}{3} T_F n_f 
  \,, \\
\beta_1 &= \frac{34}{3} C_A^2 - \frac{20}{3} C_A T_F n_f - 4 C_F T_F n_f 
  \nn \,,\\
\beta_2 &= n_f^2 T_F^2 \left(\frac{158 }{27}C_A+\frac{44
   }{9}C_F\right)+n_f T_F \left(2
    C_F^2-\frac{205 }{9}C_FC_A-\frac{1415 }{27}C_A^2\right)+\frac{2857 }{54}C_A^3
\,.\nn
\end{align}
It is also useful sometimes to have an expression for the coupling at one scale in terms of the coupling at a different scale. For two-loop running, such an expression is
\begin{equation}
	\alpha_s(\mu) = \alpha_s(Q)\left[X+\alpha_s(Q)\frac{\beta_1}{4\pi \beta_0}\ln X\right]^{-1}, \quad X\equiv 1+\frac{\alpha_s(Q)}{2\pi}\beta_0\ln\frac{\mu}{Q}
\,,\end{equation}
and at three loops
\begin{equation}
	\alpha_s(\mu) = \alpha_s(Q)\left\{X+\alpha_s(Q)\frac{\beta_1}{4\pi \beta_0}\ln X+\frac{\alpha_s^2(Q)}{16\pi^2}\left[\frac{\beta_2}{\beta_0}\left(1-\frac{1}{X}\right)+\frac{\beta_1^2}{\beta_0^2}\left(\frac{1}{X}-1+\frac{\ln X}{X}\right)\right]\right\}^{-1}
\,.\end{equation}

For reference, we also provide explicit expressions~\cite{Bosch:2003fc,Neubert:2004dd} for the Sudakov RG kernels appearing in the evolution factors. Their definitions are:
\begin{align}
	S_\Gamma(\nu,\mu)&=-\int_{\alpha_s(\nu)}^{\alpha_s(\mu)} d\alpha\ \frac{\Gamma_{\text{cusp}}(\alpha)}{\beta(\alpha)}\int_{\alpha_s(\nu)}^{\alpha} \frac{d\alpha^\prime}{\beta(\alpha^\prime)}
	\,, \notag \\
	A_\Gamma(\nu,\mu)&=-\int_{\alpha_s(\nu)}^{\alpha_s(\mu)} d\alpha\ \frac{\Gamma_{\text{cusp}}(\alpha)}{\beta(\alpha)}
	\,, \notag \\
	A_{\gamma_j}(\nu,\mu)&=-\int_{\alpha_s(\nu)}^{\alpha_s(\mu)} d\alpha\ \frac{\gamma(\alpha)}{\beta(\alpha)}
\,.\end{align}
For $S_\Gamma(\nu,\mu)$, we have
\begin{equation}
	S_\Gamma(\nu,\mu)=S_\Gamma^{\text{LL}}(\nu,\mu)+S_\Gamma^{\text{NLL}}(\nu,\mu)+S_\Gamma^{\text{NNLL}}(\nu,\mu)+\cdots
\end{equation}
with 
\begin{align}
	S_\Gamma^{\text{LL}}(\nu,\mu)&=\frac{\Gamma_0\pi}{\beta_0^2 \alpha_s(\nu)}\left(1-\frac{1}{r}-\ln r\right)
    \,, \notag\\
	S_\Gamma^{\text{NLL}}(\nu,\mu)&=\frac{\Gamma_0}{4\beta_0^2}\left[\left(\frac{\Gamma _1}{\Gamma _0}-\frac{\beta
   _1}{\beta _0}\right) (1-r+\ln r)+\frac{\beta
   _1 }{2 \beta _0}\ln^2(r)\right]
   \,, \notag \\
   S_\Gamma^{\text{NNLL}}(\nu,\mu)&=\frac{\Gamma_0 \alpha_s(\nu)}{32\pi\beta_0^2}\bigg[\left(\frac{\beta _1^2}{\beta _0^2}-\frac{\beta
   _2}{\beta _0}\right) \left(1-r^2+2 \ln r\right)+ \left(\frac{\beta _1 \Gamma
   _1}{\beta _0 \Gamma _0}-\frac{\Gamma _2}{\Gamma
   _0}\right)(1-r)^2
   \notag \\
   &+2 \left(\frac{\beta _1 \Gamma
   _1}{\beta _0 \Gamma _0}-\frac{\beta _1^2}{\beta
   _0^2}\right) (1-r+r \ln r)\bigg]
\,.\end{align}
Here we define $r\equiv \frac{\alpha_s(\mu)}{\alpha_s(\nu)}$. Similarly, we have
\begin{align}
	A_\Gamma(\nu,\mu)=A_\Gamma^{\text{LL}}(\nu,\mu)+A_\Gamma^{\text{NLL}}(\nu,\mu)+A_\Gamma^{\text{NNLL}}(\nu,\mu)+\cdots
	\,, \notag \\
	A_{\gamma_j}(\nu,\mu)=A_{\gamma_j}^{\text{LL}}(\nu,\mu)+A_{\gamma_j}^{\text{NLL}}(\nu,\mu)+A_{\gamma_j}^{\text{NNLL}}(\nu,\mu)+\cdots
\,.\end{align}
The corresponding term at each order is
\begin{align}
	A_\Gamma^{\text{LL}}(\nu,\mu)&=\frac{\Gamma_0}{2\beta_0}\ln r
	\,, \notag \\
	A_\Gamma^{\text{NLL}}(\nu,\mu)&=\frac{\Gamma_0\alpha_s(\nu)}{8\pi\beta_0}\left(\frac{\beta_1}{\beta_0}-\frac{\Gamma_1}{\Gamma_0}\right)(1-r)
	\,, \notag \\
	A_\Gamma^{\text{NNLL}}(\nu,\mu)&=\frac{\Gamma _0  \alpha (\nu )^2}{64 \pi ^2 \beta _0}\left(\frac{\beta _1^2}{\beta _0^2}-\frac{\beta
   _2}{\beta _0}+\frac{\Gamma _2}{\Gamma_0}-\frac{\beta _1 \Gamma _1}{\beta _0 \Gamma_0}\right)\left(r^2-1\right)
\,.\end{align}

\section{Dictionary between position and momentum space logs}

In this section, we provide the analytic Fourier transformation rules that can be used to relate singular cross sections
and soft and jet functions in position ($z$) and momentum ($r$) space,
which is required e.g.\ to extract the correct $\ord{\as}$ soft constant term in $z$ space.
These rules are distinct from the commonly considered case of plus distributions with support only on the positive reals,
and further require care due to the additional double integral
over the second derivative $\frac{d^3\sigma}{d\rho^3}$ of the spectrum.
For example, the singular cross section in position space at a fixed scale $\mu = Q$ is given by
\begin{align}
   \tilde\sigma(z,\mu_{h,j,s}=Q) 
  &=
   3 + \Bigl(\frac{\alpha_s}{4\pi}\Bigr)\biggl\{-6(C_A+2C_F)\ln^2\left(|z|e^{\gamma_E}\right)
   \\ 
  & \quad
   +\Bigl[
      -4n_f T_F+C_A(11+12\ln 3)+6C_F(3+4\ln 3)
   \nn \\ 
  & \quad \quad
      -2 (C_A+2C_F) \, i\pi \, \text{sgn}(z)\Bigl
   ]\ln \left(|z|e^{\gamma_E}\right)
   \nn \\ 
  & \quad
   +\frac{1}{6}\Bigl(-4n_f T_F+C_A(11+12 \ln 48)+6C_F(3+4\ln 48)\Bigl) \, i\pi \, \text{sgn}(z)+\text{const}
   \biggr\}
   \label{eq:singular_position}
\nn .
\end{align}
To obtain the singular expression in momentum ($r$) space,
we need to take the inverse Fourier transform Eq.~\eqref{eq:singular_position}
and integrate over $r$ twice.

To achieve this, we derive the map between logs in the momentum $z$ space and position $r$ space.
Specifically, we consider the dictionary between plus distributions of the form
\begin{align}
\delta(r),\quad \left(\frac{\theta(r)\ln^m r}{r}\right)_{+},\quad \left(\frac{\theta(-r)\ln^m (-r)}{-r}\right)_{+}
\,,\quad
m=0,1,2, \dots
\,,\end{align}
which appear in $\frac{d^3\sigma}{d\rho^3}$,
and Fourier-space logarithms
\begin{align} \label{eq:log_basis_fourier_space}
   \ln^m (|z|e^{\gamma_E}),\quad \ln^m (|z|e^{\gamma_E})\, \text{sgn}(z)
   \,, \quad m=0,1,2, \dots
\end{align}
appearing in $\tilde{\sigma}(z)$.
When integrating over $r$ twice in addition, which is accounted for by the kernel $K(z,r)$ in Eq.(\ref{eq:LOkernel}),
we also obtain the transformation rules between position logs and the spectrum logs:
\begin{align}
   \theta(r) r \ln^m r,\quad \theta(-r)(-r)\ln^m (-r)
   \,, \quad m=0,1,2, \dots
\,.\end{align}

To derive the dictionary for the first part of the basis in Eq.~\eqref{eq:log_basis_fourier_space},
we consider the generating functional
\begin{equation}
    f_1(z,\epsilon)=(-i z e^{\gamma_E})^\epsilon(i z e^{\gamma_E})^\epsilon
  \,,
\end{equation}
and its inverse Fourier transformation
\begin{equation}\label{eq:gen_1}
    \mathcal{F}^{-1}[f_1(z,\epsilon)]=-\frac{e^{2\epsilon \gamma_E}\Gamma(1+2 \epsilon)\sin(\pi\epsilon)}{\pi}|r|^{-1-2\epsilon}
\,,\end{equation}
where $|r|^{-1-2\epsilon}$ is expanded in terms of plus distributions as
\begin{equation}
    |r|^{-1-2\epsilon}=-\frac{\delta\left(|r|\right)}{2\epsilon}+\left(\frac{1}{|r|}\right)_{+}-2\epsilon \left(\frac{\ln|r|}{|r|}\right)_{+}+\cdots
\,.\end{equation}
We remind the reader that eqn~\eqref{eq:gen_1} is simply a special case of eqn~\eqref{eq:fspace_dist} with $a=b=-\epsilon$. The distribution $\delta\left(|r|\right)$ needs to be defined carefully, and we choose the following definition 
\begin{equation}
    \delta(|r|)\equiv 2\frac{d^2}{dr^2}\left[\xi r \theta(r)+r (\xi-1)\theta(-r)\right]\overset{\xi=1}{=}2\delta(r)
\,.\end{equation}
As explained in Sec.~\ref{sec:integration}, we choose the LO condition $\xi=1$.
Then if we integrate $f_2(r,\epsilon)$ over $r$ twice subject to the $\xi=1$ condition,
which is equivalent to using the integral kernel Eq.~\eqref{eq:LOkernel}
(let us call this combined operation $\mathcal{I}$),
we find
\begin{equation}
    \mathcal{I}[f_1(z,\epsilon)]=-\frac{e^{2\epsilon\gamma_E} \Gamma(-1+2\epsilon)\sin(\pi\epsilon)}{\pi}\left(r+|r|^{1-2\epsilon}\right)
\,.\end{equation}

Expanding $f_{1}(z, \epsilon)$ and $\mathcal{F}^{-1}[f_1(z,\epsilon)]$ in $\epsilon$,
we find the inverse Fourier transformation rules to any power of $\ln^m\left(|z|e^{\gamma_E}\right)$. The first few orders read,
\begin{align}
    \mathcal{F}^{-1}\left[1\right]&=\frac{1}{2}\delta\left(|r|\right)
    \,, \notag \\
    \mathcal{F}^{-1}\left[\ln(|z|e^{\gamma_E})\right]&=-\frac{1}{2}\left(\frac{1}{|r|}\right)_{+}=-\frac{1}{2}\left[\left(\frac{\theta(r)}{r}\right)_{+}+\left(\frac{\theta(-r)}{-r}\right)_{+}\right]
    \,, \notag \\
    \mathcal{F}^{-1}\left[\ln^2(|z|e^{\gamma_E})\right]&=\left(\frac{\ln |r|}{|r|}\right)_{+}+\frac{\pi^2}{24}\delta\left(|r|\right)=\left[\left(\frac{\theta(r)\ln r}{r}\right)_{+}+\left(\frac{\theta(-r)\ln(-r)}{-r}\right)_{+}\right]+\frac{\pi^2}{24}\delta\left(|r|\right)
    \,, \notag \\
    \mathcal{F}^{-1}\left[\ln^3(|z|e^{\gamma_E})\right]&=-\frac{\pi^2}{8}\left(\frac{1}{|r|}\right)_{+}-\frac{3}{2}\left(\frac{\ln^2 |r|}{|r|}\right)_{+}+\frac{1}{2}\psi^{(2)}(1)\delta\left(|r|\right)
    \,, \notag \\
    &=-\frac{\pi^2}{8} \left[\left(\frac{\theta(r)}{r}\right)_{+}+\left(\frac{\theta(-r)}{-r}\right)_{+}\right]-\frac{3}{2}\left[\left(\frac{\theta(r)\ln^2 r}{r}\right)_{+}+\left(\frac{\theta(-r)\ln^2(-r)}{-r}\right)_{+}\right]\notag\\
    &\quad +\frac{1}{2}\psi^{(2)}(1)\delta\left(|r|\right)
    \label{eq:fourier_rule_1}
\,.\end{align}
where the digamma function is defined as $\psi(z)\equiv \frac{\Gamma^\prime(z)}{\Gamma(z)}$ and $\psi^{(n)}(z)\equiv \frac{d^n \psi(z)}{d z^n}$. From the RHS, we can see that the first log basis $\ln^m(|z|e^{\gamma_E})$ is mapped to the {\bf sum} of both left and right shoulder r space logs, which is symmetric.
Furthermore, we can also expand $\mathcal{I}\left[f_{1}(z, \epsilon)\right]$ in $\epsilon$ and read off the rules between $\ln^m(|z|e^{\gamma_E})$ and spectrum logs
\begin{align} \label{eq:fourier_rule_2}
    \mathcal{I} [1] &=\frac{1}{2}|r|\left(1+\text{sgn}(r)\right)=r\theta(r)
    \,, \\
    \mathcal{I} \left[\ln(|z|e^{\gamma_E})\right] &=\frac{1}{2}|r|\left(1+\text{sgn}(r)\right)-\frac{1}{2}|r|\ln|r|=\theta(r)\left(r-\frac{1}{2}r\ln r\right)+\theta(-r)\frac{1}{2}r\ln(-r)
    \,, \notag \\
    \mathcal{I}\left[\ln^2(|z|e^{\gamma_E})\right] &=\left(1+\frac{\pi^2}{24}\right)|r|\left(1+\text{sgn}(r)\right)+\frac{1}{2}|r|\ln^2|r|-|r|\ln |r|\notag\\
    &=\theta(r)\left(2r+\frac{\pi^2}{12}r-r\ln r+\frac{1}{2}r\ln^2 r\right)+\theta(-r)\left(r\ln(-r)-\frac{1}{2}r\ln^2(-r)\right)
\nn \,.\end{align}

For the second group of terms in the basis, $\ln^m (|z|e^{\gamma_E})\, \text{sgn}(z)$, we consider the generating functional
\begin{equation}
    f_2(z,\epsilon)=i\pi e^{2\epsilon \gamma_E}|z|^{2\epsilon} \text{sgn}(z)
  \,,
\end{equation}
and its inverse Fourier transformation 
\begin{equation}
    \mathcal{F}^{-1}[f_2(z,\epsilon)]=e^{2\epsilon \gamma_E}|r|^{-1-2\epsilon}\cos(\pi\epsilon) \Gamma(1+2 \epsilon) \,, \text{sgn}(r)
\end{equation}
Expanding in $\epsilon$, we find
\begin{align}
    \mathcal{F}^{-1}\left[-i\pi\text{sgn}(z)\right]
  &=-\left(\frac{1}{|r|}\right)_{+}\text{sgn}(r)=\left(\frac{\theta(-r)}{-r}\right)_{+}-\left(\frac{\theta(r)}{r}\right)_{+}
    \,, \notag\\
    \mathcal{F}^{-1}\left[-i\pi\text{sgn}(z)\ln(|z|e^{\gamma_E})\right]
   &=\left(\frac{\ln |r|}{|r|}\right)_{+}\text{sgn}(r)-\frac{\pi^2}{24}\delta(|r|)\text{sgn}(r)
     \notag\\
    &=\left(\frac{\theta(r)\ln r}{r}\right)_{+}-\left(\frac{\theta(-r)\ln (-r)}{-r}\right)_{+}-\frac{\pi^2}{24}\delta(|r|)\text{sgn}(r)
    \,, \notag\\
    \mathcal{F}^{-1}\left[-i\pi\text{sgn}(z)\ln^2(|z|e^{\gamma_E})\right]&=\frac{\pi^2}{12}\left(\frac{1}{|r|}\right)_{+}\text{sgn}(r)-\left(\frac{\ln^2 |r|}{|r|}\right)_{+}\text{sgn}(r)-\frac{2\zeta_3}{3}\delta(|r|)\text{sgn}(r)\notag\\
    &=\frac{\pi^2}{12}\left[\left(\frac{\theta(r)}{r}\right)_{+}-\left(\frac{\theta(-r)}{-r}\right)_{+}\right]-\frac{2\zeta_3}{3}\delta(|r|)\text{sgn}(r)
     \notag\\
    &\hspace{0.6cm}-\left[\left(\frac{\theta(r)\ln^2 r}{r}\right)_{+}-\left(\frac{\theta(-r)\ln^2 (-r)}{-r}\right)_{+}\right]
    \label{eq:fourier_rule_3}
\,.\end{align}
where $\delta(|r|)\text{sgn}(r)$ as a distribution is defined by the fact that its double integral vanishes. Integrating over $r$ twice, we get
\begin{align}
    \mathcal{I}\left[-i\pi\text{sgn}(z)\right]&=\theta(r)r\ln r+\theta(-r)r\ln(-r)
    \,, \notag \\
    \mathcal{I}\left[-i\pi\text{sgn}(z)\ln(|z|e^{\gamma_E})\right]&=\theta(r)\left(-r\ln r+\frac{1}{2}r \ln^2 r\right)+\theta(-r)\left(-r\ln(-r)+\frac{1}{2}r\ln^2(-r)\right)
    \,, \notag \\
    \mathcal{I}\left[-i\pi\text{sgn}(z)\ln^2(|z|e^{\gamma_E})\right]&=\theta(r)\left(-2r\ln r +\frac{\pi^2}{12}r \ln r+r \ln^2 r-\frac{1}{3}r\ln^3 r\right)\notag\\
    &+\theta(-r)\left(-2 r\ln (-r)+\frac{\pi^2}{12}r \ln(-r)+r\ln^2(-r)-\frac{1}{3}r\ln^3 (-r)\right)
    \label{eq:fourier_rule_4}
\,.\end{align}
We thus find that the second basis $\ln^m (|z|e^{\gamma_E})\, \text{sgn}(z)$ gives rise to the {\bf difference} of shoulder logs in $r$ space.

\section{Fixed order expansion}
Using Eq.~(\ref{eq:fourier_rule_2}) and Eq.~(\ref{eq:fourier_rule_4}), we can conveniently expand the NNLL distribution to predict shoulder logarithms in $\frac{d\sigma}{d\rho}$ to higher orders in $\alpha_s$ using the RGE. The logarithms at $\ord{\as^2}$ were given in~\cite{Bhattacharya:2022dtm}. Here we also include the prediction for the slope at that order. As before we define $r = \frac{1}{3}-\rho$, so $r>0$ is the left shoulder and $r<0$ is the right shoulder.
The most general form of the cross section near $r=0$ is
\begin{align} \label{eq:def_sl_sr}
    \frac{1}{\sigma_\text{LO}} \frac{d\sigma}{d\rho}
    = r \, \theta(r) \, S_L\bigl[ \alpha_s,\ln r\bigr]
    + (-r) \, \theta(-r) \, S_R\bigl[ \alpha_s,\ln (-r) \bigr]
    + \mathcal{O}(r^2)
\,,\end{align}
with $\sigma_{\text{LO}}=\frac{\alpha_s}{4\pi}48 C_F \sigma_0 \propto \alpha_s$ defined below Eq.~\eqref{dsdrform}.
The functions $S_L$ and $S_R$ have perturbative expansions in logarithms and $\alpha_s$, which we can write as
\begin{align} \label{eq:def_a_n_m}
    S_L(\alpha_s,L) = \sum_{n,m} a_L^{(n,m)} \Bigl( \frac{\alpha_s}{4\pi} \Bigr)^n L^m,
    \quad
     S_R(\alpha_s, L) = \sum_{n,m} a_R^{(n,m)} \Bigl( \frac{\alpha_s}{4\pi} \Bigr)^nL^m,
\end{align}
For the lowest few terms to the the left shoulder, we find
\begin{align}
a_{L}^{(0,0)}
&= 3
\,, \nn \\
a_{L}^{(1,2)}
&= -2(2C_F + C_A)
\,, \nn \\
a_{L}^{(1,1)}
&= 2 C_F\Bigl(1+4 \ln \frac{4}{3}\Bigr)
   + C_A \Bigl(\frac{1}{3}+4 \ln\frac{4}{3}\Bigr)+
   \frac{4}{3} n_f T_F
\,.\end{align}
For the right shoulder,
\begin{align}
a_{R}^{(0,0)}
&= 0
\,, \nn \\
a_{R}^{(1,2)}
&= -4(2C_F+ C_A)
\,, \nn \\
a_{R}^{(1,1)}
&= 4 C_F\Bigl(1-4 \ln {6}\Bigr)
   + \frac{2C_A}{3} \Bigl(1-12\ln{6}\Bigr) +
   \frac{8}{3} n_f T_F
\,.\end{align}
At $\mathcal{O}(\sigma_\mathrm{LO} \, \as) = \mathcal{O}(\alpha_s^2)$,
the difference of linear slopes to the left and the right of the shoulder threshold, which amounts to a non-analytic term,
is uniquely predicted by the effective field theory,
\begin{align}
a_{L}^{(1,0)} + a_{R}^{(1,0)}
&= c_{gL}+2c_{gH}+c_{qL}+2c_{qH}
    -\frac{32}{3}n_f T_F\notag\\
    &\quad
    -\frac{3}{8}C_F\left[34+4 \pi ^2-24 \ln^2 3 -43 \ln 3+80 \ln 2 \ln 3-80 \text{Li}_2\left(\frac{1}{3}\right)\right]\notag\\
    &\quad
    +C_A\left[\frac{283}{12}+\frac{3 \pi ^2}{4}-\frac{9}{2}\ln^2 3+\frac{105}{8}\ln 3+3 \ln 2
   \ln 3-3 \text{Li}_2\left(\frac{1}{3}\right)\right]
\,.\end{align}
The constants $c_{qL},c_{gL},c_{qH},c_{gH}$ can be found in eqn~\eqref{eq:hjm_soft_constants}.
Note that Eq.~\eqref{eq:def_sl_sr} contains a conventional sign for all of $S_R$, so we give results for the sum of $a_{L,R}^{(1,0)}$.
The sum of slopes (the difference of the $a_{L,R}^{(1,0)}$) has to be determined by fixed-order matching to the full theory as discussed in Sec.~\ref{sec:integration},
i.e., one must add a linear function $c_0 + c_1 r$ to the distribution with $c_0$ and $c_1$ sensitive to hard 4-parton and higher-order contributions.

At NNLL, the following coefficients at $\mathcal{O}(\sigma_\mathrm{LO} \, \as^2) = \mathcal{O}(\alpha_s^3)$ are predicted
by the resummation individually for the left and right shoulder,
\begin{align}
a_{L}^{(2,4)} &= 2 (C_A + 2 C_F)^2
\,, \nn \\
a_{L}^{(2,3)} &= \frac{4}{3}(C_A+2C_F) \left[-4n_f T_F+C_F\left(-3+4\ln\frac{27}{4}\right)+C_A\left(5+2\ln\frac{27}{4}\right)\right]
\,, \nn \\
a_{L}^{(2,2)} &=
\frac{16}{9}n_f^2 T_F^2-4 C_F (c_{qL} + c_{qH}) - 2 C_A (c_{gL} + c_{gH})
\,, \nn \\
&\quad +C_F n_f T_F \left(\frac{340}{9} - \frac{32}{3}\ln\frac{9}{8}\right)
+C_A n_f T_F\left(\frac{136}{9}-\frac{16}{3} \log \frac{9}{8}\right)\notag\\
&\quad +C_F^2\left(51+\frac{14}{3}\pi^2+16\ln 2-\frac{75}{2}\ln 3-24\ln 2 \ln 3+20\ln^2 3-40\text{Li}_2\left(\frac{1}{3}\right)\right)\notag\\
&\quad +C_F C_A\left(-\frac{955}{18}+12\pi^2-48\ln 2-\frac{115}{12}\ln 3-48\ln 2 \ln 3+32\ln^2 3-16 \text{Li}_2\left(\frac{1}{3}\right)\right)\notag\\
&\quad +C_A^2\left(-\frac{227}{6}-\frac{7\pi^2}{6}-28\ln 2+\frac{55}{12}\ln 3-18\ln 2\ln 3+11\ln^2 3+2\text{Li}_2\left(\frac{1}{3}\right)\right)
\,, \nn \\[0.4em]
a_{R}^{(2,4)} &= 4(C_A+2C_F)^2
\,, \nn \\
a_{R}^{(2,3)} &=
 \frac{8}{3}(C_A+2C_F) \biggl[-4n_f T_F+C_A\Bigl(5+2\ln 54\Bigl)+C_F\Bigl(-3+4\ln 54\Bigl)\biggl]
\,, \nn \\
a_{R}^{(2,2)} &=
    \frac{32}{9}n_f^2T_F^2-4\left(C_A+C_F\right)\left(c_{qH}+c_{qL}\right)-4C_F\left(c_{gH}+c_{gL}\right)\notag\\
    &-C_F n_f T_F\left(-\frac{680}{9}+32 \ln 2+\frac{128 }{3}\ln 3\right)
    -C_A n_f T_F\left(\frac{16 }{3}\ln 648-\frac{272}{9}\right)\notag\\
    &-C_A^2\left(\frac{227}{3}+\frac{7}{3}\pi^2-28\ln 2-\frac{55}{6}\ln 3-12\ln 2 \ln 3-22\ln^2 3-4\text{Li}_2\left(\frac{1}{3}\right)\right)\notag\\
    &-C_AC_F\left(\frac{955}{9}-8\pi^2-48\ln 2+\frac{115}{6}\ln 3-96\ln 2 \ln 3-64\ln^2 3+32\text{Li}_2\left(\frac{1}{3}\right)\right)\notag\\
    &-C_F^2\left(-102-\frac{4}{3}\pi^2+16\ln 2+75\ln 3-144\ln 2 \ln 3-40\ln^2 3+80\text{Li}_2\left(\frac{1}{3}\right)\right)
\,.\end{align}
On the other hand, the NNLL resummation only predicts the sum of coefficients for the single logarithms on the left and right hemispheres,
\begin{align}
    a_{L}^{(2,1)}+a_{R}^{(2,1)}&=C_A T_F n_f  \Biggl[\frac{331}{9}-\frac{\pi^2}{3}+\frac{713}{6}\ln 3+4\ln 2\ln 3 -6\ln^2 3-4\text{Li}_2\left(\frac{1}{3}\right)\Biggl]\notag\\
    &+C_F T_F n_f  \Biggl[-\frac{371}{3}-10\pi^2 +\frac{1345}{6}\ln 3-40\ln 2\ln 3+12\ln^2 3+40 \text{Li}_2\left(\frac{1}{3}\right)\Biggl]\notag\\
    &+C_A^2\Biggl[\frac{827}{36}+\frac{95}{12}\pi^2-16\pi^2\ln 2-\frac{5263}{24}\ln 3-7\pi^2\ln 3+\ln 2\ln 3-54\ln^2 3\notag\\
    &\hspace{2cm}-12\ln 2\ln^2 3+18\ln^3 3-\text{Li}_2\left(\frac{1}{3}\right)+12 \ln 3 \,\text{Li}_2\left(\frac{1}{3}\right)+12\zeta_3\Biggl]\notag\\
    &+C_F^2\Biggl[-\frac{549}{2}+9\pi^2-32\pi^2 \ln 2+\frac{1305}{4} \ln 3 +12 \pi^2 \ln 3 -60\ln 2\ln 3-111\ln^2 3\notag\\
    &\hspace{2cm}+240 \ln 2 \ln^2 3-72\ln^3 3+ 60\text{Li}_2\left(\frac{1}{3}\right)-240 \ln 3 \,\text{Li}_2\left(\frac{1}{3}\right)+48\zeta_3\Biggl]\notag\\
    &+C_F C_A \Biggl[\frac{295}{12}-5\pi^2-\frac{6449}{24}\ln 3+40\pi^2 \ln 3 -4 \ln 2 \ln 3-\frac{351}{2}\ln^2 3\notag\\
    &\hspace{2cm}+96 \ln 2 \ln^2 3+4 \text{Li}_2\left(\frac{1}{3}\right)-96 \ln 3 \,\text{Li}_2\left(\frac{1}{3}\right)+264 \zeta_3\Biggl]\notag\\
    &-\left(c_{gL}+c_{gH}+2c_{qL}+2c_{qH}\right) \Bigg[-4 n_f T_F+2C_F\left(\ln 81 -1\right)+C_A\left(7+\ln 81\right)\Bigg] \notag\\
    &-\frac{256}{9}n_f^2 T_F^2
\,.\end{align}
Computing the difference 
 $a_{L}^{(2,1)} - a_{R}^{(2,1)}$ would require knowledge of the non-global structure of the two-loop trijet hemisphere soft function. This is analogous to
how the 2-loop soft constant for the dijet heavy jet mass distribution depends on the non-global structure of the hemisphere soft function. The difference is that for the shoulder case the constant is different for the left shoulder $r>0$ and the right shoulder $r<0$.
In Fourier space, the terms multiplied by $a_{L,R}^{(2,1)}$ and $a_{L,R}^{(2,0)}$ become
\begin{multline}
    \widetilde{\sigma}(z) \supset
    - \left(\frac{\alpha_s}{4\pi}\right)^2
     \left(a_L^{(2,1)} +a_R^{(2,1)} \right) \ln (e^{\gamma_E} |z|)
  \\ 
 + 
    \left(\frac{\alpha_s}{4\pi}\right)^2\left(a_L^{(2,1)} - a_R^{(2,1)} \right)\frac{ i \pi}{2} \text{sign}(z) 
    + \left(\frac{\alpha_s}{4\pi}\right)^2 \left(a_L^{(2,0)} + a_R^{(2,0)}\right) \label{nglconst}
\,.
\end{multline}
In contrast, for heavy jet mass in the dijet limit we get a Laplace transform of the form
\begin{equation}
    \mathcal{L}(\nu) \sim 
     \left(\frac{\alpha_s}{4\pi}\right)^2
     \left(b^{(2,1)} \ln \nu  + b^{(2,0)} \right)
   \,,
\end{equation}
with $b^{(2,0)} = c_{2\rho}$ in the notation of~\cite{Chien:2010kc}. For the dijet limit, the $b^{(2,1)}$ term contributes $\alpha_s^2 L$ terms in the heavy jet mass cumulant, while the non-global $b^{(2,0)}$ piece only $\alpha_s^2$ times a constant. In the shoulder case, both terms from the second line of Eq.~\eqref{nglconst} are ``constants", in that their derivative with respect to $z$ vanishes almost everywhere. A full non-global structure calculation would be required to determine them. In this way, the non-global structure is needed to predict the difference between the $\alpha_s^2 L$ terms on the left and right shoulder, while it is only need to predict the $\alpha_s^2 L^0$ term in the dijet limit. We stress that despite the scaling $\sim \alpha_s^2 L$ of this difference term in momentum space, the logarithmic
counting for the Sudakov shoulder region must be done in position space for consistency. In position space, the extra terms in consideration are constants and thus formally N$^3$LL.

\newpage
\addcontentsline{toc}{section}{References}
\bibliographystyle{jhep}
\bibliography{refs}

\end{document}